\documentclass[12pt,reqno]{amsart}
\usepackage{amssymb}
\usepackage{graphicx}
\usepackage{amscd}
\usepackage{amsmath}

\allowdisplaybreaks
\newtheorem{theorem}{Theorem}
\theoremstyle{plain}

\numberwithin{equation}{section}

\input{tcilatex}

\begin{document}
\title[The Hahn Polynomials in the Coulomb Problems]{The Hahn Polynomials in
the Nonrelativistic and Relativistic Coulomb Problems}
\author{Sergei K. Suslov}
\address{Department of Mathematics and Statistics\\
Arizona State University\\
Tempe, AZ 85287, U.S.A.}
\email{sks@asu.edu}
\urladdr{http://hahn.la.asu.edu/\symbol{126}suslov/index.html}
\author{Benjamin Trey}
\address{Department of Physics and Astronomy\\
University of Hawaii at Manoa\\
2505 Correa Rd.\\
Honolulu, Hawaii 96822, U.S.A.}
\email{btrey@hawaii.edu}%
\date{\today }
\dedicatory{Dedicated to the memory of Arnold~F.~Nikiforov and
Vasilii~B.~Uvarov}

\begin{abstract}
We derive closed formulas for mean values of all powers of $r$ in
nonrelativistic and relativistic Coulomb problems in terms of the Hahn and
Chebyshev polynomials of a discrete variable. A short review on special
functions and solution of the Coulomb problems in quantum mechanics is given.
\end{abstract}

\maketitle


\section{Introduction}

A basic problem in quantum theory of the atom is the problem of finding
solutions of the nonrelativistic Schr\"{o}dinger and relativistic Dirac wave
equations for the motion of electron in a central attractive force field.
The only atom for which these equations can be solved explicitly is the
simplest hydrogen atom, or, in general, the one electron hydrogen-like
ionized atom with the charge of the nucleus $Ze;$ this is a classical
problem in quantum mechanics which is studied in great detail; see, for
example, \cite{Akh:Ber}, \cite{Be:Sal}, \cite{Ber:Lif:Pit}, \cite{Dav}, \cite%
{Fermi}, \cite{It:Zu}, \cite{La:Lif}, \cite{Mes}, \cite{Schiff} and
references therein. Comparison of the results of theoretical calculations
with experimental data provides accurate tests of the validity of the
quantum electrodynamics \cite{Be:Sal}, \cite{It:Zu}, \cite{Wein}. Explicit
analytical solutions for hydrogen-like atoms can be useful as the starting
point in approximate calculations of more sophisticated quantum-mechanical
systems.\smallskip

The Schr\"{o}dinger equation for the hydrogen atom can be solved explicitly
in the spherical and parabolic system of coordinates \cite{Be:Sal}, \cite%
{La:Lif} and it can be shown that the Hahn polynomials connect the wave
functions derived in the spherical and parabolic coordinates; see \cite{Su1}
and references therein for more details. In the present paper we discuss
another connection with the classical polynomials --- we derive closed
formulas for the mean values%
\begin{equation}
\left\langle r^{p}\right\rangle =\dint_{\mathbf{R}^{3}}\psi ^{\dagger }\
r^{p}\psi \ dv  \label{int1}
\end{equation}%
of all admissible powers of $r$ for both nonrelativistic and relativistic
Coulomb wave functions in the spherical coordinates in terms of the Hahn and
Chebyshev polynomials of a discrete variable; see, for example, \cite{An:As}%
, \cite{An:As:Ro}, \cite{Askey}, \cite{AskeyCH}, \cite{As:WiCH}, \cite{As:Wi}%
, \cite{At:SusCH}, \cite{Erd}, \cite{Ga:Ra}, \cite{Ko:Sw}, \cite{KoelinkCH},
\cite{Ni:Su:Uv}, \cite{Ni:Su:UvDAH}, \cite{Ni:Uv}, \cite{Sze}, and \cite%
{Chebyshev59}, \cite{Chebyshev64}, \cite{Chebyshev75} for an introduction to
the theory of the classical orthogonal polynomials. Special cases $p=0,$ $1,$
$2$ of (\ref{int1}) give the normalization of the wave functions, the
average distance between the electron and the nucleus, and the mean square
deviation of the nucleus-electron separation, respectively. Special cases $%
p=-1,$ $-2,$ $-3$ of these matrix elements are important in calculations of
the energy levels by the virial theorem, the fine structure of the energy
levels in Pauli's theory of the spin, radiative corrections and Lamb shift
in hydrogen-like atoms; see \cite{Akh:Ber}, \cite{Be:Sal}, \cite{Ber:Lif:Pit}%
, \cite{It:Zu}, \cite{La:Lif}, \cite{Mes}, \cite{Wein} for more details. The
general formulas for the corresponding relativistic Coulomb matrix elements
may also be important in developing the theory of\ spectra of heavy ions for
large values of $Z$ on the basis of the methods of quantum field theory \cite%
{Bj:Dr}, \cite{Bo:Shi}, \cite{Ticciati}.\smallskip

It is worth noting that exact solutions of certain time-dependent Schr\"{o}%
dinger equations are found in \cite{Lop:Sus} and \cite{Me:Co:Su}. An
interesting area of research in physics in general is a problem of
discretization of the space--time continuum and the concept of the
fundamental length \cite{Haw:Pen}, \cite{Penrose} and, in particular, the
discretization of the classical Maxwell, Schr\"{o}dinger and Dirac
equations; see \cite{Ey:Ruf:Su} and \cite{Su2} for some solutions of the
discrete wave, Maxwell and Dirac equations.\smallskip

The paper is organized as follows. In the next section, among other things,
we evaluate an integral of the product of two Laguerre polynomials in terms
of the Hahn polynomials, which gives a \textquotedblleft master
formula\textquotedblright\ for evaluation of the matrix elements (\ref{int1}%
) for the nonrelativistic and relativistic hydrogen-like atoms in sections 3
and 4, respectively. Some special cases are\ given explicitly and evaluation
of the effective electrostatic potential in the hydrogen-like atoms is
discussed as an application. Sections~5 and 6 are written in order to make
our presentation as self-contained as possible --- they contain a short
review of Nikiforov and Uvarov's approach to the theory of special functions
of mathematical physics and a detailed solution of the wave equation of
Dirac for Coulomb potential, respectively. We follow \cite{Ni:Uv} with
somewhat different details; for example, in Section~5 we give a different
proof of the main integral representation for the special functions of
hypergeometric type and discuss the power series method; in Section~6 we
construct the spinor spherical harmonics and separate the variables in the
spherical coordinates in detail before solving the radial equations. In
Section~7 we discuss a more general version of the method of separation of
the variables for Dirac-type systems. The relativistic Coulomb wave
functions are not well known for a \textquotedblleft general
audience\textquotedblright\ and this discussion might be useful for the
reader who is not an expert in theoretical physics; our paper is written for
those who study quantum mechanics and would like to see more details than in
the classical textbooks \cite{Akh:Ber}, \cite{Be:Sal}, \cite{Ber:Lif:Pit},
\cite{La:Lif}; it is motivated by a course in quantum mechanics which one of
the authors (SKS) has been teaching at Arizona State University for many
years. Appendix contains some formulas which are widely used throughout the
paper.\smallskip

We use the absolute cgs system of units throughout the paper in order to
make the corresponding nonrelativistic limits as explicit as possible. The
fundamental constants are speed of light in vacuum $c=2.99792458\times
10^{10}~\mathrm{cm}\unit{s}^{-1},$ Planck's constant $\hbar =6.6260755\times
10^{-34}\unit{J}\unit{s},$ mass $m=m_{e}=9.1093897\times 10^{-28}~\mathrm{gm}
$ and electric charge $e=\left\vert e\right\vert =1.60217733\times 10^{-19}%
\unit{C}$ of the electron, Bohr radius $a_{0}=\hbar
^{2}/me^{2}=0.529177249\times 10^{-8}~\mathrm{cm},$ Sommerfeld's fine
structure constant $\alpha =e^{2}/\hbar c=7.29735308\times 10^{-3},$ and
Compton wave length $\lambda /2\pi =\hbar /mc=2.42631058\times 10^{-10}~%
\mathrm{cm}.$\smallskip

We dedicate this paper to the memory of Professors A.~F.~Nikiforov and
V.~B.~Uvarov in a hope that those masters would appreciate our effort to
make their method more available for the beginners.

\section{Some Integrals of the Products of Laguerre Polynomials}

\subsection{Evaluation of an Integral}

Let us compute the following integral%
\begin{equation}
J_{nms}^{\alpha \beta }=\int_{0}^{\infty }e^{-x}x^{\alpha +s}L_{n}^{\alpha
}\left( x\right) L_{m}^{\beta }\left( x\right) \ dx,  \label{i1}
\end{equation}%
where $n\geq m$ and $\alpha -\beta =0,\pm 1,\pm 2,...\ .$ Similar integrals
were evaluated in \cite{Be:Sal}, \cite{Davis} and \cite{La:Lif}, see also
references therein, but an important relation with the Hahn polynomials
seems to be missing.\smallskip

It is convenient to assume at the beginning that parameter $s$ takes some
continuous values such that $\alpha +s>-1$ for convergence of the integral.
Using the Rodrigues formula for the Laguerre polynomials \cite{Ni:Su:Uv},
\cite{Ni:Uv}, \cite{Sze}%
\begin{equation}
L_{n}^{\alpha }\left( x\right) =\frac{1}{n!}e^{x}x^{-\alpha }\left(
x^{\alpha +n}e^{-x}\right) ^{\left( n\right) },  \label{i3}
\end{equation}%
see the proof in Section~5 of the present paper, and integrating by parts%
\begin{eqnarray*}
J_{nms}^{\alpha \beta } &=&\frac{1}{n!}\int_{0}^{\infty }\left( x^{\alpha
+n}e^{-x}\right) ^{\left( n\right) }\left( x^{s}L_{m}^{\beta }\left(
x\right) \right) \ dx \\
&=&\frac{1}{n!}\left. \left( \left( x^{\alpha +n}e^{-x}\right) ^{\left(
n-1\right) }\left( x^{s}L_{m}^{\beta }\left( x\right) \right) \right)
\right\vert _{0}^{\infty }-\frac{1}{n!}\int_{0}^{\infty }\left( x^{\alpha
+n}e^{-x}\right) ^{\left( n-1\right) }\left( x^{s}L_{m}^{\beta }\left(
x\right) \right) ^{\prime }\ dx \\
&&\vdots \\
&=&\frac{\left( -1\right) ^{n}}{n!}\int_{0}^{\infty }\left( x^{\alpha
+n}e^{-x}\right) \left( x^{s}L_{m}^{\beta }\left( x\right) \right) ^{\left(
n\right) }\ dx.
\end{eqnarray*}%
But, in view of (\ref{a2}),%
\begin{eqnarray}
&&\left( x^{s}L_{m}^{\beta }\left( x\right) \right) ^{\left( n\right) }=%
\frac{\Gamma \left( \beta +m+1\right) }{m!\;\Gamma \left( \beta +1\right) }%
\sum_{k}\frac{\left( -m\right) _{k}}{k!\left( \beta +1\right) _{k}}\left(
x^{k+s}\right) ^{\left( n\right) }  \label{ider} \\
&&\quad =\frac{\Gamma \left( \beta +m+1\right) \Gamma \left( s+1\right) }{%
m!\;\Gamma \left( \beta +1\right) \Gamma \left( s-n+1\right) }\sum_{k}\frac{%
\left( -m\right) _{k}\left( s+1\right) _{k}}{k!\left( \beta +1\right)
_{k}\left( s-n+1\right) _{k}}\ x^{k+s-n}  \notag
\end{eqnarray}%
and with the help of Euler's integral representation for the gamma function
\cite{An:As:Ro}, \cite{Ni:Uv}%
\begin{equation*}
\int_{0}^{\infty }x^{\alpha +k+s}e^{-x}\ dx=\Gamma \left( \alpha
+k+s+1\right) =\left( \alpha +s+1\right) _{k}\Gamma \left( \alpha
+s+1\right) ,
\end{equation*}%
see also (\ref{gamma1}) below, one gets%
\begin{eqnarray}
J_{nms}^{\alpha \beta } &=&\left( -1\right) ^{n}\frac{\Gamma \left( \alpha
+s+1\right) \Gamma \left( \beta +m+1\right) \Gamma \left( s+1\right) }{%
n!\;m!\;\Gamma \left( \beta +1\right) \Gamma \left( s-n+1\right) }  \notag \\
&&\times ~_{3}F_{2}\left(
\begin{array}{c}
-m,\ s+1,\ \alpha +s\medskip +1 \\
\beta +1,\quad s-n+1%
\end{array}%
\right) .  \label{i4}
\end{eqnarray}%
See\ \cite{Ba} or equation (\ref{a1}) below for the definition of the
generalized hypergeometric series $_{3}F_{2}\left( 1\right) .$ Thomae's
transformation (\ref{a4}), see also \cite{Ba} or \cite{Ga:Ra}, results in%
\begin{align}
J_{nms}^{\alpha \beta }& =\int_{0}^{\infty }e^{-x}x^{\alpha +s}\
L_{n}^{\alpha }\left( x\right) L_{m}^{\beta }\left( x\right) \ dx  \label{i2}
\\
& =\left( -1\right) ^{n-m}\frac{\Gamma \left( \alpha +s+1\right) \Gamma
\left( \beta +m+1\right) \Gamma \left( s+1\right) }{m!\left( n-m\right)
!\;\Gamma \left( \beta +1\right) \Gamma \left( s-n+m+1\right) }  \notag \\
& \quad \times \ ~_{3}F_{2}\left(
\begin{array}{c}
-m,\ s+1,\ \beta -\alpha -s\medskip \\
\beta +1,\quad n-m+1%
\end{array}%
\right) ,\quad n\geq m,  \notag
\end{align}%
where parameter $s$ may take some integer values. This establishes a
connection with the Hahn polynomials given by equation (\ref{a3}) below; one
can also rewrite this integral in terms of the dual Hahn polynomials \cite%
{Ni:Su:Uv}.

\subsection{Special Cases}

Letting $s=0$ and $\alpha =\beta $ in (\ref{i2}) results in the
orthogonality relation for the Laguerre polynomials. Two special cases%
\begin{equation}
I_{1}=J_{nn1}^{\alpha \alpha }=\int_{0}^{\infty }e^{-x}x^{\alpha +1}\left(
L_{n}^{\alpha }\left( x\right) \right) ^{2}\ dx=\left( \alpha +2n+1\right)
\frac{\Gamma \left( \alpha +n+1\right) }{n!}  \label{i5}
\end{equation}%
and%
\begin{equation}
I_{2}=J_{n,\ n-1,\ 2}^{\alpha -2,\ \alpha }=\int_{0}^{\infty
}e^{-x}x^{\alpha }L_{n-1}^{\alpha }\left( x\right) L_{n}^{\alpha -2}\left(
x\right) \ dx=-2\frac{\Gamma \left( \alpha +n\right) }{\left( n-1\right) !}
\label{i6}
\end{equation}%
are convenient for normalization of the wave functions of the discrete
spectra in the nonrelativistic and relativistic Coulomb problems \cite%
{Be:Sal}, \cite{Ni:Uv}.\medskip

Two other special cases of a particular interest in this paper are%
\begin{eqnarray}
&&J_{k}=J_{nnk}^{\alpha \alpha }=\int_{0}^{\infty }e^{-x}x^{\alpha +k}\
\left( L_{n}^{\alpha }\left( x\right) \right) ^{2}\ dx  \label{i7} \\
&&\quad =\frac{\Gamma \left( \alpha +k+1\right) \Gamma \left( \alpha
+n+1\right) }{n!\;\Gamma \left( \alpha +1\right) }\ ~_{3}F_{2}\left(
\begin{array}{c}
-k,\ k+1,\ -n\medskip \\
1,\quad \alpha +1%
\end{array}%
\right)  \notag
\end{eqnarray}%
and%
\begin{eqnarray}
&&J_{-k-1}=J_{nn,\ -k-1}^{\alpha \alpha }=\int_{0}^{\infty }e^{-x}x^{\alpha
-k-1}\ \left( L_{n}^{\alpha }\left( x\right) \right) ^{2}\ dx  \label{i8} \\
&&\quad \quad \ \ =\frac{\Gamma \left( \alpha -k\right) \Gamma \left( \alpha
+n+1\right) }{n!\;\Gamma \left( \alpha +1\right) }\ ~_{3}F_{2}\left(
\begin{array}{c}
-k,\ k+1,\ -n\medskip \\
1,\quad \alpha +1%
\end{array}%
\right) .  \notag
\end{eqnarray}%
The Chebyshev polynomials of a discrete variable $t_{k}\left( x\right) $ are
special case of the Hahn polynomials $t_{k}\left( x,N\right) =h_{k}^{\left(
0,\ 0\right) }\left( x,N\right) $ \cite{Chebyshev59}, \cite{Chebyshev64} and
\cite{Chebyshev75}. Thus from (\ref{i7})--(\ref{i8}) and (\ref{a3}) one
finally gets%
\begin{eqnarray}
J_{k} &=&J_{nnk}^{\alpha \alpha }=\int_{0}^{\infty }e^{-x}x^{\alpha +k}\
\left( L_{n}^{\alpha }\left( x\right) \right) ^{2}\ dx  \label{i9} \\
&=&\frac{\Gamma \left( \alpha +n+1\right) }{n!}\ t_{k}\left( n,-\alpha
\right)  \notag
\end{eqnarray}%
and%
\begin{eqnarray}
J_{-k-1} &=&J_{nn,\ -k-1}^{\alpha \alpha }=\int_{0}^{\infty }e^{-x}x^{\alpha
-k-1}\ \left( L_{n}^{\alpha }\left( x\right) \right) ^{2}\ dx  \label{i10} \\
&=&\frac{\Gamma \left( \alpha -k\right) \Gamma \left( \alpha +n+1\right) }{%
n!\;\Gamma \left( \alpha +k+1\right) }\ t_{k}\left( n,-\alpha \right)  \notag
\end{eqnarray}%
for $0\leq k<\alpha .$ One can see that the positivity of these integrals is
related to a nonstandard orthogonality relation for the corresponding
Chebyshev polynomials of a discrete variable $t_{k}\left( x,N\right) $ when
the parameter takes negative integer values $N=-\alpha .$ Indeed, according
to the method of \cite{Ni:Su:Uv} and \cite{Ni:Uv}, these polynomials are
orthogonal with the discrete uniform distribution on the interval $\left[
-\alpha ,-1\right] $ which contains all their zeros and, therefore, they are
positive for all nonnegative values of their argument. The explicit
representation (\ref{a3}) gives also a positive sum for all positive $x$ and
negative $N.$

\subsection{Connection Relation and Linearization Formula}

The connection relation \cite{An:As:Ro}, \cite{Askey}%
\begin{equation}
L_{n}^{\alpha }\left( x\right) =\sum_{m=0}^{n}\frac{\left( \alpha -\beta
\right) _{n-m}}{\left( n-m\right) !}\;L_{m}^{\beta }\left( x\right)
\label{i10a}
\end{equation}%
is an easy consequence of the integral (\ref{i2}), we leave the details to
the reader.\smallskip

A linearization formula%
\begin{equation}
L_{n}^{\alpha }\left( x\right) L_{m}^{\beta }\left( x\right)
=\sum_{p=0}^{n+m}c_{nmp}\left( \alpha ,\beta ,\gamma \right) \;L_{p}^{\gamma
}\left( x\right)  \label{lf1}
\end{equation}%
gives a product of two Laguerre polynomials as a linear combination of other
polynomials of the same kind; see \cite{An:As:Ro}, \cite{Askey}, \cite{As:Wi}%
, \cite{Rahman81a}, \cite{Rahman81b} and references therein for a review on
the linearization of products of classical orthogonal polynomials. In view
of the orthogonality relation, the corresponding linearization coefficients
are given by%
\begin{equation}
\frac{\Gamma \left( \gamma +p+1\right) }{p!}\;c_{nmp}\left( \alpha ,\beta
,\gamma \right) =\int_{0}^{\infty }L_{n}^{\alpha }\left( x\right)
L_{m}^{\beta }\left( x\right) L_{p}^{\gamma }\left( x\right) x^{\gamma
}e^{-x}\;dx,  \label{lf2}
\end{equation}%
where one can assume that $n\geq m$ and use the expansion%
\begin{equation}
L_{p}^{\gamma }\left( x\right) =\frac{\Gamma \left( \gamma +p+1\right) }{%
p!\;\Gamma \left( \gamma +1\right) }\ \sum_{k=0}^{p}\frac{\left( -p\right)
_{k}}{k!\left( \gamma +1\right) _{k}}\;x^{k}  \label{lf3}
\end{equation}%
by (\ref{a2}). Then the integral of the product of three Laguerre
polynomials is%
\begin{eqnarray}
c_{nmp}\left( \alpha ,\beta ,\gamma \right) &=&\frac{1}{\Gamma \left( \gamma
+1\right) }\ \sum_{k=0}^{p}\frac{\left( -p\right) _{k}}{k!\left( \gamma
+1\right) _{k}}\;\int_{0}^{\infty }e^{-x}x^{\gamma +k}L_{n}^{\alpha }\left(
x\right) L_{m}^{\beta }\left( x\right) \;dx  \notag \\
&=&\frac{1}{\Gamma \left( \gamma +1\right) }\ \sum_{k=0}^{p}\frac{\left(
-p\right) _{k}}{k!\left( \gamma +1\right) _{k}}\;J_{nm,\;\gamma -\alpha
+k}^{\alpha \beta }  \label{lf4}
\end{eqnarray}%
and the remaining integral can be evaluated in terms of the Hahn or dual
Hahn polynomials with the aid of (\ref{i2}). Positivity of the linearization
coefficients is related to the orthogonality property of these
polynomials.\smallskip

On the other hand, in the most important special case,%
\begin{equation}
L_{n}^{\alpha }\left( x\right) L_{m}^{\alpha }\left( x\right)
=\sum_{p=n-m\geq 0}^{n+m}c_{nmp}\;L_{p}^{\alpha }\left( x\right)  \label{lf5}
\end{equation}%
with $\alpha =\beta =\gamma ,$ the linearization coeffitients $c_{nmp}\left(
\alpha \right) =c_{nmp}\left( \alpha ,\alpha ,\alpha \right) $ can be found
as a single sum%
\begin{equation}
c_{nmp}=\left( -1\right) ^{p}\frac{\left( -p\right) _{n-m}\left( \alpha
+1\right) _{m}}{p!\;m!}\;\sum_{k}\frac{\left( -p\right) _{k}\left( -m\right)
_{k}\;\Gamma \left( 2k+1\right) }{k!\left( \alpha +1\right) _{k}\;\Gamma
\left( n-m-p+2k+1\right) }.  \label{lf6}
\end{equation}%
The summation is to be taken over all integer values of $k$ such that $0\leq
p-n+m\leq 2k\leq \min \left( 2p,2m\right) .$\smallskip

Indeed, substituting (\ref{i2}) into (\ref{lf4})%
\begin{eqnarray}
\Gamma \left( \alpha +1\right) \;c_{nmp} &=&\left( -1\right) ^{n-m}\frac{%
\Gamma \left( \alpha +m+1\right) }{m!\;\left( n-m\right) !}\;\sum_{s=n-m\geq
0}^{p}\frac{\left( -p\right) _{s}}{\Gamma \left( s-n+m+1\right) }  \notag \\
&&\times \sum_{k=0}^{\min \left( m,s\right) }\frac{\left( -m\right)
_{k}\left( -s\right) _{k}\left( s+1\right) _{k}}{k!\left( \alpha +1\right)
_{k}\left( n-m+1\right) _{k}}  \label{lf7}
\end{eqnarray}%
with $\alpha =\beta =\gamma ,$ then replacing the index $s=n-m+l,$ $0\leq
l\leq p-n+m$ and changing the order of summation one gets%
\begin{eqnarray}
\Gamma \left( \alpha +1\right) \;c_{nmp} &=&\left( -1\right) ^{n-m}\frac{%
\Gamma \left( \alpha +m+1\right) }{m!\;\left( n-m\right) !}\;\left(
-p\right) _{n-m}  \notag \\
&&\times \sum_{k}\frac{\left( -m\right) _{k}\left( -n+m\right) _{k}}{%
k!\left( \alpha +1\right) _{k}}  \notag \\
&&\times \sum_{l}\frac{\left( -p+n-m\right) _{l}\left( n-m+k+1\right) _{l}}{%
l!\left( n-m-k+1\right) _{l}}  \label{lf8}
\end{eqnarray}%
with the help of%
\begin{eqnarray*}
\left( -p\right) _{n-m+l} &=&\left( -p\right) _{n-m}\left( -p+n-m\right)
_{l}, \\
\left( -n+m-l\right) _{k} &=&\left( -n+m\right) _{k}\frac{\left(
n-m+1\right) _{l}}{\left( n-m-k+1\right) _{l}}, \\
\left( n-m+l+1\right) _{k} &=&\left( n-m+1\right) _{k}\frac{\left(
n-m+k+1\right) _{l}}{\left( n-m+1\right) _{l}}.
\end{eqnarray*}%
The$\;_{2}F_{1}\left( 1\right) $ function in (\ref{lf8}) is evaluated by a
limining case of the Gauss summation formula (\ref{a4a}) as%
\begin{eqnarray}
&&\left( -n+m\right) _{k}\ _{2}F_{1}\left(
\begin{array}{c}
-p+n-m\medskip ,\ n-m+k+1 \\
n-m-k+1\medskip%
\end{array}%
;\ 1\right)  \label{lf9} \\
&&\qquad =\left( -1\right) ^{n-m-p}\left( -p\right) _{k}\frac{\left(
n-m\right) !\;\Gamma \left( 2k+1\right) }{p!\;\Gamma \left(
n-m-p+2k+1\right) }.  \notag
\end{eqnarray}%
This results in (\ref{lf6}) and our proof is complete.\smallskip \smallskip

Equations (\ref{lf2}) and (\ref{lf6}) imply the following positivity property%
\begin{equation}
\left( -1\right) ^{m+n+p}\int_{0}^{\infty }L_{n}^{\alpha }\left( x\right)
L_{m}^{\alpha }\left( x\right) L_{p}^{\alpha }\left( x\right) x^{\alpha
}e^{-x}\;dx\geq 0,  \label{lf10}
\end{equation}%
when $\alpha >-1$ and $m,n,p=0,1,2,\;...\;;$ see problem~33 on p.~400 of
\cite{An:As:Ro}.\smallskip

The single sum in (\ref{lf6}) can be rewritten as a $_{4}F_{3}.$ There are
distinct representations for even and odd values of $\epsilon =\left(
p-n+m\right) /2.$ When $\epsilon =0$ the summation formula (\ref{a4a}) gives%
\begin{equation}
c_{nm,\;n-m}\left( \alpha \right) =\frac{\Gamma \left( \alpha +n+1\right) }{%
m!\;\Gamma \left( \alpha +n-m+1\right) }  \label{lf11}
\end{equation}%
and the case $n=m$ corresponds to correct value of the squared norm of the
Laguerre polynomials. If $\epsilon =\left( p-n+m\right) /2$ is a positive
even number the result is%
\begin{eqnarray}
c_{nmp}\left( \alpha \right)  &=&\left( -1\right) ^{p}\frac{\left( -p\right)
_{n-m}\left( -p\right) _{\epsilon }\left( -m\right) _{\epsilon }}{p!\;m!}\;%
\frac{\Gamma \left( 2\epsilon +1\right) \Gamma \left( \alpha +m+1\right) }{%
\Gamma \left( \epsilon \right) \Gamma \left( \alpha +\epsilon +1\right) }
\notag \\
&&\times \;_{4}F_{3}\left(
\begin{array}{c}
-p+\epsilon \medskip ,\ -m+\epsilon ,\ \epsilon +1/2,\ \epsilon +1 \\
1/2\medskip ,\ \ \alpha +\epsilon +1\medskip ,\ \ \epsilon
\end{array}%
\right)   \label{lf12}
\end{eqnarray}%
and%
\begin{eqnarray}
c_{nmp}\left( \alpha \right)  &=&\left( -1\right) ^{p}\frac{\left( -p\right)
_{n-m}\left( -p\right) _{\epsilon +1/2}\left( -m\right) _{\epsilon +1/2}}{%
p!\;m!}\;\frac{\Gamma \left( 2\epsilon +2\right) \Gamma \left( \alpha
+m+1\right) }{\Gamma \left( \epsilon +1/2\right) \Gamma \left( \alpha
+\epsilon +3/2\right) }  \notag \\
&&\times \;_{4}F_{3}\left(
\begin{array}{c}
-p+\epsilon +1/2\medskip ,\ -m+\epsilon +1/2,\ \epsilon +1,\ \epsilon +3/2
\\
3/2\medskip ,\ \ \alpha +\epsilon +3/2\medskip ,\ \ \epsilon +1/2%
\end{array}%
\right) ,  \label{lf13}
\end{eqnarray}%
if $\epsilon =\left( p-n+m\right) /2$ is an odd.\smallskip

Although we have not been able to find the linearization formula for
Laguerre polynomials in explicit form in the literature, a closed expression
for the linearization coefficients should follow as a limiting case of
equations (1.7)--(1.8) for the linearization coefficients of the Jacobi
polynomials in Rahman's paper \cite{Rahman81a}; see also \cite{Rahman81b}
for $q$-extension of his result. Dick Askey has told us that he had computed
the corresponding integral of the product of three Laguerre polynomials with
the help of the generating function for these polynomials as a different
single sum. In his opinion, this may have been found in the 1930's by
Erdelyi or someone else.

\subsection{An Extension}

The integral (\ref{i1}) has a somewhat useful extension%
\begin{equation}
J_{nms}^{\alpha \beta }\left( z\right) =\int_{z}^{\infty }e^{-x}x^{\alpha
+s}L_{n}^{\alpha }\left( x\right) L_{m}^{\beta }\left( x\right) \ dx,
\label{i11}
\end{equation}%
where $n\geq m$ and $\alpha -\beta =0,\pm 1,\pm 2,...\ $and $J_{nms}^{\alpha
\beta }\left( 0\right) =J_{nms}^{\alpha \beta }.$ This integral is evaluated
in the following fashion. Upon integrating by parts%
\begin{eqnarray}
J_{nms}^{\alpha \beta }\left( z\right) &=&e^{-z}\sum_{k=1}^{l}\left(
-1\right) ^{k}\frac{\left( n-k\right) !}{n!}\;z^{\alpha +k}L_{n-k}^{\alpha
+k}\left( z\right) \left( z^{s}L_{m}^{\beta }\left( z\right) \right)
^{\left( k-1\right) }  \label{i11a} \\
&&+\;\frac{\left( -1\right) ^{l}}{n!}\int_{z}^{\infty }\left( x^{\alpha
+n}e^{-x}\right) ^{\left( n-l\right) }\left( x^{s}L_{m}^{\beta }\left(
x\right) \right) ^{\left( l\right) }\ dx  \notag
\end{eqnarray}%
and putting $l=n$ we use (\ref{ider}) and the integral representation for
incomplete gamma function \cite{Erd}, \cite{Ni:Uv}:%
\begin{equation}
\Gamma \left( \alpha ,z\right) =\int_{z}^{\infty }e^{-x}x^{\alpha -1}\
dx,\qquad \text{Re\ }\alpha >0.  \label{i12}
\end{equation}%
As a result%
\begin{eqnarray}
J_{nms}^{\alpha \beta }\left( z\right) &=&\int_{z}^{\infty }e^{-x}x^{\alpha
+s}L_{n}^{\alpha }\left( x\right) L_{m}^{\beta }\left( x\right) \ dx
\label{i13} \\
&=&z^{\alpha }e^{-z}\sum_{k=1}^{n}\left( -1\right) ^{k}\frac{\left(
n-k\right) !}{n!}\;z^{k}L_{n-k}^{\alpha +k}\left( z\right) \left(
z^{s}L_{m}^{\beta }\left( z\right) \right) ^{\left( k-1\right) }  \notag \\
&&+\left( -1\right) ^{n}\frac{\left( \beta +1\right) _{m}}{n!\;m!}%
\sum_{k=0}^{m}\frac{\left( -m\right) _{k}\Gamma \left( s+k+1\right) }{%
k!\left( \beta +1\right) _{k}\Gamma \left( s-n+k+1\right) }\ \Gamma \left(
\alpha +k+s+1,z\right)  \notag
\end{eqnarray}%
with $n\geq m$ and Re\ $\left( \alpha +s\right) >-1.$ Here one can use the
standard relations \cite{Erd}, \cite{Ni:Uv}:%
\begin{eqnarray}
\Gamma \left( \alpha ,z\right) &=&e^{-z}G\left( 1-\alpha ,1-\alpha ,z\right)
\label{i13a} \\
&=&\Gamma \left( \alpha \right) -\frac{1}{\alpha }\;z^{\alpha }F\left(
\alpha ,1+\alpha ,-z\right)  \notag \\
&=&\Gamma \left( \alpha \right) -z^{\alpha }\sum_{k=0}^{\infty }\frac{\left(
-1\right) ^{k}z^{k}}{k!\left( \alpha +k\right) },\qquad \text{Re\ }\alpha >0
\notag
\end{eqnarray}%
with the confluent hypergeometric functions $F\left( \alpha ,\gamma
,z\right) $ and $G\left( \alpha ,\gamma ,z\right) ,$ respectively. The last
expression gives the asymptotic of the incomplete gamma function as $%
z\rightarrow 0$ and $\left| \arg \left( z\right) \right| <\pi .$ The
asymptotic for large values of $z$ is%
\begin{eqnarray}
\Gamma \left( \alpha ,z\right) &=&e^{-z}z^{\alpha }G\left( 1,1+\alpha
,z\right)  \label{i13b} \\
&=&e^{-z}z^{\alpha -1}\left( \sum_{k=0}^{n-1}\left( -1\right) ^{k}\frac{%
\left( 1-\alpha \right) _{k}}{z^{k}}+\text{O}\left( \frac{1}{z^{n}}\right)
\right)  \notag
\end{eqnarray}%
as $z\rightarrow \infty $ and $\left| \arg \left( z\right) \right| <\pi ;$
see \cite{Erd} and \cite{Ni:Uv}. This allows to find asymptotic expansions
of the integral (\ref{i13}) as $z\rightarrow 0$ and $z\rightarrow \infty .$
Integration by parts in (\ref{i12}) results in the functional relation%
\begin{equation}
\Gamma \left( \alpha +1,z\right) =\alpha \Gamma \left( \alpha ,z\right)
+z^{\alpha -1}e^{-z},  \label{i14}
\end{equation}%
which implies%
\begin{equation}
\Gamma \left( n+1,z\right) =n!\;e^{-z}\sum_{k=0}^{n}\frac{z^{k}}{k!}
\label{i15}
\end{equation}%
for positive integers $n.$

\section{Nonrelativistic Coloumb Problem}

\subsection{Wave Functions and Energy Levels}

The nonrelativistic Coulomb wave functions obtained by the method of
separation of the variables in spherical coordinates, see Section 7.1, are
\begin{equation}
\psi =\psi _{nlm}\left( \mathbf{r}\right) =R_{nl}\left( r\right) \
Y_{lm}\left( \theta ,\varphi \right) ,  \label{nrc1}
\end{equation}%
where $Y_{lm}\left( \theta ,\varphi \right) $ are the spherical harmonics,
the radial functions $R_{nl}\left( r\right) $ are given in terms the
Laguerre polynomials \cite{Be:Sal}, \cite{La:Lif}, \cite{Ni:Uv}, \cite%
{Schiff}%
\begin{equation}
R\left( r\right) =R_{nl}\left( r\right) =\frac{2}{n^{2}}\left( \frac{Z}{a_{0}%
}\right) ^{3/2}\sqrt{\frac{\left( n-l-1\right) !}{\left( n+l\right) !}}\
e^{-\eta /2}\eta ^{l}\ L_{n-l-1}^{2l+1}\left( \eta \right)  \label{nrc2}
\end{equation}%
with%
\begin{equation}
\eta =\frac{2Z}{n}\left( \frac{r}{a_{0}}\right) ,\qquad a_{0}=\dfrac{\hbar
^{2}}{me^{2}}  \label{nrc2a}
\end{equation}%
and the normalization is%
\begin{equation}
\int_{0}^{\infty }R_{nl}^{2}\left( r\right) r^{2}\ dr=1.  \label{nrc2b}
\end{equation}%
Here $n=1,2,3,\ ...\ $ is the principal quantum number of the hydrogen-like
atom in the nonrelativistic Schr\"{o}dinger theory; $l=0,1,\ ...\ ,n-1$ and $%
m=-l,-l+1,\ ...\ ,l-1,l$ are the quantum numbers of the angular momentum and
its projection on the $z$-axis, respectively. The corresponding discrete
energy levels in the cgs units are given by Bohr's formula%
\begin{equation}
E=E_{n}=-\frac{mZ^{2}e^{4}}{2\hbar ^{2}n^{2}},  \label{nrc2c}
\end{equation}%
where $n=1,2,3,\ ...\ $ is the principal quantum number; they do not depend
on the quantum number of the orbital angular momenta $l$ due to a
``hidden''\ $SO\left( 4\right) -$symmetry of the Hamiltonian of the
nonrelativistic hydrogen atom; see, for example, \cite{Dul:McIn}, \cite%
{La:Lif} and references therein and the original paper by Fock \cite{Fock}
and Bargmann \cite{Bargm}.

\subsection{Matrix Elements}

In this section we evaluate the mean values%
\begin{equation}
\left\langle r^{p}\right\rangle =\dfrac{\dint_{\mathbf{R}^{3}}\left\vert
\psi _{nlm}\left( \mathbf{r}\right) \right\vert ^{2}\ r^{p}\ dv}{\dint_{%
\mathbf{R}^{3}}\left\vert \psi _{nlm}\left( \mathbf{r}\right) \right\vert
^{2}\ dv}=\dfrac{\dint_{0}^{\infty }R_{nl}^{2}\left( r\right) r^{p+2}\ dr}{%
\dint_{0}^{\infty }R_{nl}^{2}\left( r\right) r^{2}\ dr},\quad
dv=r^{2}drd\omega   \label{nrc4}
\end{equation}%
in terms of the Chebyshev polynomials of a discrete variable $t_{k}\left(
x,N\right) =h_{k}^{\left( 0,\ 0\right) }\left( x,N\right) $ \cite%
{Chebyshev59}, \cite{Chebyshev64} and \cite{Chebyshev75}. Here we have used
the orthogonality relation for the spherical harmonics \cite{Ni:Uv}, \cite%
{Var:Mos:Kher},%
\begin{equation}
\int_{S^{2}}Y_{lm}^{\ast }\left( \theta ,\varphi \right) Y_{l^{\prime
}m^{\prime }}\left( \theta ,\varphi \right) \ d\omega =\delta _{ll^{\prime
}}\delta _{mm^{\prime }}  \label{scr2}
\end{equation}%
with $d\omega =\sin \theta \ d\theta d\varphi $ and $0\leq \theta \leq \pi
,0\leq \varphi \leq 2\pi .$ The end results are%
\begin{equation}
\left\langle r^{k-1}\right\rangle =\frac{1}{2n}\left( \frac{na_{0}}{2Z}%
\right) ^{k-1}t_{k}\left( n-l-1,-2l-1\right) ,  \label{nrc5}
\end{equation}%
when $k=0,1,2,...$ and%
\begin{equation}
\left\langle \frac{1}{r^{k+2}}\right\rangle =\frac{1}{2n}\left( \frac{2Z}{%
na_{0}}\right) ^{k+2}t_{k}\left( n-l-1,-2l-1\right) ,  \label{nrc6}
\end{equation}%
when $k=0,1,...,\;2l.$\smallskip

Although a connection of the mean values (\ref{nrc4}) with a family of the
hypergeometric polynomials was established by Pasternack \cite{Past}, the
relation with the Chebyshev polynomials of a discrete variable was missing.
This is a curious but fruitful case of \ a ``mistaken identity'' in the
theory of classical orthogonal polynomials. The so-called Hahn polynomials
of a discrete variable were originally introduced by Chebyshev \cite%
{Chebyshev75}, they have a discrete measure on the finite equidistant set of
points. Bateman, in a series of papers \cite{BatCH1}, \cite{BatCH2}, \cite%
{BatCH3}, \cite{BatCH4}, and Hardy \cite{HardCH} were the first who studied
a continuous measure on the entire real line for the simplest special case
of these polynomials of Chebyshev. Pasternack gave an extension of the
results of Bateman to a one parameter family of the continuous orthogonal
polynomials \cite{PastCH}. After investigation of these Bateman--Pasternack
polynomials in the fifties by several authors; see \cite{Touch}, \cite%
{Wy:Mos}, \cite{Braf}, \cite{Carl57} and \cite{Carl58}; Askey and Wilson
\cite{As:WiCH} introduced what nowadays known as the symmetric continuous
Hahn polynomials, they have two free parameters --- but one parameter had
been yet missing! Finally, Suslov \cite{SusCH}, Atakishiyev and Suslov \cite%
{At:SusCH} and Askey \cite{AskeyCH} have introduced the continuous Hahn
polynomials in their full generality in the mid of eighties. More details on
the discovery the continuous Hahn polynomials and their properties are given
in \cite{KoelinkCH} among other things.\smallskip \smallskip

Indeed, in view of the normalization condition of the Coulomb wave functions
(\ref{nrc2b}) one gets%
\begin{eqnarray}
\left\langle r^{p}\right\rangle &=&\dint_{0}^{\infty }R_{nl}^{2}\left(
r\right) r^{p+2}\ dr  \label{nrc6a} \\
&=&\frac{4}{n^{4}}\left( \frac{na_{0}}{2Z}\right) ^{p+3}\left( \frac{Z}{a_{0}%
}\right) ^{3}\frac{\left( n-l-1\right) !}{\left( n+l\right) !}%
\int_{0}^{\infty }e^{-\eta }\eta ^{p+2l+2}\left( L_{n-l-1}^{2l+1}\left( \eta
\right) \right) ^{2}\ d\eta \   \notag
\end{eqnarray}%
and the last integral can be evaluated with the help of (\ref{i9}) or (\ref%
{i10}) giving rise to (\ref{nrc5}) and (\ref{nrc6}), respectively.\smallskip

A convenient ``inversion'' relation for the Coulomb matrix elements,%
\begin{equation}
\left\langle \frac{1}{r^{k+2}}\right\rangle =\left( \frac{2Z}{na_{0}}\right)
^{2k+1}\frac{\left( 2l-k\right) !}{\left( 2l+k+1\right) !}\ \left\langle
r^{k-1}\right\rangle  \label{nrc7}
\end{equation}%
with $0\leq k\leq 2l,$ follows directly from (\ref{nrc5}) and (\ref{nrc6}).
This relation is contained in an implicit form in \cite{La:Lif}, it was
given explicitly in \cite{Past}.

\subsection{Special Cases}

The explicit expression (\ref{nrc5}) for the matrix elements $\left\langle
r^{p}\right\rangle $ and the familiar three term recurrence relation for the
Hahn polynomials $h_{k}^{\left( \alpha ,\ \beta \right) }\left( x,N\right) $
\cite{Ni:Su:Uv}, \cite{Ni:Uv},%
\begin{equation}
xh_{k}^{\left( \alpha ,\ \beta \right) }\left( x,N\right) =\alpha
_{k}h_{k+1}^{\left( \alpha ,\ \beta \right) }\left( x,N\right) +\beta
_{k}h_{k}^{\left( \alpha ,\ \beta \right) }\left( x,N\right) +\gamma
_{k}h_{k-1}^{\left( \alpha ,\ \beta \right) }\left( x,N\right)  \label{nrc7a}
\end{equation}%
with%
\begin{eqnarray*}
\alpha _{k} &=&\frac{\left( n+1\right) \left( \alpha +\beta +n+1\right) }{%
\left( \alpha +\beta +2n+1\right) \left( \alpha +\beta +2n+2\right) }, \\
\beta _{k} &=&\frac{\alpha -\beta +2N-2}{4}+\frac{\left( \beta ^{2}-\alpha
^{2}\right) \left( \alpha +\beta +2N\right) }{4\left( \alpha +\beta
+2n\right) \left( \alpha +\beta +2n+2\right) }, \\
\gamma _{k} &=&\frac{\left( \alpha +n\right) \left( \beta +n\right) \left(
\alpha +\beta +N+n\right) \left( N-n\right) }{\left( \alpha +\beta
+2n\right) \left( \alpha +\beta +2n+1\right) },
\end{eqnarray*}%
imply the following three term recurrence relation for the matrix elements (%
\ref{nrc4}):%
\begin{eqnarray}
\left\langle r^{k}\right\rangle &=&\frac{2n\left( 2k+1\right) }{k+1}\left(
\frac{na_{0}}{2Z}\right) \left\langle r^{k-1}\right\rangle  \label{nrc8} \\
&&-\frac{k\left( \left( 2l+1\right) ^{2}-k^{2}\right) }{k+1}\left( \frac{%
na_{0}}{2Z}\right) ^{2}\left\langle r^{k-2}\right\rangle  \notag
\end{eqnarray}%
with the ``initial conditions''%
\begin{equation}
\left\langle \frac{1}{r}\right\rangle =\frac{Z}{a_{0}n^{2}},\qquad
\left\langle 1\right\rangle =1  \label{nrc9}
\end{equation}%
which is convenient for evaluation of the mean values $\left\langle
r^{k}\right\rangle $ for $k\geq 1$ \cite{Past}. The inversion relation (\ref%
{nrc7}) can be used then for all possible negative values of $k.$ One can
easily find the following matrix elements%
\begin{equation}
\left\langle r\right\rangle =\frac{a_{0}}{2Z}\left( 3n^{2}-l\left(
l+1\right) \right) ,  \label{nrc10}
\end{equation}%
\begin{equation}
\left\langle r^{2}\right\rangle =2\left( \frac{na_{0}}{2Z}\right) ^{2}\left(
5n^{2}+1-3l\left( l+1\right) \right) ,  \label{nrc11}
\end{equation}%
\begin{equation}
\left\langle \frac{1}{r}\right\rangle =\frac{Z}{a_{0}n^{2}},  \label{nrc11a}
\end{equation}%
\begin{equation}
\left\langle \frac{1}{r^{2}}\right\rangle =\frac{2Z^{2}}{a_{0}^{2}n^{3}%
\left( 2l+1\right) },  \label{nrc12}
\end{equation}%
\begin{equation}
\left\langle \frac{1}{r^{3}}\right\rangle =\frac{Z^{3}}{a_{0}^{3}n^{3}\left(
l+1\right) \left( l+1/2\right) l},  \label{nrc13}
\end{equation}%
\begin{equation}
\left\langle \frac{1}{r^{4}}\right\rangle =\frac{Z^{4}\left( 3n^{2}-l\left(
l+1\right) \right) }{2a_{0}^{4}n^{5}\left( l+3/2\right) \left( l+1\right)
\left( l+1/2\right) l\left( l-1/2\right) },  \label{nrc14}
\end{equation}%
which are important in many calculations in quantum mechanics and quantum
electrodynamics \cite{Akh:Ber}, \cite{Be:Sal}, \cite{Ber:Lif:Pit}, \cite%
{It:Zu}, \cite{Wein}; see \cite{Be:Sal} for more examples.\smallskip

Equations (\ref{nrc2c}) and (\ref{nrc11a}) show that the total energy of the
electron in the hydrogen-like atom is equal to half the average potential
energy:%
\begin{equation}
\left\langle U\right\rangle =-Ze^{2}\left\langle \frac{1}{r}\right\rangle =-%
\frac{Z^{2}e^{2}}{a_{0}n^{2}}=2E.  \label{nrc15}
\end{equation}%
This is the statement of so-called virial theorem in nonrelativistic quantum
mechanics; see, for example, \cite{Be:Sal}, p.~165 and \cite{La:Lif}%
.\smallskip

The average distance between the electron and the nucleus $\overline{r}%
=\left\langle r\right\rangle $ is given by (\ref{nrc10}). The mean square
deviation of the nucleus-electron separation is%
\begin{equation}
\overline{\left( r-\overline{r}\right) ^{2}}=\overline{r^{2}}-\left.
\overline{r}\right. ^{2}=\left( \frac{a_{0}}{2Z}\right) ^{2}\left(
n^{2}\left( n^{2}+2\right) -l^{2}\left( l+1\right) ^{2}\right) .
\label{nrc16}
\end{equation}%
The quantum mechanical analogue to Bohr orbits of large eccentricity
corresponds to large values of this number (small $l$).

\subsection{Screening}

Let us evaluate the effective electrostatic potential $V\left( \mathbf{r}%
\right) $ created by motion of the electron in a hydrogen-like atom with the
nucleus of charge $Ze.$ This result is well known in the nonrelativistic Schr%
\"{o}dinger theory --- see, for example, \cite{Hart} and \cite{Ni:Uv} ---
but we emphasize the connection with the Hahn polynomials in order to obtain
similar results in the relativistic Dirac theory in the next section. For
the electron in the stationary state with the wave function (\ref{nrc1}) and
the quantum numbers $n,$ $l$ and $m$ this potential is%
\begin{equation}
V\left( \mathbf{r}\right) =\frac{Ze}{r}-e\int_{\mathbf{R}^{3}}\frac{%
\left\vert \psi _{nlm}\left( \mathbf{r}^{\prime }\right) \right\vert ^{2}}{%
\left\vert \mathbf{r}-\mathbf{r}^{\prime }\right\vert }\ \left( r^{\prime
}\right) ^{2}dr^{\prime }d\omega ^{\prime },  \label{scr1}
\end{equation}%
where $e\rho \left( \mathbf{r}\right) =e\left\vert \psi _{nlm}\left( \mathbf{%
r}\right) \right\vert ^{2}$ is the average charge distribution of the
electron in the atom. In order to evaluate the integral one can use the
generating relation (\ref{a5}) in the form%
\begin{equation}
\frac{1}{\left\vert \mathbf{r}-\mathbf{r}^{\prime }\right\vert }%
=\sum_{s=0}^{\infty }\frac{r_{<}^{s}}{r_{>}^{s+1}}\left( \frac{4\pi }{2s+1}\
\sum_{m^{\prime }=-s}^{s}Y_{sm^{\prime }}^{\ast }\left( \theta ^{\prime
},\varphi ^{\prime }\right) Y_{sm^{\prime }}\left( \theta ,\varphi \right)
\right) ,  \label{scr1a}
\end{equation}%
where $r_{<}=\min \left( r,r^{\prime }\right) $ and $r_{>}=\max \left(
r,r^{\prime }\right) ;$ see \cite{Ni:Uv}, \cite{Rose}, \cite{Var:Mos:Kher}
for the proof of this expansion formula. In view of (\ref{nrc1}) one gets%
\begin{eqnarray}
&&\dint_{\mathbf{R}^{3}}\dfrac{\left\vert \psi _{nlm}\left( \mathbf{r}%
^{\prime }\right) \right\vert ^{2}}{\left\vert \mathbf{r}-\mathbf{r}^{\prime
}\right\vert }\ \left( r^{\prime }\right) ^{2}dr^{\prime }d\omega ^{\prime
}=\sum_{s=0}^{\infty }\frac{4\pi }{2s+1}\dint_{0}^{\infty }\frac{r_{<}^{s}}{%
r_{>}^{s+1}}\ R_{nl}^{2}\left( r^{\prime }\right) \left( r^{\prime }\right)
^{2}dr^{\prime }  \label{scr3} \\
&&\quad \ \times \sum_{m^{\prime }=-s}^{s}Y_{sm^{\prime }}\left( \theta
,\varphi \right) \dint_{S^{2}}Y_{sm^{\prime }}^{\ast }\left( \theta ^{\prime
},\varphi ^{\prime }\right) Y_{lm}^{\ast }\left( \theta ^{\prime },\varphi
^{\prime }\right) Y_{lm}\left( \theta ^{\prime },\varphi ^{\prime }\right) \
d\omega ^{\prime }.  \notag
\end{eqnarray}%
The $\varphi ^{\prime }$ integration reduces the sum over $m^{\prime }$ to a
single term with $m^{\prime }=0$ and we arrive at%
\begin{eqnarray}
V\left( \mathbf{r}\right)  &=&\frac{Ze}{r}-e\sum_{s=0}^{\infty }\frac{4\pi }{%
2s+1}\ \dint_{0}^{\infty }\frac{r_{<}^{s}}{r_{>}^{s+1}}\ R_{nl}^{2}\left(
r^{\prime }\right) \left( r^{\prime }\right) ^{2}dr^{\prime }  \notag \\
&&\times Y_{s0}\left( \theta ,0\right) \ \dint_{S^{2}}Y_{s0}^{\ast }\left(
\theta ^{\prime },\varphi ^{\prime }\right) Y_{lm}^{\ast }\left( \theta
^{\prime },\varphi ^{\prime }\right) Y_{lm}\left( \theta ^{\prime },\varphi
^{\prime }\right) \ d\omega ^{\prime }.  \label{scr4}
\end{eqnarray}%
The integral of the product of three spherical harmonics can be evaluated in
terms of the Clebsch--Gordan coefficients $C_{l_{1}m_{1}l_{2}m_{2}}^{lm}$ of
the quantum theory of angular momentum \cite{Edm}, \cite{Ni:Su:Uv}, \cite%
{Ni:Su:Uv}, \cite{Rose}, \cite{Var:Mos:Kher}, \cite{Vil}, \cite{Wigner} with
the help of the product formula (\ref{a6}) and the orthogonality property (%
\ref{scr2}). The result is%
\begin{equation}
\dint_{S^{2}}Y_{s0}^{\ast }\left( \theta ^{\prime },\varphi ^{\prime
}\right) Y_{lm}^{\ast }\left( \theta ^{\prime },\varphi ^{\prime }\right)
Y_{lm}\left( \theta ^{\prime },\varphi ^{\prime }\right) \ d\omega ^{\prime
}=\sqrt{\frac{2s+1}{4\pi }}\ C_{lms0}^{lm}\ C_{l0s0}^{l0}.  \label{scr6}
\end{equation}%
In view of the symmetry property of the Clebsch--Gordan coefficients \cite%
{Ni:Su:Uv}, \cite{Rose}, \cite{Var:Mos:Kher}%
\begin{equation}
C_{l_{1}m_{1}l_{2}m_{2}}^{lm}=\left( -1\right)
^{l_{1}+l_{2}-l}C_{l_{1},-m_{1},l_{2},-m_{2}}^{l,-m}  \label{scr7}
\end{equation}%
and the selection rule $\left\vert l_{1}-l_{2}\right\vert \leq l\leq
l_{1}+l_{2}$ of the addition of two angular momenta in quantum mechanics,
the integral (\ref{scr6}) is not zero only for $s=0,2,\ ...\ ,2l.$ As a
result%
\begin{eqnarray}
V\left( \mathbf{r}\right)  &=&\frac{Ze}{r}-e\sum_{s=0}^{l}\sqrt{\frac{4\pi }{%
4s+1}}\ C_{lm,\;2s,0}^{lm}\ C_{l0,\;2s,0}^{l0}\ Y_{2s,\;0}\left( \theta
,\varphi \right)   \notag \\
&&\times \dint_{0}^{\infty }\frac{r_{<}^{2s}}{r_{>}^{2s+1}}\
R_{nl}^{2}\left( r^{\prime }\right) \left( r^{\prime }\right) ^{2}dr^{\prime
}.  \label{scr8}
\end{eqnarray}%
It is known that the Clebsch--Gordan coefficients are simply the Hahn
polynomials up to a normalization factor; see \cite{Ni:Su:Uv} and references
therein for more details on the relation between the Clebsch--Gordan
coefficients and the Hahn polynomials, which was overlooked on the early
stage of developing of the quantum theory of angular momentum \cite{Con:Sh},
\cite{Edm}, \cite{Rose}, \cite{Var:Mos:Kher}, \cite{Vil}, \cite{Wig}, \cite%
{Wigner} and had been established much later independently by Koornwinder
\cite{Koo} and Smorodinskii and Suslov \cite{Smor:Sus}. It is worth noting
that before that Wilson \cite{Wil} found that the next \textquotedblleft
building blocks\textquotedblright\ of the quantum theory of angular
momentum, the so-called $6j$-symbols, are orthogonal polynomials of a
discrete variable, see also \cite{Smorod:Susl}, \cite{Sus} and \cite{Susl};
Askey and Wilson \cite{As:Wi} studied this new orthogonal polynomials and
their $q$-extensions in details; see also \cite{An:As}, \cite{An:As:Ro},
\cite{Ga:Ra}, \cite{Ko:Sw}, \cite{Ni:Su:Uv} and references therein for the
current status of the theory of Askey--Wilson polynomials and their special
and/or limiting cases. Thus the Hahn polynomials appear in the expression (%
\ref{scr8}) for the effective electrostatic potential $V\left( \mathbf{r}%
\right) $ in the nonrelativistic hydrogen-like atom.\smallskip

The special Clebsch--Gordan coefficients $C_{l0,\;2s,0}^{l0}$ in (\ref{scr8}%
) are \cite{Var:Mos:Kher}%
\begin{equation}
C_{l0,\;2s,0}^{l0}=\left( -1\right) ^{s}\frac{\left( l+s\right) !\left(
2s\right) !}{\left( l-s\right) !\left( s!\right) ^{2}}\sqrt{\frac{\left(
2l+1\right) \left( 2l-2s\right) !}{\left( 2l+2s+1\right) !}}  \label{scr8a}
\end{equation}%
and%
\begin{equation}
Y_{2s,\;0}\left( \theta ,\varphi \right) =\sqrt{\frac{4s+1}{4\pi }}\
P_{2s}\left( \cos \theta \right) ,  \label{scr8b}
\end{equation}%
where $P_{n}\left( x\right) $ are the Legendre polynomials. \smallskip

The integral over the radial functions in (\ref{scr8}) can be rewritten\ in
the form%
\begin{eqnarray}
&&\dint_{0}^{\infty }\frac{r_{<}^{2s}}{r_{>}^{2s+1}}\ R_{nl}^{2}\left(
r^{\prime }\right) \left( r^{\prime }\right) ^{2}dr^{\prime }\qquad
\label{scr5} \\
&&\quad =\frac{1}{r^{2s+1}}\int_{0}^{r}\ \left( r^{\prime }\right)
^{2s+2}R_{nl}^{2}\left( r^{\prime }\right) dr^{\prime
}+r^{2s}\int_{r}^{\infty }\ \left( r^{\prime }\right)
^{1-2s}R_{nl}^{2}\left( r^{\prime }\right) dr^{\prime }  \notag \\
&&\quad =\frac{1}{r^{2s+1}}\left( \int_{0}^{\infty }\ \left( r^{\prime
}\right) ^{2s+2}R_{nl}^{2}\left( r^{\prime }\right) dr^{\prime
}-\int_{r}^{\infty }\ \left( r^{\prime }\right) ^{2s+2}R_{nl}^{2}\left(
r^{\prime }\right) dr^{\prime }\right)  \notag \\
&&\qquad +r^{2s}\int_{r}^{\infty }\ \left( r^{\prime }\right)
^{1-2s}R_{nl}^{2}\left( r^{\prime }\right) dr^{\prime }  \notag \\
&&\quad =\dfrac{1}{r^{2s+1}}\ J_{2s}-\dfrac{1}{r^{2s+1}}\ J_{2s}\left(
r\right) +r^{2s}\ J_{-2s-1}\left( r\right) ,  \notag
\end{eqnarray}%
where the first integral $J_{2s}=\left\langle r^{2s}\right\rangle $ is given
by (\ref{nrc5}) in terms of \ the Chebyshev polynomials of a discrete
variable. The other two integral are of the form%
\begin{equation}
J_{k}\left( r\right) =\int_{r}^{\infty }\ \left( r^{\prime }\right)
^{k+2}R_{nl}^{2}\left( r^{\prime }\right) dr^{\prime }.  \label{scr9a}
\end{equation}%
They are special cases of our integral (\ref{i11}) and can be evaluated by (%
\ref{i13}) in terms of the incomplete gamma function; they have simple
asymptotics at infinity. The reader can work out the details.

For the electron in the ground state $n=1$ and $l=m=0$ all the integral are
easily evaluated and the result is%
\begin{equation}
V\left( r\right) =\frac{\left( Z-1\right) e}{r}+\left( \frac{e}{r}+\frac{Ze}{%
a_{0}}\right) e^{-2Zr/a_{0}}.\allowbreak  \label{scr9}
\end{equation}%
For small distances $r\rightarrow 0$ the effective potential $V\left(
r\right) \rightarrow eZ/r$ as expected and as $r\rightarrow \infty $ the
potential $V\left( r\right) \rightarrow e\left( Z-1\right) /r$ which is the
potential of the nucleus of charge $Ze$ screened by the electron.

\section{Relativistic Coulomb Problem}

\subsection{Dirac Equation}

The relativistic wave equation of Dirac \cite{Dirac}, \cite{DiracII}, \cite%
{DiracQM}, \cite{FermiRad}, \cite{Fermi}%
\begin{equation}
i\hbar \frac{\partial }{\partial t}\psi =H\psi  \label{d1}
\end{equation}%
for the electron in an external central field with the potential energy $%
U\left( r\right) $ has the Hamiltonian of the form%
\begin{equation}
H=c\mathbf{\alpha p}+mc^{2}\beta +U\left( r\right) ,  \label{d2}
\end{equation}%
where $\mathbf{\alpha p}=\alpha _{1}p_{1}+\alpha _{2}p_{2}+\alpha _{3}p_{3}$
with the momentum operator $\mathbf{p}=-i\hbar \mathbf{\nabla }$ and%
\begin{equation}
\mathbf{\alpha }=\left(
\begin{array}{cc}
\mathbf{0} & \mathbf{\sigma } \\
\mathbf{\sigma } & \mathbf{0}%
\end{array}%
\right) ,\qquad \beta =\left(
\begin{array}{cc}
\mathbf{1} & \mathbf{0} \\
\mathbf{0} & -\mathbf{1}%
\end{array}%
\right) ,\qquad \psi =\left(
\begin{array}{c}
\mathbf{u} \\
\mathbf{v}%
\end{array}%
\right) .  \label{d3}
\end{equation}%
We use the standard representation of the Pauli matrices%
\begin{equation}
\sigma _{1}=\left(
\begin{array}{cc}
0 & 1 \\
1 & 0%
\end{array}%
\right) ,\qquad \sigma _{2}=\left(
\begin{array}{cc}
0 & -i \\
i & 0%
\end{array}%
\right) ,\qquad \sigma _{3}=\left(
\begin{array}{cc}
1 & 0 \\
0 & -1%
\end{array}%
\right)  \label{d4}
\end{equation}%
and%
\begin{equation*}
\mathbf{0}=\left(
\begin{array}{cc}
0 & 0 \\
0 & 0%
\end{array}%
\right) ,\qquad \mathbf{1}=\left(
\begin{array}{cc}
1 & 0 \\
0 & 1%
\end{array}%
\right) .
\end{equation*}%
The relativistic electron has a four component wave function%
\begin{equation}
\psi =\psi \left( \mathbf{r},t\right) =\left(
\begin{array}{c}
\mathbf{u}\left( \mathbf{r},t\right) \\
\mathbf{v}\left( \mathbf{r},t\right)%
\end{array}%
\right) =\left(
\begin{array}{c}
\psi _{1}\left( \mathbf{r},t\right) \\
\psi _{2}\left( \mathbf{r},t\right) \\
\psi _{3}\left( \mathbf{r},t\right) \\
\psi _{4}\left( \mathbf{r},t\right)%
\end{array}%
\right)  \label{d4a}
\end{equation}%
and the Dirac equation (\ref{d1}) is a matrix equation that is equivalent to
a system of four first order partial differential equations. The inner
product for two Dirac (bispinor) wave functions%
\begin{equation*}
\psi =\left(
\begin{array}{c}
\mathbf{u}_{1} \\
\mathbf{v}_{1}%
\end{array}%
\right) =\left(
\begin{array}{c}
\psi _{1} \\
\psi _{2} \\
\psi _{3} \\
\psi _{4}%
\end{array}%
\right) ,\qquad \phi =\left(
\begin{array}{c}
\mathbf{u}_{2} \\
\mathbf{v}_{2}%
\end{array}%
\right) =\left(
\begin{array}{c}
\phi _{1} \\
\phi _{2} \\
\phi _{3} \\
\phi _{4}%
\end{array}%
\right)
\end{equation*}%
is defined as a scalar quantity%
\begin{eqnarray}
\left\langle \psi ,\ \phi \right\rangle &=&\int_{\mathbf{R}^{3}}\psi
^{\dagger }\phi \ dv=\int_{\mathbf{R}^{3}}\left( \mathbf{u}_{1}{}^{\dagger }%
\mathbf{u}_{2}+\mathbf{v}_{1}{}^{\dagger }\mathbf{v}_{2}\right) \ dv
\label{d5} \\
&=&\int_{\mathbf{R}^{3}}\left( \overset{\ast }{\psi _{1}}\phi _{1}+\overset{%
\ast }{\psi _{2}}\phi _{2}\overset{\ast }{+\psi _{3}}\phi _{3}\overset{\ast }%
{+\psi _{4}}\phi _{4}\right) \ dv  \notag
\end{eqnarray}%
with the squared norm%
\begin{eqnarray}
\left\vert \left\vert \psi \right\vert \right\vert ^{2} &=&\left\langle \psi
,\ \psi \right\rangle =\int_{\mathbf{R}^{3}}\psi ^{\dagger }\psi \ dv=\int_{%
\mathbf{R}^{3}}\left( \mathbf{u}_{1}{}^{\dagger }\mathbf{u}_{1}+\mathbf{v}%
_{1}{}^{\dagger }\mathbf{v}_{1}\right) \ dv  \label{d6} \\
&=&\int_{\mathbf{R}^{3}}\left( \left\vert \psi _{1}\right\vert
^{2}+\left\vert \psi _{2}\right\vert ^{2}+\left\vert \psi _{3}\right\vert
^{2}+\left\vert \psi _{4}\right\vert ^{2}\right) \ dv  \notag
\end{eqnarray}%
and the wave functions are usually normalized so that $\left\vert \left\vert
\psi \right\vert \right\vert =\left\langle \psi ,\ \psi \right\rangle
^{1/2}=1.$\smallskip

The substitution%
\begin{equation}
\psi \left( \mathbf{r},t\right) =e^{-i\left( E\ t\right) /\hbar }\ \psi
\left( \mathbf{r}\right) ,  \label{d7}
\end{equation}%
gives the stationary Dirac equation%
\begin{equation}
H\psi \left( \mathbf{r}\right) =E\psi \left( \mathbf{r}\right) ,  \label{d8}
\end{equation}%
where $E$ is the total energy of the electron.\smallskip

According to Steven Weinberg (\cite{Wein}, vol.~I, p.~565), physicists learn
in kindergarten how to solve problems related to the wave equation of Dirac
in the presence of external fields. In Section~6 of this paper, for the
benefits of the reader who is not an expert in theoretical physics, we
outline a procedure of separation of the variables and solve the
corresponding first order system of radial equations for the Dirac equation
in the Coulomb field $U\left( r\right) =-Ze^{2}/r.$ The end results are
presented in the next section; see also \cite{Akh:Ber}, \cite{Be:Sal}, \cite%
{Ber:Lif:Pit}, \cite{Fermi}, \cite{It:Zu}, \cite{Mes}, \cite{Ni:Uv}, \cite%
{Schiff} and references therein for more information.

\subsection{Relativistic Coulomb Wave Functions and Discrete Energy Levels}

The exact solutions of the stationary Dirac equation%
\begin{equation}
H\psi =E\psi  \label{sde}
\end{equation}%
for the Coulomb potential can be obtained in the spherical coordinates a
result of a rather lengthy calculation. Remarkably the energy levels (\ref%
{rc6}) were discovered in 1916 by Sommerfeld from the ``old'' Bohr quantum
theory and the corresponding (bispinor) Dirac wave functions were originally
found by Darwin \cite{Dar} and Gordon \cite{Gor} at early age of discovery
of the ``new'' wave mechanics; see also \cite{Bie} for a modern discussion
of ``Sommerfeld's puzzle''. These classical results are nowadays included in
all textbooks on relativistic quantum mechanics, quantum field theory and
advanced texts on mathematical physics; see, for example, \cite{Akh:Ber},
\cite{Ber:Lif:Pit}, \cite{Be:Sal}, \cite{Dav},\cite{It:Zu}, \cite{Mes} and
\cite{Ni:Uv}. More details are given in the last but one section of this
paper. The end result is%
\begin{equation}
\psi =\left(
\begin{array}{c}
\mathbf{\varphi \medskip } \\
\mathbf{\chi }%
\end{array}%
\right) =\left(
\begin{array}{c}
\mathcal{Y}_{jm}^{\pm }\left( \mathbf{n}\right) \ F\left( r\right) \medskip
\\
i\mathcal{Y}_{jm}^{\mp }\left( \mathbf{n}\right) \ G\left( r\right)%
\end{array}%
\right) ,  \label{rc1}
\end{equation}%
where the spinor spherical harmonics $\mathcal{Y}_{jm}^{\pm }\left( \mathbf{n%
}\right) =\mathcal{Y}_{jm}^{\left( j\pm 1/2\right) }\left( \mathbf{n}\right)
$ are given explicitly in terms of the ordinary spherical functions $%
Y_{lm}\left( \mathbf{n}\right) ,$ $\mathbf{n}=\mathbf{n}\left( \theta
,\varphi \right) =\mathbf{r}/r$ and the special Clebsch--Gordan coefficients
with the spin $1/2$ as\ follows \cite{Akh:Ber}, \cite{Ber:Lif:Pit}, \cite%
{Rose}, \cite{Var:Mos:Kher}%
\begin{equation}
\mathcal{Y}_{jm}^{\pm }\left( \mathbf{n}\right) =\left(
\begin{array}{c}
\mp \sqrt{\dfrac{\left( j+1/2\right) \mp \left( m-1/2\right) }{2j+\left(
1\pm 1\right) }}\ Y_{j\pm 1/2,\ m-1/2}\left( \mathbf{n}\right) \medskip \\
\sqrt{\dfrac{\left( j+1/2\right) \pm \left( m+1/2\right) }{2j+\left( 1\pm
1\right) }}\ Y_{j\pm 1/2,\ m+1/2}\left( \mathbf{n}\right)%
\end{array}%
\right)  \label{rc2}
\end{equation}%
with the total angular momentum $j=1/2,3/2,5/2,\ ...$ and $m=-j,-j+1,\
...,j-1,j;$ see Section~6.1 for discussion of properties of the spinor
spherical harmonics in detail.\smallskip\

The radial functions $F\left( r\right) $ and $G\left( r\right) $ can be
presented as \cite{Ni:Uv}%
\begin{eqnarray}
\left(
\begin{array}{c}
F\left( r\right) \medskip \mathbf{\medskip \bigskip } \\
G\left( r\right)%
\end{array}%
\right) &=&\frac{a^{2}\beta ^{3/2}}{\nu }\sqrt{\frac{\left( \varepsilon
\kappa -\nu \right) n!}{\mu \left( \kappa -\nu \right) \Gamma \left( n+2\nu
\right) }}\ \xi ^{\nu -1}e^{-\xi /2}  \notag \\
&&\times \left(
\begin{array}{c}
f_{1}\qquad f_{2}\mathbf{\medskip } \\
g_{1}\qquad g_{2}%
\end{array}%
\right) \left(
\begin{array}{c}
\xi L_{n-1}^{2\nu +1}\left( \xi \right) \bigskip \mathbf{\medskip } \\
L_{n}^{2\nu -1}\left( \xi \right)%
\end{array}%
\right) .  \label{rc3}
\end{eqnarray}%
Here $L_{k}^{\alpha }\left( \xi \right) $ are the Laguerre polynomials and
we use the following notations $\kappa =\pm \left( j+1/2\right) ,$ $\nu =%
\sqrt{\kappa ^{2}-\mu ^{2}},$ $\mu =\alpha Z=Ze^{2}/\hbar c,$ $a=\sqrt{%
1-\varepsilon ^{2}},$ $\varepsilon =E/mc^{2},$ $\beta =mc/\hbar =2\pi
/\lambda $ and%
\begin{equation}
\xi =2a\beta r=2\sqrt{1-\varepsilon ^{2}}\ \frac{mc}{\hbar }\ r.  \label{rc4}
\end{equation}%
The elements of $2\times 2$--transition matrix in (\ref{rc3}) are given by%
\begin{equation}
f_{1}=\frac{a\mu }{\varepsilon \kappa -\nu },\quad f_{2}=\kappa -\nu ,\quad
g_{1}=\frac{a\left( \kappa -\nu \right) }{\varepsilon \kappa -\nu },\quad
g_{2}=\mu .  \label{rc5}
\end{equation}%
This particular form of the relativistic radial functions (\ref{rc3}) is due
to Nikiforov and Uvarov \cite{Ni:Uv}; it is very convenient for taking the
nonrelativistic limit $c\rightarrow \infty ,$ see more details in
Section~6.4; we shall use this form throughout the paper instead of the
traditional one given in \cite{Akh:Ber}, \cite{Ber:Lif:Pit}, \cite{Be:Sal},
\cite{Davis} and elsewhere; cf. equation~(\ref{rrc44}) below.\smallskip

The relativistic discrete energy levels $\varepsilon =\varepsilon
_{n}=E_{n}/E_{0}$ with the rest mass energy $E_{0}=mc^{2}$ are given by the
famous Sommerfeld fine structure formula%
\begin{equation}
E_{n}=\frac{mc^{2}}{\sqrt{1+\mu ^{2}/\left( n+\nu \right) ^{2}}},\quad \mu
=\alpha Z=\frac{Ze^{2}}{\hbar c},\quad \nu =\sqrt{\kappa ^{2}-\mu ^{2}}.
\label{rc6}
\end{equation}%
Here $n=n_{r}=0,1,2,\ ...$ is the radial quantum number and $\kappa =\pm
\left( j+1/2\right) =\pm 1,\pm 2,\pm 3,\ ...\ .$ In the nonrelativistic
limit $c\rightarrow \infty $ one can expand the exact Sommerfeld--Dirac
formula (\ref{rc6}) in ascending powers of $\mu ^{2}=\left( \alpha Z\right)
^{2},$ the first terms in this expansion are%
\begin{equation}
\frac{E}{mc^{2}}=1-\frac{\mu ^{2}}{2n^{2}}-\frac{\mu ^{4}}{2n^{4}}\left(
\frac{n}{j+1/2}-\frac{3}{4}\right) +\text{O}\left( \mu ^{6}\right) ,\quad
\mu \rightarrow 0.  \label{rc6a}
\end{equation}%
Here $n=n_{r}+j+1/2$ is the principal quantum number of the nonrelativistic
hydrogen-like atom. The first term in this expansion is simply the rest mass
energy $E_{0}=mc^{2}$ of the electron, the second term coincides with the
energy eigenvalue in the nonrelativistic Schr\"{o}dinger theory (\ref{nrc2c}%
) and the third term gives the so-called fine structure of the energy levels
-- the correction obtained for the energy in the Pauli approximation which
includes interaction of the spin of the electron with its orbital angular
momentum; see \cite{Be:Sal} and \cite{Schiff} for further discussion of the
hydrogen-like energy levels including comparison with the experimental data.
One can show that in the same limit $\mu \rightarrow 0$ the relativistic
Coulomb wave functions (\ref{rc1}) tend to the nonrelativistic wave
functions of the Pauli theory; see, for example, \cite{Ni:Uv} for more
details; we shall elaborate more on this limit in Section~6.4.\smallskip

We give below the explicit form \ of the radial wave functions (\ref{rc3})
for the $1S_{1/2}$-state, when $n=n_{r}=0,$ $l=0,$ $j=1/2$ and $\kappa =-1:$%
\begin{equation}
\left(
\begin{array}{c}
F\left( r\right) \medskip \mathbf{\medskip \bigskip } \\
G\left( r\right)%
\end{array}%
\right) =\left( \frac{2Z}{a_{0}}\right) ^{3/2}\sqrt{\frac{\nu _{1}+1}{%
2\Gamma \left( 2\nu _{1}+1\right) }}\ \left(
\begin{array}{c}
-1\medskip \mathbf{\medskip \bigskip } \\
\sqrt{\dfrac{1-\nu _{1}}{1+\nu _{1}}}%
\end{array}%
\right) \xi _{1}^{\nu _{1}-1}e^{-\xi _{1}/2}.  \label{rc6b}
\end{equation}%
Here $\nu _{1}=\sqrt{1-\mu ^{2}}=\varepsilon _{1}$ and $\xi _{1}=2\sqrt{%
1-\varepsilon _{1}^{2}}\beta r=2Z\left( r/a_{0}\right) .$ One can see also
\cite{Akh:Ber}, \cite{Be:Sal}, \cite{Ber:Lif:Pit}, \cite{Dar}, \cite{Dav},
\cite{Gor}, \cite{It:Zu}, \cite{Mes}, \cite{Schiff} and references therein
for more information on the relativistic Coulomb problem.

\subsection{Matrix Elements}

In this section we evaluate the mean values%
\begin{equation}
\left\langle r^{p}\right\rangle =\dfrac{_{\dint_{\mathbf{R}^{3}}\psi
^{\dagger }\ r^{p}\psi \ dv}}{_{\dint_{\mathbf{R}^{3}}\psi ^{\dagger }\ \psi
\ dv}}=\dfrac{\dint_{0}^{\infty }\left( F^{2}\left( r\right) +G^{2}\left(
r\right) \right) r^{p+2}\ dr}{\dint_{0}^{\infty }\left( F^{2}\left( r\right)
+G^{2}\left( r\right) \right) r^{2}\ dr}  \label{rc7}
\end{equation}%
of all possible powers of $r$ with respect to the relativistic Coulomb
functions given by (\ref{rc1})--(\ref{rc3}) in terms of the Hahn polynomials
(\ref{a3}). First we use the orthogonality relation of the spinor spherical
harmonics;%
\begin{equation}
\int_{S^{2}}\left( \mathcal{Y}_{jm}^{\left( l\right) }\left( \mathbf{n}%
\right) \right) ^{\dagger }\ \mathcal{Y}_{j^{\prime }m^{\prime }}^{\left(
l^{\prime }\right) }\left( \mathbf{n}\right) \ d\omega =\delta _{jj^{\prime
}}\delta _{ll^{\prime }}\delta _{mm^{\prime }},  \label{rc7a}
\end{equation}%
see \cite{Akh:Ber}, \cite{Ber:Lif:Pit}, \cite{Var:Mos:Kher} or (\ref{ang5a}%
); and the explicit form of the wave functions (\ref{rc1}) in order to
simplify%
\begin{eqnarray}
\dint_{\mathbf{R}^{3}}\psi ^{\dagger }\ r^{p}\psi \ dv &=&\dint_{\mathbf{R}%
^{3}}\left( \mathcal{Y}^{\dagger }\left( \mathbf{n}\right) \mathcal{Y}\left(
\mathbf{n}\right) \right) \ r^{p}\left( F^{2}\left( r\right) +G^{2}\left(
r\right) \right) \ r^{2}drd\omega  \notag \\
&=&\dint_{S^{2}}\left( \mathcal{Y}^{\dagger }\left( \mathbf{n}\right)
\mathcal{Y}\left( \mathbf{n}\right) \right) \ d\omega \ \int_{0}^{\infty }\
r^{p+2}\left( F^{2}\left( r\right) +G^{2}\left( r\right) \right) \ dr  \notag
\\
&=&\int_{0}^{\infty }\ r^{p+2}\left( F^{2}\left( r\right) +G^{2}\left(
r\right) \right) \ dr.  \label{rc7b}
\end{eqnarray}%
More details on construction of the spinor spherical harmonics $\mathcal{Y}%
_{jm}^{\pm }\left( \mathbf{n}\right) =\mathcal{Y}_{jm}^{\left( j\pm
1/2\right) }\left( \mathbf{n}\right) $ and study of their properties are
given in Section~6.4. The radial wave functions in (\ref{rc3}) are
normalized as follows%
\begin{equation}
\int_{0}^{\infty }\ r^{2}\left( F^{2}\left( r\right) +G^{2}\left( r\right)
\right) \ dr=1.  \label{rc7c}
\end{equation}%
Thus one needs to evaluate the integral%
\begin{equation}
\left\langle r^{p}\right\rangle =\int_{0}^{\infty }\ r^{p+2}\left(
F^{2}\left( r\right) +G^{2}\left( r\right) \right) \ dr  \label{rc7d}
\end{equation}%
only, and our final result with the notations from the previous section can
be presented in the following closed form%
\begin{eqnarray}
&&4\mu \nu ^{2}\left( 2a\beta \right) ^{p}\ \left\langle r^{p}\right\rangle
\label{rc8} \\
&&\quad =a\kappa \left( \varepsilon \kappa +\nu \right) \frac{\Gamma \left(
2\nu +p+3\right) }{\Gamma \left( 2\nu +2\right) }~_{3}F_{2}\left(
\begin{array}{c}
1-n,\ p+2,\ -p-1\medskip \\
2\nu +2,\quad 1%
\end{array}%
\right)  \notag \\
&&\quad -2\left( p+2\right) \mu \left( \varepsilon ^{2}\kappa ^{2}-\nu
^{2}\right) \frac{\Gamma \left( 2\nu +p+2\right) }{\Gamma \left( 2\nu
+2\right) }~_{3}F_{2}\left(
\begin{array}{c}
1-n,\ p+3,\ -p\medskip \\
2\nu +2,\quad 2%
\end{array}%
\right)  \notag \\
&&\quad +a\kappa \left( \varepsilon \kappa -\nu \right) \frac{\Gamma \left(
2\nu +p+1\right) }{\Gamma \left( 2\nu \right) }~_{3}F_{2}\left(
\begin{array}{c}
-n,\ p+2,\ -p-1\medskip \\
2\nu ,\quad 1%
\end{array}%
\right) ,  \notag
\end{eqnarray}%
where the terminating generalized hypergeometric series $_{3}F_{2}\left(
1\right) $ are related to the Hahn and Chebyshev polynomials of a discrete
variable due to (\ref{a3}).\smallskip

Substituting (\ref{rc3}) into (\ref{rc7d}) one gets%
\begin{eqnarray}
C\left\langle r^{p}\right\rangle &=&\int_{0}^{\infty }e^{-\xi }\xi ^{2\nu
+p}\left( \left( f_{1}\xi L_{n-1}^{2\nu +1}\left( \xi \right)
+f_{2}L_{n}^{2\nu -1}\left( \xi \right) \right) ^{2}\right.  \label{rc9} \\
&&+\left. \left( g_{1}\xi L_{n-1}^{2\nu +1}\left( \xi \right)
+g_{2}L_{n}^{2\nu -1}\left( \xi \right) \right) ^{2}\right) \ d\xi  \notag \\
&=&\left( f_{1}^{2}+g_{1}^{2}\right) \int_{0}^{\infty }e^{-\xi }\xi ^{2\nu
+p+2}\left( L_{n-1}^{2\nu +1}\left( \xi \right) \right) ^{2}\ d\xi  \notag \\
&&+2\left( f_{1}f_{2}+g_{1}g_{2}\right) \int_{0}^{\infty }e^{-\xi }\xi
^{2\nu +p+1}L_{n-1}^{2\nu +1}\left( \xi \right) L_{n}^{2\nu -1}\left( \xi
\right) \ d\xi  \notag \\
&&+\left( f_{2}^{2}+g_{2}^{2}\right) \int_{0}^{\infty }e^{-\xi }\xi ^{2\nu
+p}\left( L_{n}^{2\nu -1}\left( \xi \right) \right) ^{2}\ d\xi ,  \notag
\end{eqnarray}%
where%
\begin{equation}
C=8\mu \nu ^{2}\left( 2a\beta \right) ^{p}\dfrac{\left( \kappa -\nu \right)
\Gamma \left( 2\nu +n\right) }{a\left( \varepsilon \kappa -\nu \right) n!}
\label{rc10}
\end{equation}%
and, in view of (\ref{rc5}),%
\begin{equation}
f_{1}^{2}+g_{1}^{2}=\dfrac{2a^{2}\kappa \left( \kappa -\nu \right) }{\left(
\varepsilon \kappa -\nu \right) ^{2}},\ \ \ f_{1}f_{2}+g_{1}g_{2}=\dfrac{%
2a\mu \left( \kappa -\nu \right) }{\varepsilon \kappa -\nu },\ \ \
f_{2}^{2}+g_{2}^{2}=2\kappa \left( \kappa -\nu \right) .  \label{rc10a}
\end{equation}%
The two types of the integrals occuring in the calculation are given by our
``master'' formula (\ref{i2}) as follows%
\begin{eqnarray}
&&\dint_{0}^{\infty }e^{-\xi }\xi ^{2\nu +p+2}\left( L_{n-1}^{2\nu +1}\left(
\xi \right) \right) ^{2}  \label{rc11} \\
&&\quad =\frac{\Gamma \left( 2\nu +p+3\right) \Gamma \left( 2\nu +n+1\right)
}{\left( n-1\right) !\Gamma \left( 2\nu +2\right) }~_{3}F_{2}\left(
\begin{array}{c}
1-n,\ p+2,\ -p-1\medskip \\
2\nu +2,\quad 1%
\end{array}%
\right) ,  \notag
\end{eqnarray}%
\begin{eqnarray}
&&\int_{0}^{\infty }e^{-\xi }\xi ^{2\nu +p+1}L_{n-1}^{2\nu +1}\left( \xi
\right) L_{n}^{2\nu -1}\left( \xi \right) \ d\xi  \label{rc12} \\
&&\quad =-\frac{\left( p+2\right) \Gamma \left( 2\nu +p+2\right) \Gamma
\left( 2\nu +n+1\right) }{\left( n-1\right) !\Gamma \left( 2\nu +2\right) }%
~_{3}F_{2}\left(
\begin{array}{c}
1-n,\ p+3,\ -p\medskip \\
2\nu +2,\quad 2%
\end{array}%
\right)  \notag
\end{eqnarray}%
and%
\begin{eqnarray}
&&\int_{0}^{\infty }e^{-\xi }\xi ^{2\nu +p}\left( L_{n}^{2\nu -1}\left( \xi
\right) \right) ^{2}\ d\xi  \label{rc12a} \\
&&\quad =\frac{\Gamma \left( 2\nu +p+1\right) \Gamma \left( 2\nu +n\right) }{%
n!\Gamma \left( 2\nu \right) }~_{3}F_{2}\left(
\begin{array}{c}
-n,\ p+2,\ -p-1\medskip \\
2\nu ,\quad 1%
\end{array}%
\right) .  \notag
\end{eqnarray}%
Substituting these integrals into (\ref{rc9}) and using the identity%
\begin{equation}
a^{2}n\left( 2\nu +n\right) =\varepsilon ^{2}\kappa ^{2}-\nu ^{2},
\label{rc12b}
\end{equation}%
see Section~6.3 for the proof, one finally arrives at (\ref{rc8}).\smallskip

In view of (\ref{a3}), the relations with the Hahn and Chebyshev polynomials
of a discrete variable are%
\begin{eqnarray}
&&4\mu \nu ^{2}\left( 2a\beta \right) ^{p}\ \left\langle r^{p}\right\rangle
=a\kappa \left( \varepsilon \kappa +\nu \right) \;h_{p+1}^{\left(
0,\;0\right) }\left( n-1,-1-2\nu \right)  \label{rc13} \\
&&\qquad \qquad \qquad \qquad \ -2\frac{p+2}{p+1}\mu \left( \varepsilon
^{2}\kappa ^{2}-\nu ^{2}\right) \;h_{p}^{\left( 1,\;1\right) }\left(
n-1,-1-2\nu \right)  \notag \\
&&\qquad \qquad \qquad \qquad \quad +a\kappa \left( \varepsilon \kappa -\nu
\right) \;h_{p+1}^{\left( 0,\;0\right) }\left( n,1-2\nu \right) ,\qquad
p\geq 0  \notag
\end{eqnarray}%
and%
\begin{eqnarray}
&\dfrac{4\mu \nu ^{2}}{\left( 2a\beta \right) ^{p+3}}&\left\langle \dfrac{1}{%
r^{p+3}}\right\rangle =a\kappa \left( \varepsilon \kappa +\nu \right) \frac{%
\Gamma \left( 2\nu -p\right) }{\Gamma \left( 2\nu +p+3\right) }%
\;h_{p+1}^{\left( 0,\;0\right) }\left( n-1,-1-2\nu \right)  \label{rc14} \\
&&\qquad \qquad \quad +2\mu \left( \varepsilon ^{2}\kappa ^{2}-\nu
^{2}\right) \frac{\Gamma \left( 2\nu -p-1\right) }{\Gamma \left( 2\nu
+p+2\right) }\;h_{p}^{\left( 1,\;1\right) }\left( n-1,-1-2\nu \right)  \notag
\\
&&\qquad \qquad \quad \ +a\kappa \left( \varepsilon \kappa -\nu \right)
\frac{\Gamma \left( 2\nu -p-2\right) }{\Gamma \left( 2\nu +p+1\right) }%
\;h_{p+1}^{\left( 0,\;0\right) }\left( n,1-2\nu \right) ,\qquad p\geq 0.
\notag
\end{eqnarray}%
The averages of $r^{p}$ for the relativistic hydrogen atom were evaluated by
Davis \cite{Davis} in a form which is slightly different from our equation (%
\ref{rc8}). It does not appear to have been noticed that the corresponding $%
_{3}F_{2}$-functions can be expressed in terms of Hahn polynomials. The ease
of handling of these matrix elements for the discrete levels is greatly
increased if use is made of the known properties of these polynomials \cite%
{Erd}, \cite{Ko:Sw}, \cite{Ni:Su:Uv}, \cite{Ni:Uv} and \cite{Chebyshev59},
\cite{Chebyshev64}, \cite{Chebyshev75}. For instance, the
difference-differentiation formula%
\begin{equation}
\Delta h_{m}^{\left( \alpha ,\;\beta \right) }\left( x,N\right) =\left(
\alpha +\beta +m+1\right) h_{m-1}^{\left( \alpha +1,\;\beta +1\right)
}\left( x,N-1\right) ,  \label{rc15}
\end{equation}%
where $\Delta f\left( x\right) =f\left( x+1\right) -f\left( x\right) ,$ in
the form%
\begin{equation}
h_{p+1}^{\left( 0,\;0\right) }\left( n,-2\nu \right) -h_{p+1}^{\left(
0,\;0\right) }\left( n-1,-2\nu \right) =\left( p+2\right) h_{p}^{\left(
1,\;1\right) }\left( n-1,-1-2\nu \right)  \label{tc16}
\end{equation}%
allows to rewrite formulas (\ref{rc13})--(\ref{rc15}) in terms of the
Chebyshev polynomials of a discrete variable $t_{m}\left( x\right)
=h_{m}^{\left( 0,\;0\right) }\left( x,N\right) $ only. Use of the three term
recurrence relation (\ref{nrc7a}) simplifies evaluation of the special cases
of these averages. Equation (\ref{a3ha}) gives asymptotic formulas for the
matrix elements as $\left\vert \kappa \right\vert \rightarrow \infty .$

\subsection{Nonrelativistic Limit}

In the limit $c\rightarrow \infty $ the relativistic Coulomb matrix elements
given by (\ref{rc7}) and (\ref{rc8}) tend to the nonrelativistic ones (\ref%
{nrc4}). This can be easily shown with the help of the asymptotic formulas%
\begin{equation}
\varepsilon \kappa \pm \nu =\left( \kappa \pm \left| \kappa \right| \right) -%
\frac{\kappa \left| \kappa \right| \pm \left( n_{r}+\left| \kappa \right|
\right) ^{2}}{2\left| \kappa \right| \left( n_{r}+\left| \kappa \right|
\right) ^{2}}\;\mu ^{2}+\text{O}\left( \mu ^{4}\right)
\end{equation}%
as $\mu =Ze^{2}/\hbar c\rightarrow 0;$ see Section 6.4 for more details.

\subsection{Special Cases}

Some important special cases of the relativistic matrix elements are%
\begin{equation}
\left\langle r^{2}\right\rangle =\frac{5n\left( n+2\nu \right) +4\nu
^{2}+1-\varepsilon \kappa \left( 2\varepsilon \kappa +3\right) }{2\left(
a\beta \right) ^{2}}  \label{sp1}
\end{equation}%
\begin{equation}
\left\langle r\right\rangle =\frac{a_{0}}{2Z}\left( 3\varepsilon n\left(
n+2\nu \right) +\kappa \left( 2\varepsilon \kappa -1\right) \right) ,
\label{sp2}
\end{equation}%
\begin{equation}
\left\langle 1\right\rangle =1,  \label{sp3}
\end{equation}%
\begin{equation}
\left\langle \dfrac{1}{r}\right\rangle =\frac{\beta }{\mu \nu }\left(
1-\varepsilon ^{2}\right) \left( \varepsilon \nu +\mu \sqrt{1-\varepsilon
^{2}}\right) ,  \label{sp4}
\end{equation}%
\begin{equation}
\left\langle \dfrac{1}{r^{2}}\right\rangle =\frac{2a^{3}\beta ^{2}\kappa
\left( 2\varepsilon \kappa -1\right) }{\mu \nu \left( 4\nu ^{2}-1\right) },
\label{sp5}
\end{equation}%
\begin{equation}
\left\langle \frac{1}{r^{3}}\right\rangle =2\left( a\beta \right) ^{3}\frac{%
3\varepsilon ^{2}\kappa ^{2}-3\varepsilon \kappa -\nu ^{2}+1}{\nu \left( \nu
^{2}-1\right) \left( 4\nu ^{2}-1\right) }.  \label{sp6}
\end{equation}%
Derivation of these closed forms requires more work than in the
nonrelativistic case because of a more complicated structure of the general
expression (\ref{rc8}). Here are some calculation details.\smallskip\

The special case $p=0$ implies the normalization of the radial wave
functions (\ref{rc7c}) and (\ref{sp3}). Indeed,%
\begin{eqnarray}
4\mu \nu ^{2}\ \left\langle 1\right\rangle &=&-4\mu \left( \varepsilon
^{2}\kappa ^{2}-\nu ^{2}\right) +a\kappa \left( \varepsilon \kappa +\nu
\right) \left( 2\nu +2\right) ~_{3}F_{2}\left(
\begin{array}{c}
1-n,\ 2,\ -1\medskip \\
2\nu +2,\quad 1%
\end{array}%
\right)  \label{sp7} \\
&&+a\kappa \left( \varepsilon \kappa -\nu \right) 2\nu ~_{3}F_{2}\left(
\begin{array}{c}
-n,\ 2,\ -1\medskip \\
2\nu ,\quad 1%
\end{array}%
\right)  \notag \\
&=&-4\mu \left( \varepsilon ^{2}\kappa ^{2}-\nu ^{2}\right) +2a\kappa \left(
\varepsilon \kappa +\nu \right) \left( \nu +n\right) +2a\kappa \left(
\varepsilon \kappa -\nu \right) \left( \nu +n\right)  \notag \\
&=&4\varepsilon \kappa ^{2}a\left( \nu +n\right) -4\mu \left( \varepsilon
^{2}\kappa ^{2}-\nu ^{2}\right) \;=\;4\mu \nu ^{2},\   \notag
\end{eqnarray}%
in view of the quantization rule%
\begin{equation}
\varepsilon \mu =a\left( \nu +n\right) ,  \label{sp8}
\end{equation}%
which leads to the Sommerfeld--Dirac formula for the discrete energy levels (%
\ref{rc6}); see Section~6.3 for more details on (\ref{sp8}).\smallskip

The special case $p=-1$ of (\ref{rc8}) reads%
\begin{eqnarray}
\frac{2\mu \nu ^{2}}{a\beta }\ \left\langle \frac{1}{r}\right\rangle
&=&a\kappa \left( \varepsilon \kappa +\nu \right) +a\kappa \left(
\varepsilon \kappa -\nu \right)  \label{sp9} \\
&&-2\mu \left( \varepsilon ^{2}\kappa ^{2}-\nu ^{2}\right) \frac{\Gamma
\left( 2\nu +1\right) }{\Gamma \left( 2\nu +2\right) }~_{2}F_{1}\left(
\begin{array}{c}
1-n,\ 1\medskip \\
2\nu +2%
\end{array}%
;1\right) ,  \notag
\end{eqnarray}%
where by the summation formula of Gauss (\ref{a4a})%
\begin{equation}
~_{2}F_{1}\left(
\begin{array}{c}
1-n,\ 1\medskip \\
2\nu +2%
\end{array}%
;1\right) =\frac{\Gamma \left( 2\nu +2\right) \Gamma \left( 2\nu +n\right) }{%
\Gamma \left( 2\nu +1\right) \Gamma \left( 2\nu +n+1\right) }.  \label{sp10}
\end{equation}%
Thus, in view of (\ref{rc12b})%
\begin{equation}
\frac{2\mu \nu ^{2}}{a\beta }\ \left\langle \frac{1}{r}\right\rangle
=2a\varepsilon \kappa ^{2}-2\mu a^{2}n,  \label{sp11}
\end{equation}%
and\ the final use of the quantization rule (\ref{sp8}) results in (\ref{sp5}%
). Our calculation of $\left\langle r^{-1}\right\rangle $ shows that there
is no simple form of the virial theorem in the Dirac theory of relativistic
electron moving in the central field of the Coulomb potential.\smallskip

In the case $p=-2$ we get%
\begin{equation}
\frac{\mu \nu ^{2}}{\left( a\beta \right) ^{2}}\left\langle \frac{1}{r^{2}}%
\right\rangle =\frac{a\kappa \left( \varepsilon \kappa +\nu \right) }{2\nu +1%
}+\frac{a\kappa \left( \varepsilon \kappa -\nu \right) }{2\nu -1},
\label{sp11a}
\end{equation}%
which leads to (\ref{sp5}).\smallskip

In a similar fashion, for $p=-3:$%
\begin{eqnarray}
&&\frac{4\mu \nu ^{2}}{\left( 2a\beta \right) ^{3}}\left\langle \frac{1}{%
r^{3}}\right\rangle =2\mu \left( \varepsilon ^{2}\kappa ^{2}-\nu ^{2}\right)
\frac{\Gamma \left( 2\nu -1\right) }{\Gamma \left( 2\nu +2\right) }
\label{sp11b} \\
&&\quad ~\ \ \ +a\kappa \left( \varepsilon \kappa +\nu \right) \frac{\Gamma
\left( 2\nu \right) }{\Gamma \left( 2\nu +2\right) }~_{3}F_{2}\left(
\begin{array}{c}
1-n,\ 2,\ -1\medskip \\
2\nu +2,\quad 1%
\end{array}%
\right)  \notag \\
&&\quad \quad ~+a\kappa \left( \varepsilon \kappa -\nu \right) \frac{\Gamma
\left( 2\nu -2\right) }{\Gamma \left( 2\nu \right) }~_{3}F_{2}\left(
\begin{array}{c}
-n,\ 2,\ -1\medskip \\
2\nu ,\quad 1%
\end{array}%
\right)  \notag \\
&&\quad ~=\frac{a\kappa \left( \varepsilon \kappa +\nu \right) \left( 2\nu
+2n\right) }{\left( 2\nu +2\right) \left( 2\nu +1\right) 2\nu }+\frac{%
a\kappa \left( \varepsilon \kappa -\nu \right) \left( 2\nu +2n\right) }{2\nu
\left( 2\nu -1\right) \left( 2\nu -2\right) }+\frac{2\mu \left( \varepsilon
^{2}\kappa ^{2}-\nu ^{2}\right) }{\left( 2\nu +1\right) 2\nu \left( 2\nu
-1\right) },  \notag
\end{eqnarray}%
which is simplified to (\ref{sp6}).\smallskip

The special case $p=1$ gives the average distance $\overline{r}=\left\langle
r\right\rangle $ between the electron and the nucleus in the relativistic
hydrogen-like atom. One gets%
\begin{eqnarray*}
&&4\mu \nu ^{2}\left( 2a\beta \right) \ \left\langle r\right\rangle =-6\mu
\left( \varepsilon ^{2}\kappa ^{2}-\nu ^{2}\right) \left( 2\nu +2\right)
~_{3}F_{2}\left(
\begin{array}{c}
1-n,\ 4,\ -1\medskip \\
2\nu +2,\quad 2%
\end{array}%
\right) \\
&&\qquad \qquad \qquad \qquad +a\kappa \left( \varepsilon \kappa +\nu
\right) \left( 2\nu +2\right) _{2}~_{3}F_{2}\left(
\begin{array}{c}
1-n,\ 3,\ -2\medskip \\
2\nu +2,\quad 1%
\end{array}%
\right) \\
&&\qquad \qquad \qquad \qquad \quad +a\kappa \left( \varepsilon \kappa -\nu
\right) \left( 2\nu \right) _{2}~_{3}F_{2}\left(
\begin{array}{c}
-n,\ 3,\ -2\medskip \\
2\nu ,\quad 1%
\end{array}%
\right) ,
\end{eqnarray*}%
where%
\begin{equation*}
\left( 2\nu +2\right) ~_{3}F_{2}\left(
\begin{array}{c}
1-n,\ 4,\ -1\medskip \\
2\nu +2,\quad 2%
\end{array}%
\right) =2\left( n+\nu \right) ,
\end{equation*}%
\begin{equation*}
\left( 2\nu \right) _{2}~_{3}F_{2}\left(
\begin{array}{c}
-n,\ 3,\ -2\medskip \\
2\nu ,\quad 1%
\end{array}%
\right) =6n^{2}+12\nu n+4\nu ^{2}+2\nu
\end{equation*}%
Thus%
\begin{equation}
2\mu \nu ^{2}a\beta \ \left\langle r\right\rangle =a\varepsilon \kappa
^{2}\left( 3n\left( n+2\nu \right) +2\nu ^{2}\right) -a\kappa \nu ^{2}-3\mu
\left( \varepsilon ^{2}\kappa ^{2}-\nu ^{2}\right) \left( \nu +n\right) ,
\label{sp12}
\end{equation}%
which can be simplified to (\ref{sp2}) by a straightforward calculation with
the aid of (\ref{rc12b}) and (\ref{sp8}); we leave the details to the
reader.\smallskip

For $p=2:$%
\begin{eqnarray}
&&4\mu \nu ^{2}\left( 2a\beta \right) ^{2}\left\langle r^{2}\right\rangle
=-8\mu \left( \varepsilon ^{2}\kappa ^{2}-\nu ^{2}\right) \frac{\Gamma
\left( 2\nu +4\right) }{\Gamma \left( 2\nu +2\right) }~_{3}F_{2}\left(
\begin{array}{c}
1-n,\ 5,\ -2\medskip \\
2\nu +2,\quad 2%
\end{array}%
\right)  \label{sp13} \\
&&\quad ~\ \ \ \qquad \quad \quad \qquad \qquad +a\kappa \left( \varepsilon
\kappa +\nu \right) \frac{\Gamma \left( 2\nu +5\right) }{\Gamma \left( 2\nu
+2\right) }~_{3}F_{2}\left(
\begin{array}{c}
1-n,\ 4,\ -3\medskip \\
2\nu +2,\quad 1%
\end{array}%
\right)  \notag \\
&&\quad \quad ~\qquad \quad \quad \qquad \qquad \ +a\kappa \left(
\varepsilon \kappa -\nu \right) \frac{\Gamma \left( 2\nu +3\right) }{\Gamma
\left( 2\nu \right) }~_{3}F_{2}\left(
\begin{array}{c}
-n,\ 4,\ -3\medskip \\
2\nu ,\quad 1%
\end{array}%
\right) ,  \notag
\end{eqnarray}%
where%
\begin{equation*}
\left( 2\nu +2\right) _{2}~_{3}F_{2}\left(
\begin{array}{c}
1-n,\ 5,\ -2\medskip \\
2\nu +2,\quad 2%
\end{array}%
\right) =\allowbreak 5n^{2}+10n\nu +4\nu ^{2}+1
\end{equation*}%
and%
\begin{equation*}
\left( 2\nu \right) _{3}~_{3}F_{2}\left(
\begin{array}{c}
-n,\ 4,\ -3\medskip \\
2\nu ,\quad 1%
\end{array}%
\right) =4\left( n+\nu \right) \left( 5n^{2}+10n\nu +2\nu ^{2}+3\nu
+1\right) .
\end{equation*}%
This can be transformed to (\ref{sp1}) with the help of (\ref{sp8}).

\subsection{Screening}

Let us evaluate the effective electrostatic potential $V\left( \mathbf{r}%
\right) $ for the relativistic hydro\-gen-like atom. For the electron in the
stationary state with the wave functions (\ref{rc1}) corresponding to the
total angular momentum $j,$ its projection $m$ and the radial quantum number
$n=n_{r}$ this potential is%
\begin{equation}
V\left( \mathbf{r}\right) =\frac{Ze}{r}-e\int_{\mathbf{R}^{3}}\frac{\rho
\left( \mathbf{r}^{\prime }\right) }{\left\vert \mathbf{r}-\mathbf{r}%
^{\prime }\right\vert }\ dv^{\prime },  \label{rscr1}
\end{equation}%
where%
\begin{eqnarray}
&&e\rho \left( \mathbf{r}\right) =e\psi ^{\dagger }\left( \mathbf{r}\right)
\psi \left( \mathbf{r}\right) =eQ_{jm}\left( \mathbf{n}\right) \left(
F^{2}\left( r\right) +G^{2}\left( r\right) \right) ,  \label{rscr1aa} \\
&&\qquad \qquad Q_{jm}\left( \mathbf{n}\right) =\left( \mathcal{Y}_{jm}^{\pm
}\left( \mathbf{n}\right) \right) ^{\dagger }\ \mathcal{Y}_{jm}^{\pm }\left(
\mathbf{n}\right)   \notag
\end{eqnarray}%
is the charge distribution of the electron in the atom. We evaluate the
integral with the help of the generating relation (\ref{scr1a}). Indeed,%
\begin{eqnarray}
&&\dint_{\mathbf{R}^{3}}\dfrac{\rho \left( \mathbf{r}^{\prime }\right) }{%
\left\vert \mathbf{r}-\mathbf{r}^{\prime }\right\vert }\ \left( r^{\prime
}\right) ^{2}dr^{\prime }d\omega ^{\prime }=\sum_{s=0}^{\infty }\frac{4\pi }{%
2s+1}\dint_{0}^{\infty }\frac{r_{<}^{s}}{r_{>}^{s+1}}\ \left( F^{2}\left(
r^{\prime }\right) +G^{2}\left( r^{\prime }\right) \right) \left( r^{\prime
}\right) ^{2}dr^{\prime }  \notag \\
&&\quad \ \qquad \qquad \qquad \qquad \qquad \times \sum_{m^{\prime
}=-s}^{s}Y_{sm^{\prime }}\left( \mathbf{n}\right) \
\dint_{S^{2}}Y_{sm^{\prime }}^{\ast }\left( \mathbf{n}^{\prime }\right)
Q_{jm}\left( \mathbf{n}^{\prime }\right) \ d\omega ^{\prime }.
\label{rscr1a}
\end{eqnarray}%
In view of (\ref{ang14}) and (\ref{scr8b}),
\begin{equation}
Q_{jm}\left( \theta ^{\prime }\right) =\sum_{p=0}^{j-1/2}\sqrt{\frac{4\pi }{%
4p+1}}\ a_{p}\left( j,m\right) \;Y_{2p,\;0}\left( \mathbf{n}^{\prime
}\right) ,  \label{rscr2}
\end{equation}%
and with the help of the orthogonality property (\ref{scr2}) one gets%
\begin{equation*}
\dint_{S^{2}}Y_{sm^{\prime }}^{\ast }\left( \mathbf{n}^{\prime }\right)
Q_{jm}\left( \mathbf{n}^{\prime }\right) \ d\omega ^{\prime }=\delta
_{m^{\prime }0}\sum_{p=0}^{j-1/2}\sqrt{\frac{4\pi }{4p+1}}\;a_{p}\left(
j,m\right) \;\delta _{s,\;2p}.
\end{equation*}%
Substitution to (\ref{rscr1a}) gives%
\begin{eqnarray}
V\left( \mathbf{r}\right)  &=&\frac{Ze}{r}-e\sum_{s=0}^{j-1/2}\ P_{2s}\left(
\cos \theta \right) \ C_{jm,\;2s0}^{jm}  \label{rscr6} \\
&&\times \left( -1\right) ^{s}\sqrt{\frac{\left( 2j+2s+1\right) \left(
2j-2s\right) !}{\left( 2j+1\right) \left( 2j+2s\right) !}}\frac{\left(
j+s-1/2\right) !\left( 2s\right) !}{\left( j-s-1/2\right) !\left( s!\right)
^{2}}  \notag \\
&&\quad \quad \qquad \times \dint_{0}^{\infty }\frac{r_{<}^{2s}}{r_{>}^{2s+1}%
}\ \left( F^{2}\left( r^{\prime }\right) +G^{2}\left( r^{\prime }\right)
\right) \left( r^{\prime }\right) ^{2}dr^{\prime }.  \notag
\end{eqnarray}%
with the help of (\ref{ang15}).\smallskip

The integral over the radial functions in (\ref{rscr6}) can be rewritten\ in
the form%
\begin{eqnarray}
&&\dint_{0}^{\infty }\frac{r_{<}^{2s}}{r_{>}^{2s+1}}\ \left( F^{2}\left(
r^{\prime }\right) +G^{2}\left( r^{\prime }\right) \right) \;\left(
r^{\prime }\right) ^{2}dr^{\prime }\qquad  \label{rscr7} \\
&&\ =\frac{1}{r^{2s+1}}\int_{0}^{r}\ \left( r^{\prime }\right) ^{2s+2}\left(
F^{2}\left( r^{\prime }\right) +G^{2}\left( r^{\prime }\right) \right)
\;dr^{\prime }  \notag \\
&&\ \ \ \ \ +r^{2s}\int_{r}^{\infty }\ \left( r^{\prime }\right)
^{1-2s}\left( F^{2}\left( r^{\prime }\right) +G^{2}\left( r^{\prime }\right)
\right) \;dr^{\prime }  \notag \\
&&\ =\dfrac{1}{r^{2s+1}}\ \int_{0}^{\infty }\ \left( r^{\prime }\right)
^{2s+2}\left( F^{2}\left( r^{\prime }\right) +G^{2}\left( r^{\prime }\right)
\right) \;dr^{\prime }  \notag \\
&&\ \ \ \ -\dfrac{1}{r^{2s+1}}\ \int_{r}^{\infty }\ \left( r^{\prime
}\right) ^{2s+2}\left( F^{2}\left( r^{\prime }\right) +G^{2}\left( r^{\prime
}\right) \right) \;dr^{\prime }  \notag \\
&&\ \ \ \ \ \ +r^{2s}\ \int_{r}^{\infty }\ \left( r^{\prime }\right)
^{1-2s}\left( F^{2}\left( r^{\prime }\right) +G^{2}\left( r^{\prime }\right)
\right) \;dr^{\prime },  \notag
\end{eqnarray}%
where the first integral is given by (\ref{rc8}). The next two can be
evaluated with the help of (\ref{i13}).\smallskip

For the electron in the $1S_{1/2}$-state with the radial functions (\ref%
{rc6b}) the result is%
\begin{eqnarray}
V\left( r\right) &=&\frac{\left( Z-1\right) e}{r}+\frac{e\left(
2Z/a_{0}\right) ^{2\nu _{1}}}{\Gamma \left( 2\nu _{1}+1\right) }r^{2\nu
_{1}-1}e^{-2Zr/a_{0}}  \label{rscr8} \\
&&+\;\frac{\Gamma \left( 2\nu _{1},2Zr/a_{0}\right) }{\Gamma \left( 2\nu
_{1}+1\right) }\left( \frac{2\nu _{1}e}{r}-\frac{2Ze}{a_{0}}\right) ,  \notag
\end{eqnarray}%
where $\nu _{1}=\sqrt{1-\mu ^{2}}.$ For small distances $r\rightarrow 0$ the
effective potential $V\left( r\right) \rightarrow eZ/r$ and as $r\rightarrow
\infty $ the potential $V\left( r\right) \rightarrow e\left( Z-1\right) /r$
which is the potential of the nucleus of charge $Ze$ screened by the
electron. In the limit $c\rightarrow \infty $ one gets the nonrelativistic
formula (\ref{scr9}).

\section{Special Functions and Quantum Mechanics}

In this section we give a short summary of Nikiforov and Uvarov's approach
to special functions of mathematical physics and their applications in
quantum mechanics \cite{Ni:Uv}.

\subsection{Generalized Equation of Hypergeometric Type}

The second order differential equation of the form%
\begin{equation}
u^{\prime \prime }+\frac{\widetilde{\tau }\left( z\right) }{\sigma \left(
z\right) }\ u^{\prime }+\frac{\widetilde{\sigma }\left( z\right) }{\sigma
^{2}\left( z\right) }\ u=0,  \label{nu1}
\end{equation}%
where $\sigma \left( z\right) $ and $\widetilde{\sigma }\left( z\right) $
are polynomials of degree at most $2$ and $\widetilde{\tau }\left( z\right) $
is a polynomial of degree at most $1$ of a complex variable $z,$ is called
the \textit{generalized equation of hypergeometric type}. By the
substitution $u=\varphi \left( z\right) y$ equation (\ref{nu1}) can be
reduced to the \textit{equation of hypergeometric type}%
\begin{equation}
\sigma \left( z\right) y^{\prime \prime }+\tau \left( z\right) y^{\prime
}+\lambda y=0,  \label{nu2}
\end{equation}%
where $\tau \left( z\right) $ is a polynomial of degree at most $1,$ and $%
\lambda $ is a constant. The factor $\varphi \left( z\right) $ here satisfies%
\begin{equation}
\frac{\varphi ^{\prime }}{\varphi }=\frac{\pi \left( z\right) }{\sigma
\left( z\right) },  \label{nu3}
\end{equation}%
where $\pi \left( z\right) $ is a polynomial of degree at most $1$ given by
a quadratic formula%
\begin{equation}
\pi \left( z\right) =\frac{\sigma ^{\prime }-\widetilde{\tau }}{2}\pm \sqrt{%
\left( \frac{\sigma ^{\prime }-\widetilde{\tau }}{2}\right) ^{2}-\widetilde{%
\sigma }+k\sigma }  \label{nu4}
\end{equation}%
and constant $k$ is determined by the condition that the discriminant of the
quadratic polynomial under the square root sign is zero. Then $\tau \left(
z\right) $ and $\lambda $ are determined by%
\begin{equation}
\tau \left( z\right) =\widetilde{\tau }\left( z\right) +2\pi \left( z\right)
,\qquad \lambda =k+\pi ^{\prime }\left( z\right) .  \label{nu5}
\end{equation}%
Two exceptions are \cite{Ni:Uv}:

\begin{enumerate}
\item If $\sigma \left( z\right) $ has a double root, $\sigma \left(
z\right) =\left( z-a\right) ^{2},$ the original equation can be carried out
into a generalized equation of hypergeometric type with $\sigma \left(
s\right) =s,$ by a substitution $s=\left( z-a\right) ^{-1}.$

\item If $\sigma \left( z\right) =1$ and $\left( \widetilde{\tau }\left(
z\right) /2\right) ^{2}-\widetilde{\sigma }\left( z\right) $ is a polynomial
of degree $1,$ the substitution $\pi \left( z\right) =-\widetilde{\tau }%
\left( z\right) /2$ reduces the original equation to the form%
\begin{equation}
y^{\prime \prime }+\left( az+b\right) y=0.  \label{nu6}
\end{equation}%
The linear transformation $s=az+b$ takes this into a Lommel equation (\ref%
{lomm}).
\end{enumerate}

Solutions of (\ref{nu1})--(\ref{nu2}) are known as special functions of
hypergeometric type; they include classical orthogonal polynomials,
hypergeometric and confluent hypergeometric functions, Hermite functions,
Bessel functions and spherical harmonics. These functions are often called
special functions of mathematical physics.

\subsection{Classical Orthogonal Polynomials}

The Jacobi, Laguerre and Hermite polynomials are the simplest solutions of
the equation of hypergeometric type. By differentiating (\ref{nu2}) we
verify that the function $v_{1}=y^{\prime }\left( z\right) $ satisfy the
equation of the same type%
\begin{equation}
\sigma \left( z\right) v_{1}{}^{\prime \prime }+\tau _{1}\left( z\right)
v_{1}{}^{\prime }+\mu _{1}v_{1}=0,  \label{nup1}
\end{equation}%
where $\tau _{1}\left( z\right) =$ $\tau \left( z\right) +\sigma ^{\prime
}\left( z\right) $ is a polynomial of degree at most $1,$ and $\mu
_{1}=\lambda +\tau ^{\prime }\left( z\right) $ is a constant.

The converse is also true: any solution of (\ref{nup1}) is the derivative of
a solution of (\ref{nu2}) if $\lambda =\mu _{1}-\tau ^{\prime }\neq 0.$ Let $%
v_{1}\left( z\right) $ be a solution of (\ref{nup1}) and define the function%
\begin{equation*}
y\left( z\right) =-\frac{1}{\lambda }\left( \sigma \left( z\right)
v_{1}^{\prime }+\tau \left( z\right) v_{1}\right) .
\end{equation*}%
Then%
\begin{equation*}
\lambda y^{\prime }=-\left( \sigma v_{1}^{\prime \prime }+\tau
_{1}v_{1}^{\prime }+\tau ^{\prime }v_{1}\right) =\lambda v_{1}
\end{equation*}%
or $v_{1}=y^{\prime }\left( z\right) $ and, therefore, $y\left( z\right) $
satisfy (\ref{nu2}).

By differentiating (\ref{nu2}) $n$ times we obtain an equation of
hypergeometric type for the function $v_{n}=y^{\left( n\right) }\left(
z\right) ,$%
\begin{equation}
\sigma \left( z\right) v_{n}{}^{\prime \prime }+\tau _{n}\left( z\right)
v_{n}{}^{\prime }+\mu _{n}v_{n}=0,  \label{nup2}
\end{equation}%
where%
\begin{eqnarray}
&&\tau _{n}\left( z\right) =\tau \left( z\right) +n\sigma ^{\prime }\left(
z\right) ,  \label{nup3} \\
&&\mu _{n}=\lambda +n\tau ^{\prime }+\frac{1}{2}n\left( n-1\right) \sigma
^{\prime \prime }.  \label{nup4}
\end{eqnarray}%
This property lets us construct the simplest solutions of (\ref{nu2})
corresponding to some values of $\lambda .$ Indeed, when $\mu _{n}=0$
equation (\ref{nup2}) has the trivial solution $v_{n}=\ $constant$.$ Since $%
v_{n}\left( z\right) =y^{\left( n\right) }\left( z\right) ,$ the equation (%
\ref{nu2}) has a particular solution $y=y_{n}\left( z\right) $ which is a
polynomial of degree $n$ if%
\begin{equation}
\lambda =\lambda _{n}=-n\tau ^{\prime }-\frac{1}{2}n\left( n-1\right) \sigma
^{\prime \prime }\qquad \left( n=0,1,2,\ ...\ \right) .  \label{nup5}
\end{equation}

To find these polynomials explicitly let us rewrite equations (\ref{nu2})
and (\ref{nup2}) in the self-adjoint forms%
\begin{equation}
\left( \sigma \rho y^{\prime }\right) ^{\prime }+\lambda \rho y=0,
\label{nup6}
\end{equation}%
\begin{equation}
\left( \sigma \rho _{n}v_{n}^{\prime }{}\right) ^{\prime }+\mu _{n}\rho
_{n}v_{n}=0.  \label{nup7}
\end{equation}%
Functions $\rho \left( z\right) $ and $\rho _{n}\left( z\right) $ satisfy
the first order differential equations%
\begin{equation}
\left( \sigma \rho \right) ^{\prime }=\tau \rho ,  \label{nup8}
\end{equation}%
\begin{equation}
\left( \sigma \rho _{n}\right) ^{\prime }=\tau _{n}\rho _{n}.  \label{nup9}
\end{equation}%
So,%
\begin{equation*}
\dfrac{\left( \sigma \rho _{n}\right) ^{\prime }}{\rho _{n}}=\tau +n\sigma
^{\prime }=\dfrac{\left( \sigma \rho \right) ^{\prime }}{\rho }+n\sigma
^{\prime },
\end{equation*}%
whence%
\begin{equation*}
\dfrac{\rho _{n}^{\prime }}{\rho _{n}}=\dfrac{\rho ^{\prime }}{\rho }+n%
\dfrac{\sigma ^{\prime }}{\sigma }
\end{equation*}%
and, consequently,%
\begin{equation}
\rho _{n}\left( z\right) =\sigma ^{n}\left( z\right) \rho \left( z\right) .
\label{nup10}
\end{equation}%
Since $\sigma \rho _{n}=\rho _{n+1}$ and $v_{n}^{\prime }=v_{n+1}$ one can
rewrite (\ref{nup7}) in the form%
\begin{equation*}
\rho _{n}v_{n}=-\frac{1}{\mu _{n}}\left( \rho _{n+1}v_{n+1}\right) ^{\prime
}.
\end{equation*}%
Hence we obtain successively%
\begin{eqnarray*}
\rho y &=&\rho _{0}v_{0}=-\frac{1}{\mu _{0}}\left( \rho _{1}v_{1}\right)
^{\prime } \\
&=&\left( -\frac{1}{\mu _{0}}\right) \left( -\frac{1}{\mu _{1}}\right)
\left( \rho _{2}v_{2}\right) ^{\prime \prime } \\
&&\vdots \\
&=&\frac{1}{A_{n}}\left( \rho _{n}v_{n}\right) ^{\left( n\right) },
\end{eqnarray*}%
where%
\begin{equation}
A_{0}=1,\qquad A_{n}=\left( -1\right) ^{n}\prod_{k=0}^{n-1}\mu _{k}.
\label{nup11}
\end{equation}%
If $y=y_{n}\left( z\right) $ is a polynomial of degree $n,$ then $%
v_{n}=y_{n}^{\left( n\right) }\left( z\right) =\ $constant and we arrive at
the \textit{Rodrigues formula} for polynomial solutions of (\ref{nu2}),%
\begin{equation}
y_{n}\left( z\right) =\frac{B_{n}}{\rho \left( z\right) }\left( \sigma
^{n}\left( z\right) \rho \left( z\right) \right) ^{\left( n\right) },
\label{nup12}
\end{equation}%
where $B_{n}=A_{n}^{-1}y_{n}^{\left( n\right) }$ is a constant. These
solutions correspond to the eigenvalues (\ref{nup5}).

The polynomial solutions of (\ref{nu2}) obey an orthogonality property. Let
us write equations for polynomials $y_{n}\left( x\right) $ and $y_{m}\left(
x\right) $ in the self-adjoint form%
\begin{equation*}
\left( \sigma \left( x\right) \rho \left( x\right) y_{n}^{\prime }\left(
x\right) \right) ^{\prime }+\lambda _{n}\rho \left( x\right) y_{n}\left(
x\right) =0,
\end{equation*}%
\begin{equation*}
\left( \sigma \left( x\right) \rho \left( x\right) y_{m}^{\prime }\left(
x\right) \right) ^{\prime }+\lambda _{m}\rho \left( x\right) y_{m}\left(
x\right) =0,
\end{equation*}%
multiply the first equation by $y_{m}\left( x\right) $ and the second by $%
y_{n}\left( x\right) ,$ subtract the second equality from the first one and
then integrate the result over $x$ on the interval $\left( a,b\right) .$
Since%
\begin{eqnarray*}
&&y_{m}\left( x\right) \left( \sigma \left( x\right) \rho \left( x\right)
y_{n}^{\prime }\left( x\right) \right) ^{\prime }-y_{n}\left( x\right)
\left( \sigma \left( x\right) \rho \left( x\right) y_{m}^{\prime }\left(
x\right) \right) ^{\prime } \\
&&\quad =\frac{d}{dx}\left[ \sigma \left( x\right) \rho \left( x\right)
W\left( y_{m}\left( x\right) ,y_{n}\left( x\right) \right) \right] ,
\end{eqnarray*}%
where $W\left( u,v\right) =uv^{\prime }-vu^{\prime }$ is the Wronskian, we
get%
\begin{equation}
\left( \lambda _{m}-\lambda _{n}\right) \int_{a}^{b}y_{m}\left( x\right)
y_{n}\left( x\right) \ \rho \left( x\right) dx=\left. \left[ \sigma \left(
x\right) \rho \left( x\right) W\left( y_{m}\left( x\right) ,y_{n}\left(
x\right) \right) \right] \right| _{x=a}^{b}.  \label{nup13}
\end{equation}%
If the conditions%
\begin{equation}
\left. \sigma \left( x\right) \rho \left( x\right) x^{k}\right|
_{x=a,b}=0,\qquad k=0,1,2,\ ...  \label{nup14}
\end{equation}%
are satisfied for some points $a$ and $b,$ then the right hand side of (\ref%
{nup13}) vanishes because the Wronskian is a polynomial in $x.$ Therefore,
we arrive at the \textit{orthogonality property}%
\begin{equation}
\int_{a}^{b}y_{m}\left( x\right) y_{n}\left( x\right) \ \rho \left( x\right)
dx=0  \label{nup15}
\end{equation}%
provided that $\lambda _{m}\neq \lambda _{n}.$ One can replace this
condition by $m\neq n$ due to the relation $\lambda _{m}-\lambda _{n}=\left(
m-n\right) \left( \tau ^{\prime }+\left( n+m-1\right) \sigma ^{\prime \prime
}/2\right) $ if $\tau ^{\prime }+\left( n+m-1\right) \sigma ^{\prime \prime
}/2\neq 0.$\smallskip

We shall refer to polynomial solutions of (\ref{nu2}) obeying the
orthogonality property (\ref{nup15}) with respect to a positive weight
function $\rho \left( x\right) $ on a real interval $\left( a,b\right) $ as
\textit{classical orthogonal polynomials}.\smallskip

Equation%
\begin{equation}
\left( \sigma \left( x\right) \rho \left( x\right) \right) ^{\prime }=\tau
\left( x\right) \rho \left( x\right)  \label{nup16}
\end{equation}%
for the weight function $\rho \left( x\right) $ is usually called the
\textit{Pearson equation}. By using the linear transformations of
independent variable $x$ one can reduce solutions of (\ref{nup16}) to the
following canonical forms%
\begin{equation}
\rho \left( x\right) =\left\{
\begin{array}{l}
\left( 1-x\right) ^{\alpha }\left( 1+x\right) ^{\beta }\qquad \text{%
for\qquad }\sigma \left( x\right) =1-x^{2},\medskip \\
x^{\alpha }e^{-x}\qquad \text{for\qquad }\sigma \left( x\right) =x,\medskip
\\
e^{-x^{2}}\qquad \text{for\qquad }\sigma \left( x\right) =1.\medskip%
\end{array}%
\right.  \label{nup17}
\end{equation}%
The corresponding orthogonal polynomials are the \textit{Jacobi polynomials}
$P_{n}^{\left( \alpha ,\ \beta \right) }\left( x\right) ,$ the \textit{%
Laguerre polynomials} $L_{n}^{\alpha }\left( x\right) $ and the \textit{%
Hermite polynomials} $H_{n}\left( x\right) .$\smallskip

The basic information about the classical orthogonal polynomials is given in
Table, which contains also the leading coefficients in the expansion $%
y_{n}(x)=a_{n}x^{n}+b_{n}x^{n-1}+\ \ldots \ ,$ the squared norms%
\begin{equation}
d_{n}^{2}=\int_{a}^{b}y_{n}^{2}(x)\ \rho (x)dx  \label{nup18}
\end{equation}%
and the coefficients of the three-term recurrence relation%
\begin{equation}
x\,y_{n}(x)=\alpha _{n}\,y_{n+1}(x)+\beta _{n}\,y_{n}(x)+\gamma
_{n}\,y_{n-1}(x)  \label{nup19}
\end{equation}%
with%
\begin{equation}
\alpha _{n}=\frac{a_{n}}{a_{n+1}},\quad \beta _{n}=\frac{b_{n}}{a_{n}}-\frac{%
b_{n+1}}{a_{n+1}},\quad \gamma _{n}=\alpha _{n-1}\,\frac{d_{n}^{2}}{%
d_{n-1}^{2}}.  \label{nup20}
\end{equation}%
\bigskip\ \bigskip\ \newline
\
\begin{tabular}{|l|c|c|c|}
\hline
$y_{n}(x)$ & $\qquad P_{n}^{(\alpha ,\,\beta )}(x)\,(\alpha >-1,\beta
>-1)\qquad $ & $\qquad L_{n}^{\alpha }(x)\,(\alpha >-1)\qquad $ & $\quad
H_{n}(x)\quad $ \\ \hline
&  &  &  \\
$(a,b)$ & $(-1,1)$ & $(0,\infty )$ & $(-\infty ,\infty )$ \\
$\rho (x)$ & $(1-x)^{\alpha }(1+x)^{\beta }$ & $x^{\alpha }\,e^{-x}$ & $%
e^{-x^{2}}$ \\
$\sigma (x)$ & $1-x^{2}$ & $x$ & $1$ \\
$\tau (x)$ & $\beta -\alpha -(\alpha +\beta +2)\,x$ & $1+\alpha -x$ & $-2x$
\\
$\lambda _{n}$ & $n(\alpha +\beta +n+1)$ & $n$ & $2n$ \\
&  &  &  \\
$B_{n}$ & $\dfrac{(-1)^{n}}{2^{n}n!}$ & $\dfrac{1}{n!}$ & $(-1)^{n}$ \\
$a_{n}$ & $\dfrac{\Gamma (\alpha +\beta +2n+1)}{2^{n}n!\Gamma (\alpha +\beta
+n+1)}$ & $\dfrac{(-1)^{n}}{n!}$ & $2^{n}$ \\
$b_{n}$ & $\dfrac{(\alpha -\beta )\Gamma (\alpha +\beta +2n)}{%
2^{n}(n-1)!\Gamma (\alpha +\beta +n+1)}$ & $\left( -1\right) ^{n-1}\dfrac{%
\alpha +n}{(n-1)!}$ & $0$ \\
$d^{2}$ & $\dfrac{2^{\alpha +\beta +1}\Gamma (\alpha +n+1)\Gamma (\beta +n+1)%
}{n!(\alpha +\beta +2n+1)\Gamma (\alpha +\beta +n+1)}$ & $\dfrac{\Gamma
(\alpha +n+1)}{n!}$ & $2^{n}n!\sqrt{\pi }$ \\
&  &  &  \\
$\alpha _{n}$ & $\dfrac{2(n+1)(\alpha +\beta +n+1)}{(\alpha +\beta
+2n+1)(\alpha +\beta +2n+2)}$ & $-(n+1)$ & $\dfrac{1}{2}$ \\
$\beta _{n}$ & $\dfrac{\beta ^{2}-\alpha ^{2}}{(\alpha +\beta +2n)(\alpha
+\beta +2n+2)}$ & $\alpha +2n+1$ & $0$ \\
$\gamma _{n}$ & $\dfrac{2(\alpha +n)(\beta +n)}{(\alpha +\beta +2n)(\alpha
+\beta +2n+1)}$ & $-(\alpha +n)$ & $n$ \\ \hline
\end{tabular}
\bigskip\ \bigskip\ \newline
\ More details about the Jacobi, Laguerre and Hermite polynomials and their
numerous extensions can be found in \cite{An:As:Ro}, \cite{As:Wi}, \cite%
{Ga:Ra}, \cite{Ko:Sw}, \cite{Ni:Su:Uv}, \cite{Ni:Uv}, \cite{Suslov}, \cite%
{Sze} and references therein.

\subsection{Classical Orthogonal Polynomials and Eigenvalue Problems}

The following theorem is a useful tool for finding of the square integrable
solutions of basic problems in quantum mechanics \cite{Ni:Uv}.

\begin{theorem}
Let $y=y\left( x\right) $ be a solution of the equation of hypergeometric
type (\ref{nu2}) and let $\rho \left( x\right) ,$ a solution of the Pearson
equation (\ref{nup16}), be bounded on the interval $\left( a,b\right) $ and
satisfy the boundary conditions (\ref{nup14}). Then nontrivial solutions of (%
\ref{nu2}) such that $y\left( x\right) \sqrt{\rho \left( x\right) }$ is
bounded and of integrable square on $\left( a,b\right) $ exist only for the
eigenvalues given by (\ref{nup5}); they are the corresponding classical
orthogonal polynomials on $\left( a,b\right) $ and can be found by the
Rodrigues-type formula (\ref{nup12}).
\end{theorem}

The proof is given in \cite{Ni:Uv}.

\subsection{Integral Representation for Special Functions}

The differential equation of hypergeometric type (\ref{nu2}) can be
rewritten in self-adjoint form%
\begin{equation}
\left( \sigma \left( z\right) \rho \left( z\right) y^{\prime }\left(
z\right) \right) ^{\prime }+\lambda \rho \left( z\right) y\left( z\right) =0,
\label{nu7}
\end{equation}%
where $\rho \left( z\right) $ satisfies the first order equation%
\begin{equation}
\left( \sigma \left( z\right) \rho \left( z\right) \right) ^{\prime }=\tau
\left( z\right) \rho \left( z\right) .  \label{nu8}
\end{equation}%
Nikiforov and Uvarov \cite{Ni:Uv} suggested to construct particular
solutions of the differential equation of hypergeometric type (\ref{nu2}) in
a form of a general integral representation for special functions of
hypergeometric type as a refinement of the Laplace method. A slightly
modified version of their main theorem is

\begin{theorem}
Let $\rho \left( z\right) $ satisfy (\ref{nu8}) and $\nu $ be a root of the
equation%
\begin{equation}
\lambda +\nu \tau ^{\prime }+\frac{1}{2}\nu \left( \nu -1\right) \sigma
^{\prime \prime }=0.  \label{nu10}
\end{equation}%
Then the differential equation (\ref{nu2}) has a particular solution of the
form%
\begin{equation}
y\left( z\right) =y_{\nu }\left( z\right) =\frac{C_{\nu }}{\rho \left(
z\right) }\ \dint_{C}\frac{\sigma ^{\nu }\left( s\right) \rho \left(
s\right) }{\left( s-z\right) ^{\nu +1}}\ ds,  \label{nu11}
\end{equation}%
where $C_{\nu }$ is a constant and $C$ is a contour in the complex $s$%
-plane, if:

\begin{enumerate}
\item the derivative of the integral%
\begin{equation}
\varphi _{\nu \mu }\left( z\right) =\int_{C}\frac{\rho _{\nu }\left(
s\right) }{\left( s-z\right) ^{\mu +1}}\ ds\qquad \text{with\qquad }\rho
_{\nu }\left( s\right) =\sigma ^{\nu }\left( s\right) \rho \left( s\right)
\label{nu12}
\end{equation}%
can be evaluated for $\mu =\nu -1$ and $\mu =\nu $ by using the formula%
\begin{equation}
\varphi _{\nu \mu }^{\prime }\left( z\right) =\left( \mu +1\right) \ \varphi
_{\nu ,\ \mu +1}\left( z\right) ;  \label{nu12a}
\end{equation}

\item the contour $C$ is chosen so that the equality%
\begin{equation}
\left. \frac{\sigma ^{\nu +1}\left( s\right) \rho \left( s\right) }{\left(
s-z\right) ^{\nu +1}}\right| _{s_{1}}^{s_{2}}=0  \label{nu13}
\end{equation}%
holds, where $s_{1}$ and $s_{2}$ are end points of the contour $C.$
\end{enumerate}
\end{theorem}

We present here a simple proof of this theorem \cite{Suslov}, which is
different from one in \cite{Ni:Uv}.

\begin{proof}
The function $\rho _{\nu }\left( s\right) =\sigma ^{\nu }\left( s\right)
\rho \left( s\right) $ satisfy the equation%
\begin{equation}
\left( \sigma \left( s\right) \rho _{\nu }\left( s\right) \right) ^{\prime
}=\tau _{\nu }\left( s\right) \rho _{\nu }\left( s\right) ,  \label{nu14}
\end{equation}%
where $\tau _{\nu }\left( s\right) =\tau \left( s\right) +\nu \sigma
^{\prime }\left( s\right) .$ We multiply both sides of this equality by $%
\left( s-z\right) ^{-\nu -1}$ and integrate over contour $C.$ Upon
integrating by parts we obtain%
\begin{equation}
\left. \frac{\sigma \left( s\right) \rho _{\nu }\left( s\right) }{\left(
s-z\right) ^{\nu +1}}\right| _{s_{1}}^{s_{2}}+\left( \nu +1\right) \int_{C}%
\frac{\sigma \left( s\right) \rho _{\nu }\left( s\right) }{\left( s-z\right)
^{\nu +2}}\ ds=\int_{C}\frac{\tau _{\nu }\left( s\right) \rho _{\nu }\left(
s\right) }{\left( s-z\right) ^{\nu +1}}\ ds.  \label{nu15}
\end{equation}%
By hypothesis, the first term is equal to zero. We expand polynomials $%
\sigma \left( s\right) $ and $\tau _{\nu }\left( s\right) $ in powers of $%
s-z:$%
\begin{eqnarray*}
\sigma \left( s\right) &=&\sigma \left( z\right) +\sigma ^{\prime }\left(
z\right) \left( s-z\right) +\frac{1}{2}\sigma ^{\prime \prime }\ \left(
s-z\right) ^{2}, \\
\tau _{\nu }\left( s\right) &=&\tau _{\nu }\left( z\right) +\tau _{\nu
}^{\prime }\ \left( s-z\right) .
\end{eqnarray*}%
Taking into account the integral formulas for the functions $\varphi _{\nu
,\ \nu -1}\left( z\right) ,$ $\varphi _{\nu \nu }\left( z\right) $ and $%
\varphi _{\nu ,\ \nu +1}\left( z\right) ,$ we arrive at the relation%
\begin{equation*}
\left( \nu +1\right) \left( \sigma \left( z\right) \varphi _{\nu ,\ \nu
+1}+\sigma ^{\prime }\left( z\right) \varphi _{\nu \nu }+\frac{1}{2}\sigma
^{\prime \prime }\ \varphi _{\nu ,\ \nu -1}\right) =\tau _{\nu }\left(
z\right) \varphi _{\nu \nu }+\tau _{\nu }^{\prime }\ \varphi _{\nu ,\ \nu
-1}.
\end{equation*}%
Upon substituting $\tau _{\nu }=\tau +\nu \sigma ^{\prime }$ and using the
formula $\varphi _{\nu \nu }^{\prime }=\left( \nu +1\right) \varphi _{\nu ,\
\nu +1}$ one gets%
\begin{equation}
\sigma \varphi _{\nu \nu }^{\prime }+\left( \sigma ^{\prime }-\tau \right)
\varphi _{\nu \nu }=\left( \tau ^{\prime }+\frac{1}{2}\left( \nu -1\right)
\sigma ^{\prime \prime }\right) \varphi _{\nu ,\ \nu -1}.  \label{nu16}
\end{equation}

At the sane time, by differentiating the relation $\sigma \rho y^{\prime
}=C_{\nu }\sigma \varphi _{\nu \nu }$ we find that%
\begin{equation}
\frac{1}{C_{\nu }}\sigma \rho y^{\prime }=\sigma \varphi _{\nu \nu }^{\prime
}+\left( \sigma ^{\prime }-\tau \right) \varphi _{\nu \nu }.  \label{nu17}
\end{equation}%
Comparing (\ref{nu16}) and (\ref{nu17}) we obtain%
\begin{equation}
\sigma \rho y^{\prime }=\kappa _{\nu }C_{\nu }\varphi _{\nu ,\ \nu -1},
\label{nu18}
\end{equation}%
where $\kappa _{\nu }=\tau ^{\prime }+\left( \nu -1\right) \sigma ^{\prime
\prime }/2.$ Upon differentiating (\ref{nu18}) we arrive to the equation of
hypergeometric type in the self-adjoint form%
\begin{equation*}
\left( \sigma \rho y^{\prime }\right) ^{\prime }+\lambda \rho y=0,
\end{equation*}%
where $\lambda =-\nu \kappa _{\nu }=-\nu \tau ^{\prime }-\nu \left( \nu
-1\right) \sigma ^{\prime \prime }/2.$ This proves the theorem.
\end{proof}

In the proof of Theorem~2 we have, en route, deduced the formula (\ref{nu18}%
), which is a simple integral representation for the first derivative of the
function of hypergeometric type:%
\begin{equation}
y_{\nu }^{\prime }\left( z\right) =\frac{C_{\nu }^{\left( 1\right) }}{\sigma
\left( z\right) \rho \left( z\right) }\ \dint_{C}\frac{\rho _{\nu }\left(
s\right) }{\left( s-z\right) ^{\nu }}\ ds,  \label{nu19}
\end{equation}%
where $C_{\nu }^{\left( 1\right) }=\kappa _{\nu }C_{\nu }=\left( \tau
^{\prime }+\dfrac{1}{2}\left( \nu -1\right) \sigma ^{\prime \prime }\right)
C_{\nu }.$ Hence%
\begin{equation}
y_{\nu }^{\left( k\right) }\left( z\right) =\frac{C_{\nu }^{\left( k\right) }%
}{\rho _{k}\left( z\right) }\ \varphi _{\nu ,\ \nu -k}\left( z\right) =\frac{%
C_{\nu }^{\left( k\right) }}{\sigma ^{k}\left( z\right) \rho \left( z\right)
}\ \dint_{C}\frac{\rho _{\nu }\left( s\right) }{\left( s-z\right) ^{\nu -k+1}%
}\ ds,  \label{nu20}
\end{equation}%
where $C_{\nu }^{\left( k\right) }=\dprod_{p=0}^{k-1}\left( \tau ^{\prime }+%
\dfrac{1}{2}\left( \nu +p-1\right) \sigma ^{\prime \prime }\right) C_{\nu }.$

See \cite{Ni:Su:UvDAH}, \cite{Suslov} and \cite{Su2} for an extension of
this theorem to the case of the so-called difference equation of
hypergeometric type on nonuniform lattices.

\subsection{Power Series Method}

We can construct particular solutions of equation (\ref{nu2}) by using the
power series method; see, for example, the classical work of Boole \cite%
{BooleDE}.

\begin{theorem}
Let $a$ be a root of the equation $\sigma (z)=0.$ Then Eq.~\textrm{(\ref{nu2}%
)} has a particular solution of the form%
\begin{equation}
y(z)=\sum\limits_{n=0}^{\infty }c_{n}(z-a)^{n},  \label{ps1}
\end{equation}
where%
\begin{equation}
\dfrac{c_{n+1}}{c_{n}}=-\frac{\lambda +n\left( \tau ^{\prime }+(n-1)\sigma
^{\prime \prime }2\right) }{(n+1)\left( \tau (a)+n\sigma ^{\prime
}(a)\right) },  \label{ps2}
\end{equation}
if:

\begin{enumerate}
\item \quad $\lim\limits_{m\rightarrow \infty }\dfrac{d^{k}}{dx^{k}}%
\,y_{m}(x)=\dfrac{d^{k}}{dx^{k}}\,y(x)\text{ with }k=0,1,2;\medskip $
\textrm{\ }

\item \quad $\lim\limits_{m\rightarrow \infty }\left( \lambda -\lambda
_{m}\right) c_{m}(x-a)^{m}=0.$
\end{enumerate}

\noindent \textrm{(}Here $y_{m}(x)=\dsum_{n=0}^{m}c_{n}(x-a)^{n}$ and $%
\lambda _{m}=-m\tau ^{\prime }-\dfrac{1}{2}m(m-1)\sigma ^{\prime \prime }.$%
\textrm{)}

In the case $\sigma (z)=\text{constant}\neq 0$ series \textrm{(\ref{ps1})}
satisfies \textrm{(\ref{nu2})} when $a$ is a root of the equation $\tau
(z)=0,$%
\begin{equation}
\dfrac{c_{n+2}}{c_{n}}=-\frac{\lambda +n\tau ^{\prime }}{(n+1)(n+2)\sigma }
\label{ps3}
\end{equation}%
and convergence conditions \textrm{(}$1$\textrm{)}--\textrm{(}$2$\textrm{)}
are valid.
\end{theorem}

\begin{proof}
The proof\/ of the theorem relays on the identity%
\begin{align}
& \rho ^{-1}\frac{d}{dz}\left( \sigma \rho \frac{d}{dz}(z-\xi )^{n}\right)
=\left( \sigma (z)\frac{d^{2}}{dz^{2}}+\tau (z)\frac{d}{dz}\right) (z-\xi
)^{n}  \label{ps4} \\
& \qquad =n(n-1)\sigma (\xi )(z-\xi )^{n-2}\newline
+n\tau _{n-1}(\xi )(z-\xi )^{n-1}-\lambda _{n}(z-\xi )^{n},  \notag
\end{align}%
where $\tau _{m}(\xi )=\tau (\xi )+m\sigma ^{\prime }(\xi )$ and $\lambda
_{n}=-n\tau ^{\prime }-n(n-1)\sigma ^{\prime \prime }/2,$ which can be
easily verified.\smallskip

In fact, for a partial sum of the series (\ref{ps1}) we can write%
\begin{align}
& \left( \sigma (z)\frac{d^{2}}{dz^{2}}+\tau (z)\frac{d}{dz}+\lambda \right)
y_{m}(z)=\sigma (a)\sum\limits_{n=0}^{m}c_{n}n(n-1)(z-a)^{n-2}  \label{ps5}
\\
& \quad \quad \quad \quad \qquad +\sum\limits_{n=0}^{m}c_{n}n\tau
_{n-1}(a)(z-a)^{n-1}+\sum\limits_{n=0}^{m}c_{n}\left( \lambda -\lambda
_{n}\right) (z-a)^{n}.  \notag
\end{align}%
By the hypothesis $\sigma (a)=0$ and the first term in the right hand side
is equal to zero. Equating the coefficients in the next two terms with the
aid of%
\begin{equation}
\dfrac{c_{n+1}}{c_{n}}=\frac{\lambda _{n}-\lambda }{(n+1)\tau _{n}(a)},
\label{ps6}
\end{equation}%
which is equivalent to (\ref{ps2}), one gets%
\begin{equation}
\left( \sigma (z)\frac{d^{2}}{dz^{2}}+\tau (z)\frac{d}{dz}+\lambda \right)
y_{m}(z)=c_{m}\left( \lambda -\lambda _{m}\right) (z-a)^{m}.  \label{ps7}
\end{equation}%
Taking the limit $m\rightarrow \infty $ we prove the first part of the
theorem under the convergence conditions ($1$)--($2$).\smallskip

When $\sigma =$ constant we can obtain in the same manner%
\begin{align}
& \left( \sigma (z)\frac{d^{2}}{dz^{2}}+\tau (z)\frac{d}{dz}+\lambda \right)
y_{m}(z)\newline
\label{ps8} \\
& \quad =\sigma \sum\limits_{n=0}^{m}c_{n}n(n-1)(z-a)^{n-2}\newline
+\sum\limits_{n=0}^{m}c_{n}\left( \lambda -\lambda _{n}\right) (z-a)^{n}
\notag \\
& \quad \newline
\quad =c_{m}\left( \lambda -\lambda _{m}\right) (z-a)^{m},  \notag
\end{align}%
which proves the second part of the theorem in the limit $m\rightarrow
\infty .$
\end{proof}

\noindent \textbf{Corollary.\/} \ \textit{Equation \textrm{(\ref{nu2})} has
polynomial solutions }$y_{m}(x)$\textit{\ corresponding to the eigenvalues }$%
\lambda =\lambda _{m}=-m\tau ^{\prime }-m(m-1)\sigma ^{\prime \prime }/2,$ $%
m=0,1,2,\;...\;.$

\noindent This follows from (\ref{ps7}) and (\ref{ps8}).\medskip

\noindent \textbf{Examples.\/} \ With the aid of linear transformations of
the independent variable, equation (\ref{nu2}) for $\tau ^{\prime }\neq 0$
can be reduced to one of the following \textit{canonical forms\/} \cite%
{Ni:Uv}:%
\begin{equation}
z(1-z)y^{\prime \prime }+\left[ \gamma -(\alpha +\beta +1)z\right] y^{\prime
}-\alpha \beta y=0,  \label{ps9}
\end{equation}%
\begin{equation}
zy^{\prime \prime }+(\gamma -z)y^{\prime }-\alpha y=0,  \label{ps10}
\end{equation}%
\begin{equation}
y^{\prime \prime }-2zy^{\prime }+2\nu y=0.  \label{ps11}
\end{equation}%
According to (\ref{ps1})--(\ref{ps3}) the appropriate particular solutions
are:

\noindent the \textit{hypergeometric function\/},%
\begin{equation}
y(z)={}_{2}F_{1}(\alpha ,\beta ;\gamma ;z)=\sum\limits_{n=0}^{\infty }\,%
\frac{(\alpha )_{n}(\beta )_{n}}{(\gamma )_{n}n!}\,z^{n},  \label{ps12}
\end{equation}%
the \textit{confluent hypergeometric function\/},%
\begin{equation}
y(z)={}_{1}F_{1}(\alpha ;\gamma ;z)=\sum\limits_{n=0}^{\infty }\,\frac{%
(\alpha )_{n}}{(\gamma )_{n}n!}\,z^{n},  \label{ps13}
\end{equation}%
and the \textit{Hermite function},%
\begin{align}
& y(z)=H_{\nu }(z)=\frac{1}{2\Gamma (-\nu )}\,\sum\limits_{n=0}^{\infty
}\Gamma \left( \frac{n-\nu }{2}\right) \frac{(-2z)^{n}}{n!}  \label{ps14} \\
& \quad \quad =\frac{2^{\nu }\Gamma \left( \frac{1}{2}\right) }{\Gamma
\left( \frac{1-\nu }{2}\right) }\,{}_{1}F_{1}\left( -\frac{\nu }{2};\,\frac{1%
}{2};\,z^{2}\right) {}+\frac{2^{\nu }\Gamma \left( -\frac{1}{2}\right) }{%
\Gamma \left( -\frac{\nu }{2}\right) }\,z\,{}_{1}F_{1}\left( \frac{1-\nu }{2}%
;\,\frac{3}{2};\,z^{2}\right) ,  \notag
\end{align}%
respectively. Here $(a)_{n}=a(a+1)\ldots (a+n-1)=\Gamma (a+n)/\Gamma (a)$
and $\Gamma (a)$ is the gamma function of Euler.\smallskip

Generally speaking, these solutions arise under certain restrictions on the
variable and parameters; see equation (\ref{a1}) below for more details.
They can be extended to wider domains by analytic continuation.\vspace*{0.1in%
}

\noindent \textbf{Extended Power Series Method.\/} \ The solution (\ref{ps1}%
)--(\ref{ps2}) can be rewritten in the following explicit form%
\begin{equation}
y(z)=c_{0}\sum\limits_{n=0}^{\infty }\prod\limits_{k=0}^{n-1}\frac{\left(
\lambda -\lambda _{k}\right) (a-z)}{\tau _{k}(a)(k+1)},  \label{ps15}
\end{equation}%
where $c_{0}$ is a constant. Using the expansion

\begin{equation}
y(z)=\sum\limits_{n}c_{n}(z-\xi )^{\alpha +n},\quad \dfrac{c_{n+1}}{c_{n}}=%
\frac{\lambda _{\alpha +n}-\lambda }{(\alpha +n+1)\tau _{\alpha +n}(a)}
\label{ps16}
\end{equation}%
\noindent one can construct solutions of the more general form%
\begin{equation}
y(z)=c_{0}(z-a)^{\alpha }\,\sum\limits_{n=0}^{\infty
}\,\prod\limits_{k=0}^{n-1}\,\ \frac{\left( \lambda -\lambda _{\alpha
+k}\right) (a-z)}{\tau _{\alpha +k}(a)(\alpha +k+1)},  \label{ps17}
\end{equation}%
\noindent provided that $\sigma (a)=0$ and $\alpha \,\tau _{\alpha -1}(a)=0.$
In particular, putting $\alpha =0$ we recover (\ref{ps15}).\smallskip

We can also satisfy (\ref{nu2}) by using the series of the form%
\begin{equation}
y(x)=\sum\limits_{n}\frac{c_{n}}{(x-\xi )^{\alpha +n}},\quad \frac{c_{n+1}}{%
c_{n}}=\frac{(\alpha +n)\tau _{-\alpha -n-1}(a)}{\lambda -\lambda _{-\alpha
-n-1}},  \label{ps18}
\end{equation}

\noindent if $\sigma (a)=0$ and $\lambda =\lambda _{-\alpha }.$ Hence%
\begin{equation}
y(z)=\frac{c_{0}}{(z-a)^{\alpha }}\,\sum\limits_{n=0}^{\infty }\,\
\prod\limits_{k=0}^{n-1}\,\frac{(\alpha +k)\tau _{-\alpha -k-1}(a)}{\left(
\lambda -\lambda _{-\alpha -k-1}\right) (z-a)}.  \label{ps19}
\end{equation}%
\noindent When $\sigma =\text{constant}\neq 0$ one can write the solution as%
\begin{equation}
y(x)=\sum_{n}\frac{c_{n}}{(x-a)^{\alpha +n}},\quad \frac{c_{n+2}}{c_{n}}=-%
\frac{(\alpha +n)(\alpha +n+1)\sigma }{\lambda -\lambda _{-\alpha -n-2}},
\label{ps20}
\end{equation}%
\noindent if $\tau (a)=0$ and $\lambda =\lambda _{-\alpha }$ (for even
integer values of $n$) or $\lambda =\lambda _{-\alpha -1}$ (for odd integer
values of $n$). This method allows to construct the fundamental set of
solutions of equation (\ref{nu2}). Examples are given in \cite{SusSF}. See
\cite{At:SusPS} and \cite{Su2} for an extension of the power series method
to the case of the difference equation of hypergeometric type on nonuniform
lattices.

\subsection{Integrals for Hypergeometric and Bessel Functions}

Using Theorem 2 one can obtain integral representations for all the most
commonly used special functions of hypergeometric type, in particular, for
the hypergeometric functions:%
\begin{equation}
_{2}F_{1}(\alpha ,\beta ;\gamma ;z)=\frac{\Gamma (\gamma )}{\Gamma (\alpha
)\Gamma (\gamma -\alpha )}\,\int_{0}^{1}\,t^{\alpha -1}(1-t)^{\gamma -\alpha
-1}(1-zt)^{-\beta }\,dt,  \label{g1}
\end{equation}%
\begin{equation}
{}_{1}F_{1}(\alpha ;\gamma ;z)=\frac{\Gamma (\gamma )}{\Gamma (\alpha
)\Gamma (\gamma -\alpha )}\,\int_{0}^{1}\,t^{\alpha -1}(1-t)^{\gamma -\alpha
-1}e^{zt}\,dt,  \label{g2}
\end{equation}%
\begin{equation}
H_{\nu }(z)=\frac{1}{\Gamma (-\nu )}\,\int_{0}^{\infty
}\,e^{-t^{2}-2zt}t^{-\nu -1}\,dt.  \label{h1}
\end{equation}%
Here $\text{Re}\,\gamma >\text{Re}\,\alpha >0$ and $\text{Re}\,(-\nu )>0.$
These functions satisfy equations~(\ref{ps9})--(\ref{ps11}),
respectively.\smallskip

The Bessel equation is%
\begin{equation}
z^{2}u^{\prime \prime }+zu^{\prime }+\left( z^{2}-\nu ^{2}\right) u=0,
\label{bess}
\end{equation}%
where $z$ is a complex variable and parameter $\nu $ can have any real or
complex values. The solutions of (\ref{bess}) are Bessel functions $u_{\nu
}\left( z\right) $ of order $\nu .$ With the aid of the change of the
function $u=\varphi (z)y$ when $\varphi (z)=z^{\nu }e^{iz}$ equation (\ref%
{bess}) can be reduced to the hypergeometric form%
\begin{equation}
zy^{\prime \prime }+(2iz+2\nu +1)y^{\prime }+i(2\nu +1)y=0  \label{bess1}
\end{equation}%
\noindent and based on Theorem~2 one can obtain the \textit{Poisson integral
representations\/} for the Bessel function of the first kind, $J_{\nu }(z),$
and the Hankel functions of the first and second kind, $H_{\nu }^{(1)}(z)$
and $H_{\nu }^{(2)}(z):$%
\begin{equation}
J_{\nu }(z)=\frac{(z/2)^{\nu }}{\sqrt{\pi }\,\Gamma (\nu +1/2)}%
\,\int_{-1}^{1}\,\left( 1-t^{2}\right) ^{\nu -1/2}\cos zt\,dt,
\label{bessel}
\end{equation}%
\begin{equation}
H_{\nu }^{(1,\;2)}(z)=\sqrt{\frac{2}{\pi z}}\;\frac{e^{\pm i\left( z-\pi \nu
/2-\pi /4\right) }}{\Gamma (\nu +1/2)}\,\int_{0}^{\infty }\,e^{-t}t^{\nu
-1/2}\left( 1\pm \frac{it}{2z}\right) ^{\nu -1/2}dt,  \label{hankel}
\end{equation}%
\noindent where $\text{Re}\,\nu >-1/2$. It is then possible to deduce from
these integral representations all the remaining properties of these
functions. For details, see \cite{Ni:Uv}.\smallskip

The Lommel equation%
\begin{equation}
v^{\prime \prime }+\frac{1-2\alpha }{\xi }v^{\prime }+\left[ \left( \beta
\gamma \xi ^{\gamma -1}\right) ^{2}+\frac{\alpha ^{2}-\nu ^{2}\gamma ^{2}}{%
\xi ^{2}}\right] v=0,  \label{lomm}
\end{equation}%
where $\alpha ,$ $\beta $ and $\gamma $ some constants, is very convenient
in applications. Its solutions are%
\begin{equation}
v\left( \xi \right) =\xi ^{\alpha }u_{\nu }\left( \beta \xi ^{\gamma
}\right) ,  \label{lommel}
\end{equation}%
where $u_{\nu }\left( z\right) $ is a Bessel function of order $\nu .$

\section{Solution of Dirac Wave Equation for Coulomb Potential}

This section is written for the benefits of the reader who is not an expert
in relativistic quantum mechanics and quantum field theory. We separate the
variables and construct exact solutions of the Dirac equation of in
spherical coordinates for the Coulomb field. The corresponding four
component (bispinor) wave functions are given explicitly by (\ref{rc1})--(%
\ref{rc5}). We first construct the angular parts of these solutions in terms
of the so-called spinor spherical harmonics or spherical spinors.

\subsection{The Spinor Spherical Harmonics}

The vector addition $\mathbf{j}=\mathbf{l}+\mathbf{s}$ of the orbital $%
\mathbf{l}=-i\mathbf{r\times \nabla }$ and the spin $\mathbf{s}=\dfrac{1}{2}%
\mathbf{\sigma }$ angular momenta (in the units of $\hbar $)\ for the
electron in the central field gives the eigenfunctions of the total angular
momentum $\mathbf{j},$ or the spinor spherical harmonics \cite{Var:Mos:Kher}%
, in the form%
\begin{eqnarray}
&&\ \ \mathcal{Y}_{jm}^{\left( l\right) }\left( \mathbf{n}\right)
=\sum_{m_{l}+m_{s}=m}C_{lm_{l}\frac{1}{2}m_{s}}^{jm}Y_{lm_{l}}\left( \mathbf{%
n}\right) \ \chi _{\frac{1}{2}m_{s}}  \label{ang1} \\
&&\left( j=\left| l-1/2\right| ,l+1/2;\qquad m=-j,-j+1,\ ...,\;j-1,j\right)
\notag
\end{eqnarray}%
where $Y_{lm}\left( \mathbf{n}\right) $ with $\mathbf{n}=\mathbf{n}\left(
\theta ,\varphi \right) =\mathbf{r}/r$ are the spherical harmonics, $%
C_{lm_{l}\frac{1}{2}m_{s}}^{jm}$ are the special Clebsch--Gordan
coefficients, and $\chi _{m_{s}}=$ $\chi _{\frac{1}{2}m_{s}}$are
eigenfunctions of the spin $1/2$ operator $\mathbf{s}:$%
\begin{equation}
\mathbf{s}^{2}\chi _{\frac{1}{2}m_{s}}=\frac{3}{4}\chi _{\frac{1}{2}%
m_{s}},\qquad s_{3}\chi _{\frac{1}{2}m_{s}}=m_{s}\chi _{\frac{1}{2}%
m_{s}},\qquad m_{s}=\pm 1/2  \label{ang2}
\end{equation}%
given by%
\begin{equation}
\chi _{\frac{1}{2}}=\left(
\begin{array}{c}
1 \\
0%
\end{array}%
\right) ,\qquad \chi _{\ -\frac{1}{2}}=\left(
\begin{array}{c}
0 \\
1%
\end{array}%
\right) ;  \label{ang3}
\end{equation}%
see \cite{La:Lif}, \cite{Rose}, \cite{Ni:Su:Uv}, and \cite{Var:Mos:Kher}.
From (\ref{ang1})%
\begin{eqnarray}
\mathcal{Y}_{jm}^{\left( l\right) }\left( \mathbf{n}\right)
&=&\sum_{m_{s}=-1/2}^{1/2}C_{l,\ m-m_{s},\frac{1}{2}m_{s}}^{jm}Y_{l,\
m-m_{s}}\left( \mathbf{n}\right) \ \chi _{m_{s}}  \label{ang4} \\
\quad &=&C_{l,\ m+\frac{1}{2},\frac{1}{2},\ -\frac{1}{2}}^{jm}\ Y_{l,\ m+%
\frac{1}{2}}\left( \mathbf{n}\right) \ \chi _{-\frac{1}{2}}+C_{l,\ m-\frac{1%
}{2},\frac{1}{2},\ \frac{1}{2}}^{jm}\ Y_{l,\ m-\frac{1}{2}}\left( \mathbf{n}%
\right) \ \chi _{\frac{1}{2}}  \notag \\
&=&\left(
\begin{array}{c}
C_{l,\ m-\frac{1}{2},\frac{1}{2},\ \frac{1}{2}}^{jm}\ Y_{l,\ m-\frac{1}{2}%
}\left( \mathbf{n}\right) \medskip \\
C_{l,\ m+\frac{1}{2},\frac{1}{2},\ -\frac{1}{2}}^{jm}\ Y_{l,\ m+\frac{1}{2}%
}\left( \mathbf{n}\right)%
\end{array}%
\right) ,\qquad l=j\pm 1/2.  \notag
\end{eqnarray}%
Substituting the special values of the Clebsch--Gordan coefficients \cite%
{Var:Mos:Kher}, we obtain the spinor spherical harmonics $\mathcal{Y}%
_{jm}^{\pm }\left( \mathbf{n}\right) =\mathcal{Y}_{jm}^{\left( j\pm
1/2\right) }\left( \mathbf{n}\right) $ in the form%
\begin{equation}
\mathcal{Y}_{jm}^{\pm }\left( \mathbf{n}\right) =\left(
\begin{array}{c}
\mp \sqrt{\dfrac{\left( j+1/2\right) \mp \left( m-1/2\right) }{2j+\left(
1\pm 1\right) }}\ Y_{j\pm 1/2,\ m-1/2}\left( \mathbf{n}\right) \medskip \\
\sqrt{\dfrac{\left( j+1/2\right) \pm \left( m+1/2\right) }{2j+\left( 1\pm
1\right) }}\ Y_{j\pm 1/2,\ m+1/2}\left( \mathbf{n}\right)%
\end{array}%
\right)  \label{ang5}
\end{equation}%
with the total angular momentum $j=1/2,3/2,5/2,\ ...$ and its projection $%
m=-j,-j+1,\ ...,j-1,j.$ The orthogonality property for the spinor spherical
harmonics $\mathcal{Y}_{jm}^{\pm }\left( \mathbf{n}\right) =\mathcal{Y}%
_{jm}^{\left( j\pm 1/2\right) }\left( \mathbf{n}\right) $ is \cite%
{Var:Mos:Kher}
\begin{equation}
\int_{S^{2}}\left( \mathcal{Y}_{jm}^{\left( l\right) }\left( \mathbf{n}%
\right) \right) ^{\dagger }\ \mathcal{Y}_{j^{\prime }m^{\prime }}^{\left(
l^{\prime }\right) }\left( \mathbf{n}\right) \ d\omega =\delta _{jj^{\prime
}}\delta _{ll^{\prime }}\delta _{mm^{\prime }}  \label{ang5a}
\end{equation}%
with $d\omega =\sin \theta \ d\theta d\varphi $ and $0\leq \theta \leq \pi
,0\leq \varphi \leq 2\pi .$ They are common eigenfunctions of the following
set of commuting operators%
\begin{eqnarray}
\mathbf{j}^{2}\mathcal{Y}_{jm}^{\pm }\left( \mathbf{n}\right) &=&\left(
\mathbf{l}+\frac{1}{2}\mathbf{\sigma }\right) ^{2}\mathcal{Y}_{jm}^{\pm
}\left( \mathbf{n}\right) =j\left( j+1\right) \mathcal{Y}_{jm}^{\pm }\left(
\mathbf{n}\right) ,  \label{ang6} \\
j_{3}\mathcal{Y}_{jm}^{\pm }\left( \mathbf{n}\right) &=&m\mathcal{Y}%
_{jm}^{\pm }\left( \mathbf{n}\right) ,  \label{ang7} \\
\mathbf{l}^{2}\mathcal{Y}_{jm}^{\pm }\left( \mathbf{n}\right) &=&\left( j\pm
\frac{1}{2}\right) \left( j\pm \frac{1}{2}+1\right) \mathcal{Y}_{jm}^{\pm
}\left( \mathbf{n}\right) ,  \label{ang8} \\
\mathbf{\sigma }^{2}\mathcal{Y}_{jm}^{\pm }\left( \mathbf{n}\right) &=&3%
\mathcal{Y}_{jm}^{\pm }\left( \mathbf{n}\right) .  \label{ang9}
\end{eqnarray}%
But%
\begin{equation*}
\mathbf{j}^{2}=\left( \mathbf{l}+\frac{1}{2}\mathbf{\sigma }\right) ^{2}=%
\mathbf{l}^{2}+\mathbf{\sigma \cdot l}+\frac{3}{4},
\end{equation*}%
or%
\begin{equation}
\mathbf{\sigma \cdot l}=\mathbf{j}^{2}-\mathbf{l}^{2}-\frac{3}{4}.
\label{ang9a}
\end{equation}%
This implies that the spinor spherical harmonics $\mathcal{Y}_{jm}^{\pm
}\left( \mathbf{n}\right) $\ are also eigenfunctions of the operator $%
\mathbf{\sigma \cdot l}:$
\begin{equation}
\left( \mathbf{\sigma \cdot l}\right) \mathcal{Y}_{jm}^{\pm }\left( \mathbf{n%
}\right) =-\left( 1\pm \left( j+\frac{1}{2}\right) \right) \mathcal{Y}%
_{jm}^{\pm }\left( \mathbf{n}\right) ,  \label{ang10}
\end{equation}%
and it is a custom to write%
\begin{equation}
\left( \mathbf{\sigma \cdot l}\right) \mathcal{Y}_{jm}^{\pm }\left( \mathbf{n%
}\right) =-\left( 1+\kappa \right) \mathcal{Y}_{jm}^{\pm }\left( \mathbf{n}%
\right) ,  \label{ang10a}
\end{equation}%
where the quantum number $\kappa =\kappa _{\pm }=\pm \left( j+\dfrac{1}{2}%
\right) =\pm 1,\pm 2,\pm 3,\ ...\ \ $takes all positive and negative integer
values with exception of zero: $\kappa \neq 0.$\smallskip

Finally, the following relation for the spinor spherical harmonics,%
\begin{equation}
\left( \mathbf{\sigma \cdot n}\right) \mathcal{Y}_{jm}^{\pm }\left( \mathbf{n%
}\right) =-\mathcal{Y}_{jm}^{\mp }\left( \mathbf{n}\right) ,  \label{ang11}
\end{equation}%
plays an important role in the Dirac theory of relativistic electron. In
view of $\left( \mathbf{\sigma \cdot n}\right) ^{2}=\mathbf{1},$ it is
sufficient to prove only one of these relations, say%
\begin{equation*}
\left( \mathbf{\sigma \cdot n}\right) \mathcal{Y}_{jm}^{+}\left( \mathbf{n}%
\right) =-\mathcal{Y}_{jm}^{-}\left( \mathbf{n}\right) ,
\end{equation*}%
and the second will follow. A direct proof can be given by using the
recurrence relations for the spherical harmonics (\ref{a7})--(\ref{a9}), or
with the help of the Wigner--Eckart theorem; see \cite{Rose} and \cite%
{Var:Mos:Kher}, the reader can work out the details.\smallskip

The quadratic forms%
\begin{equation}
Q_{jm}=\left( \mathcal{Y}_{jm}^{\left( l\right) }\left( \mathbf{n}\right)
\right) ^{\dagger }\ \mathcal{Y}_{jm}^{\left( l\right) }\left( \mathbf{n}%
\right)  \label{ang12}
\end{equation}%
of the spinor spherical harmonics $\mathcal{Y}_{jm}^{\left( l\right) }\left(
\mathbf{n}\right) $ describe the angular distributions of the electron in
states with the total angular momentum $j,$ its projection $m$ and the
orbital angular momentum $l.$ These forms, given by \cite{Var:Mos:Kher}%
\begin{eqnarray}
Q_{jm}\left( \theta \right) &=&\frac{1}{2j}\left( \left( j+m\right) \left|
Y_{j-1/2,\ m-1/2}\left( \mathbf{n}\right) \right| ^{2}+\left( j-m\right)
\left| Y_{j-1/2,\ m+1/2}\left( \mathbf{n}\right) \right| ^{2}\right)
\label{ang13} \\
&=&\frac{1}{2j+2}\left( \left( j+m+1\right) \left| Y_{j+1/2,\ m+1/2}\left(
\mathbf{n}\right) \right| ^{2}+\left( j-m+1\right) \left| Y_{j+1/2,\
m-1/2}\left( \mathbf{n}\right) \right| ^{2}\right) ,  \notag
\end{eqnarray}%
are, in fact, independent of $l$ and $\varphi .$ There is the useful
expansion in terms of the Laguerre polynomials%
\begin{equation}
Q_{jm}\left( \theta \right) =\sum_{s=0}^{j-1/2}a_{s}\left( j,m\right) \
P_{2s}\left( \cos \theta \right)  \label{ang14}
\end{equation}%
with the coefficients of the form%
\begin{eqnarray}
&&a_{s}\left( j,m\right) =-\frac{4s+1}{4\pi }\sqrt{2j\left( 2j+1\right) }\
\left\{
\begin{array}{ccc}
j & j & 2s\medskip \\
j-1/2 & j-1/2 & 1/2%
\end{array}%
\right\} \;C_{j-1/2,0,2s0}^{j-1/2,0}\;C_{jm\;2s0}^{jm}  \notag \\
&&\qquad \quad \ \ =\left( -1\right) ^{s}\frac{4s+1}{4\pi }\sqrt{\frac{%
\left( 2j+2s+1\right) \left( 2j-2s\right) !}{\left( 2j+1\right) \left(
2j+2s\right) !}}\;\frac{\left( j+s-1/2\right) !\left( 2s\right) !}{\left(
j-s-1/2\right) !\left( s!\right) ^{2}}\;C_{jm\;2s0}^{jm}.  \label{ang15}
\end{eqnarray}%
See \cite{Var:Mos:Kher} for more information.

\subsection{Separation of Variables in Spherical Coordinates}

Using the explicit form of the $\mathbf{\alpha }$ and $\beta $ matrices (\ref%
{d3}) we rewrite the stationary Dirac equation (\ref{d8}) in a central field
with the Hamiltonian%
\begin{equation}
H=c\mathbf{\alpha p}+mc^{2}\beta +U\left( r\right) =\left(
\begin{array}{cc}
U+mc^{2} & c\mathbf{\sigma p\medskip \medskip } \\
c\mathbf{\sigma p} & U-mc^{2}%
\end{array}%
\right)  \label{ds1}
\end{equation}%
and the bispinor wave function%
\begin{equation}
\psi =\left(
\begin{array}{c}
\mathbf{\varphi \medskip } \\
\mathbf{\chi }%
\end{array}%
\right)  \label{ds2}
\end{equation}%
in a matrix form%
\begin{equation}
\left(
\begin{array}{cc}
U+mc^{2} & c\mathbf{\sigma p\medskip \medskip } \\
c\mathbf{\sigma p} & U-mc^{2}%
\end{array}%
\right) \left(
\begin{array}{c}
\mathbf{\varphi \medskip } \\
\mathbf{\chi }%
\end{array}%
\right) =E\left(
\begin{array}{c}
\mathbf{\varphi \medskip } \\
\mathbf{\chi }%
\end{array}%
\right) ,  \label{ds3}
\end{equation}%
or%
\begin{eqnarray}
c\mathbf{\sigma p\medskip \ \varphi \medskip } &=&\left( E+mc^{2}-U\right) \
\mathbf{\chi },  \label{ds4} \\
c\mathbf{\sigma p\medskip \ \mathbf{\chi }\medskip } &=&\left(
E-mc^{2}-U\right) \ \mathbf{\mathbf{\varphi }}.  \label{ds5}
\end{eqnarray}%
Here we shall use the following operator identity%
\begin{equation}
\mathbf{\sigma \cdot \nabla }=\left( \mathbf{\sigma \cdot n}\right) \left(
\mathbf{n\cdot \nabla }+i\mathbf{\sigma \cdot \left( n\mathbf{\times \nabla }%
\right) }\right)  \label{ds6}
\end{equation}%
in the form%
\begin{equation}
c\mathbf{\sigma p}=\hbar c\left( \mathbf{\sigma n}\right) \left( \frac{1}{i}%
\mathbf{n\nabla }+\frac{i}{r}\mathbf{\sigma l}\right) ,  \label{ds6a}
\end{equation}%
where $\mathbf{l}=-i\mathbf{r\times \nabla }$ is the operator of orbital
angular momentum, $\mathbf{n}=\mathbf{r}/r$ and $\mathbf{p}=-i\hbar \mathbf{%
\nabla }.$ It can be obtained as a consequence of a more general operator
identity \cite{Dirac}%
\begin{equation}
\left( \mathbf{\sigma \cdot A}\right) \left( \mathbf{\sigma \cdot B}\right) =%
\mathbf{A\cdot B}+i\mathbf{\sigma \cdot }\left( \mathbf{A\times B}\right) ,
\label{ds7}
\end{equation}%
which is valid for any vector operators $\mathbf{A}$ and $\mathbf{B}$
commuting with the Pauli $\mathbf{\sigma }$-matrices; it is not required
that $\mathbf{A}$ and $\mathbf{B}$ commute. The proof uses a familiar
property of the Pauli matrices%
\begin{equation}
\sigma _{i}\ \sigma _{k}=ie_{ikl}\ \sigma _{l}+\delta _{ik},  \label{ds8}
\end{equation}%
where $e_{ikl}$ is the completely antisymmetric Levi-Civita symbol, $\delta
_{ik}$ is the symmetric Kronecker delta symbol and we use Einstein's
summation rule over the repeating indices; it is understood that a summation
is to be taken over the three values of $l=1,2,3.$ Thus%
\begin{eqnarray*}
&&\left( \mathbf{\sigma \cdot A}\right) \left( \mathbf{\sigma \cdot B}%
\right) =\left( \sigma _{i}\ A_{i}\right) \left( \sigma _{k}\ B_{k}\right) \\
&&\ =\left( \sigma _{i}\ \sigma _{k}\right) \ A_{i}\ B_{k}=i\sigma _{l}\
e_{ikl}\ A_{i}\ B_{k}+\delta _{ik}\ A_{i}\ B_{k} \\
&&\ =i\sigma _{l}\ \left( \mathbf{A\times B}\right) _{l}+A_{k}\ B_{k}=i%
\mathbf{\sigma \cdot }\left( \mathbf{A\times B}\right) +\mathbf{A\cdot B},
\end{eqnarray*}%
where $\left( \mathbf{A\times B}\right) _{l}=e_{lik}\ A_{i}\ B_{k}=e_{ikl}\
A_{i}\ B_{k}$ in view of antisymmetry of the Levi-Civita symbol: $%
e_{ikl}=-e_{ilk}=e_{lik},$ and $\delta _{ik}\ A_{i}=A_{k}.$\smallskip

If $\mathbf{A}=\mathbf{B},$ Eq.~(\ref{ds7}) implies $\left( \mathbf{\sigma
\cdot A}\right) ^{2}=\mathbf{A}^{2}.$ In particular, $\left( \mathbf{\sigma n%
}\right) ^{2}=\mathbf{n}^{2}=\mathbf{1},$ and the proof of the ``gradient''
formula (\ref{ds6}) is%
\begin{eqnarray*}
&&\mathbf{\sigma \cdot \nabla }=\left( \mathbf{\sigma \cdot n}\right)
^{2}\left( \mathbf{\sigma \cdot \nabla }\right) =\left( \mathbf{\sigma \cdot
n}\right) \left( \left( \mathbf{\sigma \cdot n}\right) \left( \mathbf{\sigma
\cdot \nabla }\right) \right) \\
&&\quad \quad \ \ =\left( \mathbf{\sigma \cdot n}\right) \left( \mathbf{%
n\cdot \nabla }+i\mathbf{\sigma \cdot \left( n\mathbf{\times \nabla }\right)
}\right)
\end{eqnarray*}%
by (\ref{ds7}) with $\mathbf{A}=\mathbf{n}$ and $\mathbf{B}=\mathbf{\nabla }%
. $ In a similar fashion, one can derive the following ``anticommutation''
relation,%
\begin{equation}
\left( \mathbf{\sigma n}\right) \left( \mathbf{\sigma l}\right) +\left(
\mathbf{\sigma l}\right) \left( \mathbf{\sigma n}\right) =-2\left( \mathbf{%
\sigma n}\right) ,\quad \mathbf{n}=\mathbf{r}/r,  \label{ds9}
\end{equation}%
we leave details to the reader.\smallskip

The structure of operator $\mathbf{\sigma p}$ in (\ref{ds6a}) suggests to
look for solutions of the Dirac system (\ref{ds4})--(\ref{ds5}) in spherical
coordinates $\mathbf{r}=r\ \mathbf{n}\left( \theta ,\varphi \right) $ in the
form of the Ansatz:%
\begin{eqnarray}
&&\mathbf{\varphi }=\mathbf{\varphi }\left( \mathbf{r}\right) =\mathcal{Y}%
\left( \mathbf{n}\right) \ F\left( r\right) ,  \label{ds10} \\
&&\mathbf{\chi }=\mathbf{\chi }\left( \mathbf{r}\right) =-i\left( \left(
\mathbf{\sigma n}\right) \mathcal{Y}\left( \mathbf{n}\right) \right) \
G\left( r\right) ,  \label{ds11}
\end{eqnarray}%
where $\mathcal{Y}=\mathcal{Y}_{jm}^{\pm }\left( \mathbf{n}\right) $ are the
spinor spherical harmonics given by (\ref{ang5}). This substitution
preserves the symmetry properties of the wave functions under inversion $%
\mathbf{r}\rightarrow -\mathbf{r}.$ Then the radial functions $F\left(
r\right) $ and $G\left( r\right) $ satisfy the system of two first order
ordinary differential equations%
\begin{eqnarray}
\dfrac{dF}{dr}+\frac{1+\kappa }{r}\ F &=&\dfrac{mc^{2}+E-U\left( r\right) }{%
\hbar c}\ G,  \label{ds12} \\
\dfrac{dG}{dr}+\frac{1-\kappa }{r}\ G &=&\dfrac{mc^{2}-E+U\left( r\right) }{%
\hbar c}\ F,  \label{ds13}
\end{eqnarray}%
where $\kappa =\kappa _{\pm }=\pm \left( j+1/2\right) =\pm 1,\pm 2,\pm 3,\
...\ ,$ respectively.\smallskip

If $f=f\left( \mathbf{r}\right) =f\left( r\mathbf{n}\right) ,$ then%
\begin{equation*}
\frac{\partial f}{\partial r}=\frac{\partial f}{\partial \mathbf{r}}\frac{%
\partial \mathbf{r}}{\partial r}=\mathbf{n\nabla }f
\end{equation*}%
and in spherical coordinates Eq.~(\ref{ds6a}) becomes%
\begin{equation}
c\mathbf{\sigma p}=\hbar c\left( \mathbf{\sigma n}\right) \left( \frac{1}{i}%
\frac{\partial }{\partial r}+\frac{i}{r}\mathbf{\sigma l}\right) .
\label{ds14}
\end{equation}%
Thus\qquad
\begin{eqnarray*}
&&c\mathbf{\sigma p\ \varphi }=\hbar c\left( \mathbf{\sigma n}\right) \left(
\frac{1}{i}\frac{\partial }{\partial r}+\frac{i}{r}\mathbf{\sigma l}\right)
\mathcal{Y}F \\
&&\qquad \quad \ =\hbar c\left( \mathbf{\sigma n}\right) \left( \frac{1}{i}%
\mathcal{Y}\frac{dF}{dr}+\frac{i}{r}\left( \mathbf{\sigma l}\mathcal{Y}%
\right) F\right) \\
&&\qquad \quad \ =-i\hbar c\left( \mathbf{\sigma n}\mathcal{Y}\right) \left(
\frac{dF}{dr}+\frac{1+\kappa }{r}F\right)
\end{eqnarray*}%
by (\ref{ang10a}), and we arrive at (\ref{ds12}) in view of (\ref{ds4}) and (%
\ref{ds11}). Equation (\ref{ds13}) can be verified in a similar fashion with
the help of (\ref{ds9}) or (\ref{ang11}).\smallskip

Eqs.~(\ref{ds12})--(\ref{ds13}) hold in any central field with the potential
energy $U=U\left( r\right) $\smallskip $.$ For states with discrete spectra
the radial functions $rF\left( r\right) $ and $rG\left( r\right) $ should be
bounded as $r\rightarrow 0$ and satisty the normalization condition%
\begin{equation}
\dint_{\mathbf{R}^{3}}\psi ^{\dagger }\ \psi \ dv=\int_{0}^{\infty
}r^{2}\left( F^{2}\left( r\right) +G^{2}\left( r\right) \right) \ dr=1
\label{ds15}
\end{equation}%
in view of (\ref{d6}), (\ref{ds10})--(\ref{ds11}) and (\ref{ang5a}).

\subsection{Solution of Radial Equations}

For the relativistic Coulomb problem $U=-Ze^{2}/r,$ we introduce the
dimensionless quantities%
\begin{equation}
\varepsilon =\frac{E}{mc^{2}},\qquad x=\beta r=\frac{mc}{\hbar }r,\qquad \mu
=\frac{Ze^{2}}{\hbar c}  \label{rrc1}
\end{equation}%
and the radial functions%
\begin{equation}
f\left( x\right) =F\left( r\right) ,\qquad g\left( x\right) =G\left(
r\right) .  \label{rrc2}
\end{equation}%
The system (\ref{ds12})--(\ref{ds13}) becomes%
\begin{eqnarray}
\dfrac{df}{dx}+\frac{1+\kappa }{x}\ f &=&\left( 1+\varepsilon +\frac{\mu }{x}%
\right) g,  \label{rrc3} \\
\dfrac{dg}{dx}+\frac{1-\kappa }{x}\ g &=&\left( 1-\varepsilon -\frac{\mu }{x}%
\right) f.  \label{rrc4}
\end{eqnarray}%
We shall see later that in nonrelativistic limit $c\rightarrow \infty $ the
following estimate holds $\left| f\left( x\right) \right| \gg \left| g\left(
x\right) \right| .$\medskip

We follow \cite{Ni:Uv} with somewhat different details. Let us rewrite the
system (\ref{rrc3})--(\ref{rrc4}) in matrix form \cite{Gor}. If%
\begin{equation}
u=\left(
\begin{array}{c}
u_{1}\medskip \\
u_{2}%
\end{array}%
\right) =\left(
\begin{array}{c}
xf\left( x\right) \medskip \\
xg\left( x\right)%
\end{array}%
\right) ,\qquad u^{\prime }=\left(
\begin{array}{c}
u_{1}^{\prime }\medskip \\
u_{2}^{\prime }%
\end{array}%
\right) .  \label{rrc5}
\end{equation}%
Then%
\begin{equation}
u^{\prime }=Au,  \label{rrc6}
\end{equation}%
where%
\begin{equation}
A=\left(
\begin{array}{cc}
a_{11} & a_{12}\medskip \\
a_{21} & a_{22}%
\end{array}%
\right) =\left(
\begin{array}{cc}
-\dfrac{\kappa }{x} & 1+\varepsilon +\dfrac{\mu }{x}\medskip \\
1-\varepsilon -\dfrac{\mu }{x} & \dfrac{\kappa }{x}%
\end{array}%
\right) .  \label{rrc7}
\end{equation}%
To find $u_{1}\left( x\right) ,$ we eliminate $u_{2}\left( x\right) $ from
the system (\ref{rrc6}), obtaining a second order differential equation%
\begin{eqnarray}
&&u_{1}^{\prime \prime }-\left( a_{11}+a_{22}+\dfrac{a_{12}^{\prime }}{a_{12}%
}\right) u_{1}^{\prime }  \label{rrc8} \\
&&\qquad +\left( a_{11}a_{22}-a_{12}a_{21}-a_{11}^{\prime }+\dfrac{%
a_{12}^{\prime }}{a_{12}}\;a_{11}\right) u_{1}=0.  \notag
\end{eqnarray}%
Similarly, eliminating $u_{1}\left( x\right) ,$ one gets equation for $%
u_{2}\left( x\right) :$
\begin{eqnarray}
&&u_{2}^{\prime \prime }-\left( a_{11}+a_{22}+\dfrac{a_{21}^{\prime }}{a_{21}%
}\right) u_{2}^{\prime }  \label{rrc9} \\
&&\qquad +\left( a_{11}a_{22}-a_{12}a_{21}-a_{22}^{\prime }+\dfrac{%
a_{21}^{\prime }}{a_{21}}\;a_{22}\right) u_{2}=0.  \notag
\end{eqnarray}

The components of matrix $A$ have the form%
\begin{equation}
a_{ik}=b_{ik}+c_{ik}/x,  \label{me1}
\end{equation}%
where $b_{ik}$ and $c_{ik}$ are constants. Equations (\ref{rrc8}) and (\ref%
{rrc9}) are not generalized equations of hypergeometric type (\ref{nu1}).
Indeed,%
\begin{equation*}
\dfrac{a_{12}^{\prime }}{a_{12}}=-\frac{c_{12}}{c_{12}x+b_{12}x^{2}},
\end{equation*}%
and the coefficients of $u_{1}^{\prime }\left( x\right) $ and $u_{1}\left(
x\right) $ in (\ref{rrc8}) are%
\begin{eqnarray*}
&&a_{11}+a_{22}+\dfrac{a_{12}^{\prime }}{a_{12}}=\frac{p_{1}\left( x\right)
}{x}-\frac{c_{12}}{c_{12}x+b_{12}x^{2}}, \\
&&a_{11}a_{22}-a_{12}a_{21}-a_{11}^{\prime }+\dfrac{a_{12}^{\prime }}{a_{12}}%
a_{11}=\frac{p_{2}\left( x\right) }{x^{2}}-\frac{c_{12}\left(
c_{11}+b_{11}x\right) }{\left( c_{12}+b_{12}x\right) x^{2}},
\end{eqnarray*}%
where $p_{1}\left( x\right) $ and $p_{2}\left( x\right) $ are polynomials of
degrees at most one and two, respectively. Equation (\ref{rrc8}) will become
a generalized equation of hypergeometric type (\ref{nu1}) with $\sigma
\left( x\right) =x$ if either $b_{12}=0$ or $c_{12}=0.$ The following
consideration helps. By a linear transformation%
\begin{equation}
\left(
\begin{array}{c}
v_{1}\medskip \\
v_{2}%
\end{array}%
\right) =C\left(
\begin{array}{c}
u_{1}\medskip \\
u_{2}%
\end{array}%
\right)  \label{rrc10}
\end{equation}%
with a nonsingular matrix $C$ that is independent of $x$ we transform the
original system (\ref{rrc6}) to a similar one%
\begin{equation}
v^{\prime }=\widetilde{A}v,  \label{rrc11}
\end{equation}%
where%
\begin{equation*}
v=\left(
\begin{array}{c}
v_{1}\medskip \\
v_{2}%
\end{array}%
\right) ,\qquad \widetilde{A}=CAC^{-1}=\left(
\begin{array}{cc}
\widetilde{a}_{11} & \widetilde{a}_{12}\medskip \\
\widetilde{a}_{21} & \widetilde{a}_{22}%
\end{array}%
\right) .
\end{equation*}%
The new coefficients $\widetilde{a}_{ik}$ are linear combinations of the
original ones $a_{ik}.$ Hence they have a similar form%
\begin{equation}
\widetilde{a}_{ik}=\widetilde{b}_{ik}+\widetilde{c}_{ik}/x,  \label{me2}
\end{equation}%
where $\widetilde{b}_{ik}$ and $\widetilde{c}_{ik}$ are constants.\smallskip

The equations for $v_{1}\left( x\right) $ and $v_{2}\left( x\right) $ are
similar to (\ref{rrc8}) and (\ref{rrc9}):%
\begin{eqnarray}
&&v_{1}^{\prime \prime }-\left( \widetilde{a}_{11}+\widetilde{a}_{22}+\dfrac{%
\widetilde{a}_{12}^{\prime }}{\widetilde{a}_{12}}\right) v_{1}^{\prime }
\label{rrc12} \\
&&\qquad +\left( \widetilde{a}_{11}\widetilde{a}_{22}-\widetilde{a}_{12}%
\widetilde{a}_{21}-\widetilde{a}_{11}^{\prime }+\dfrac{\widetilde{a}%
_{12}^{\prime }}{\widetilde{a}_{12}}\;\widetilde{a}_{11}\right) \;v_{1}=0,
\notag
\end{eqnarray}%
\begin{eqnarray}
&&v_{2}^{\prime \prime }-\left( \widetilde{a}_{11}+\widetilde{a}_{22}+\dfrac{%
\widetilde{a}_{21}^{\prime }}{\widetilde{a}_{21}}\right) v_{2}^{\prime }
\label{rrc13} \\
&&\qquad +\left( \widetilde{a}_{11}\widetilde{a}_{22}-\widetilde{a}_{12}%
\widetilde{a}_{21}-\widetilde{a}_{22}^{\prime }+\dfrac{\widetilde{a}%
_{21}^{\prime }}{\widetilde{a}_{21}}\;\widetilde{a}_{22}\right) \;v_{2}=0.
\notag
\end{eqnarray}%
The calculation of the coefficients in (\ref{rrc12}) and (\ref{rrc13}) is
facilitated by a similarity of the matrices $A$ and $\widetilde{A}:$%
\begin{equation*}
\widetilde{a}_{11}+\widetilde{a}_{22}=a_{11}+a_{22},\qquad \widetilde{a}_{11}%
\widetilde{a}_{22}-\widetilde{a}_{12}\widetilde{a}%
_{21}=a_{11}a_{22}-a_{12}a_{21}.
\end{equation*}%
By a previous consideration, in order for (\ref{rrc12}) to be an equation of
hypergeometric type, it is sufficient to choose either $\widetilde{b}_{12}=0$
or $\widetilde{c}_{12}=0.$ Similarly, for (\ref{rrc13}): either $\widetilde{b%
}_{21}=0$ or $\widetilde{c}_{21}=0.$ These conditions impose certain
restrictions on our choice of the transformation matrix $C.$ Let%
\begin{equation}
C=\left(
\begin{array}{cc}
\alpha & \beta \smallskip \\
\gamma & \delta%
\end{array}%
\right) .  \label{rrc14}
\end{equation}%
Then%
\begin{equation*}
C^{-1}=\dfrac{1}{\Delta }\left(
\begin{array}{cc}
\delta & -\beta \smallskip \\
-\gamma & \alpha%
\end{array}%
\right) ,\qquad \Delta =\det C=\alpha \delta -\beta \gamma ,
\end{equation*}%
and%
\begin{eqnarray}
&&\widetilde{A}=CAC^{-1}  \label{rrc15} \\
&&\ \ \ =\dfrac{1}{\Delta }\left(
\begin{array}{cc}
a_{11}\alpha \delta -a_{12}\alpha \gamma +a_{21}\beta \delta -a_{22}\beta
\gamma \quad & a_{12}\alpha ^{2}-a_{12}\beta ^{2}+\left(
a_{22}-a_{11}\right) \alpha \beta \medskip \\
a_{21}\delta ^{2}-a_{12}\gamma ^{2}+\left( a_{11}-a_{22}\right) \gamma \delta
& a_{12}\alpha \gamma -a_{11}\beta \gamma +a_{22}\alpha \delta -a_{21}\beta
\delta%
\end{array}%
\right) .  \notag
\end{eqnarray}%
For the Dirac system (\ref{rrc6})--(\ref{rrc7}):%
\begin{eqnarray*}
&&a_{11}=-\dfrac{\kappa }{x},\qquad a_{12}\medskip =1+\varepsilon +\dfrac{%
\mu }{x}, \\
&&a_{21}=1-\varepsilon -\dfrac{\mu }{x},\qquad a_{22}=\dfrac{\kappa }{x}
\end{eqnarray*}%
and%
\begin{eqnarray}
\Delta \ \ \widetilde{a}_{12} &=&\alpha ^{2}-\beta ^{2}+\left( \alpha
^{2}+\beta ^{2}\right) \varepsilon +\dfrac{\left( \alpha ^{2}+\beta
^{2}\right) \mu +2\alpha \beta \kappa }{x},  \label{rrc16} \\
\Delta \ \ \widetilde{a}_{21} &=&\delta ^{2}-\gamma ^{2}-\left( \delta
^{2}+\gamma ^{2}\right) \varepsilon -\dfrac{\left( \delta ^{2}+\gamma
^{2}\right) \mu +2\gamma \delta \kappa }{x}.  \label{rrc17}
\end{eqnarray}%
\begin{equation*}
\begin{array}{ccccc}
\text{The} & \text{condition} & \widetilde{b}_{12}=0 & \text{yields} &
\left( 1+\varepsilon \right) \alpha ^{2}-\left( 1-\varepsilon \right) \beta
^{2}=0, \\
" & " & \widetilde{c}_{12}=0 & " & \left( \alpha ^{2}+\beta ^{2}\right) \mu
+2\alpha \beta \kappa =0, \\
" & " & \widetilde{b}_{21}=0 & " & \left( 1+\varepsilon \right) \gamma
^{2}-\left( 1-\varepsilon \right) \delta ^{2}=0, \\
" & " & \widetilde{c}_{21}=0 & " & \left( \delta ^{2}+\gamma ^{2}\right) \mu
+2\gamma \delta \kappa =0.%
\end{array}%
\end{equation*}%
\medskip\ We see that there are several possibilities to choose the elements
$\alpha ,$ $\beta ,$ $\gamma ,$ $\delta $ of the transition matrix $C.$ All
quantum mechanics textbooks use the original one, namely, $\widetilde{b}%
_{12}=0$ and $\widetilde{b}_{21}=0,$ due to Darwin \cite{Dar} and Gordon
\cite{Gor}; cf. equations (\ref{maa})--(\ref{mbb}) below. Nikiforov and
Uvarov \cite{Ni:Uv} take another path, they choose $\widetilde{c}_{12}=0$
and $\widetilde{c}_{21}=0$ and show that it is more convenient for taking
the nonrelativistic limit $c\rightarrow \infty .$ These conditions are
satisfied if%
\begin{equation}
C=\left(
\begin{array}{cc}
\mu & \smallskip \nu -\kappa \\
\smallskip \nu -\kappa & \mu%
\end{array}%
\right) ,  \label{rrc18}
\end{equation}%
where $\nu =\sqrt{\kappa ^{2}-\mu ^{2}},$ and we finally arrive at the
following system of the first order equations for $v_{1}\left( x\right) $
and $v_{2}\left( x\right) :$
\begin{eqnarray}
v_{1}^{\prime } &=&\left( \dfrac{\varepsilon \mu }{\nu }-\dfrac{\nu }{x}%
\right) v_{1}+\left( 1+\dfrac{\varepsilon \kappa }{\nu }\right) v_{2},
\label{rrc19} \\
v_{2}^{\prime } &=&\left( 1-\dfrac{\varepsilon \kappa }{\nu }\right)
v_{1}+\left( \dfrac{\nu }{x}-\dfrac{\varepsilon \mu }{\nu }\right) v_{2}.
\label{rrc20}
\end{eqnarray}%
Here%
\begin{equation}
\text{Tr}\ \widetilde{A}=\widetilde{a}_{11}+\widetilde{a}_{22}=0,\quad \det
\widetilde{A}=\varepsilon ^{2}-1+\dfrac{2\varepsilon \mu }{x}-\dfrac{\nu ^{2}%
}{x^{2}},\quad \nu ^{2}=\kappa ^{2}-\mu ^{2},  \label{rrc21}
\end{equation}%
which is simpler that the original choice in \cite{Ni:Uv}. The corresponding
second order differential equations (\ref{rrc12})--(\ref{rrc13}) become%
\begin{eqnarray}
v_{1}^{\prime \prime }+\dfrac{\left( \varepsilon ^{2}-1\right)
x^{2}+2\varepsilon \mu x-\nu \left( \nu +1\right) }{x^{2}}\;v_{1} &=&0,
\label{rrc22} \\
v_{2}^{\prime \prime }+\dfrac{\left( \varepsilon ^{2}-1\right)
x^{2}+2\varepsilon \mu x-\nu \left( \nu -1\right) }{x^{2}}\;v_{2} &=&0.
\label{rrc23}
\end{eqnarray}%
They are the generalized equations of hypergeometric type (\ref{nu1}) of a
simplest form $\widetilde{\tau }=0,$ thus resembling the one dimensional Schr%
\"{o}dinder equation; the second equation can be obtain from the first one
by replacing $\nu \rightarrow -\nu .$\smallskip

Let $1+\varepsilon \kappa /\nu =0,$ then $\varepsilon =-\nu /\kappa $ that
is possible only if $\kappa <0,$ since $\nu >0$ and $\varepsilon >0.$ The
corresponding solution of (\ref{rrc19}),%
\begin{equation*}
v_{1}\left( x\right) =C_{1}x^{-\nu }e^{\left( \varepsilon \mu \;x\right)
/\nu },
\end{equation*}%
satisfies the conditions of the problem only if $C_{1}=0.$ Then from (\ref%
{rrc20})%
\begin{equation*}
v_{2}\left( x\right) =C_{2}x^{\nu }e^{-\left( \varepsilon \mu \;x\right)
/\nu },
\end{equation*}%
which does satisfy the condition of the problem with $C_{2}\neq 0.$\smallskip

Let us analyze the behavior of the solutions of (\ref{rrc22}) as $%
x\rightarrow 0.$ Since%
\begin{equation*}
\left| \left( \varepsilon ^{2}-1\right) x^{2}+2\varepsilon \mu x\right| \ll
\nu \left( \nu +1\right)
\end{equation*}%
as $x\rightarrow 0,$ one can approximate this equation in the neighborhood
of $x=0$ by the corresponding Euler equation%
\begin{equation*}
x^{2}v_{1}^{\prime \prime }-\nu \left( \nu +1\right) v_{1}=0,
\end{equation*}%
whose solutions are%
\begin{equation*}
v_{1}\left( x\right) =C_{1}x^{\nu +1}+C_{2}x^{-\nu },\qquad C_{2}=0.
\end{equation*}%
Thus $v_{1}\rightarrow C_{1}x^{\nu +1}$ as $x\rightarrow 0.$ The results for
(\ref{rrc23}) are similar$:v_{2}\rightarrow C_{2}x^{\nu }$ as $x\rightarrow
0;$ one can use the symmetry $\nu \rightarrow -\nu .$\smallskip

Equation (\ref{rrc22}) is the generalized equation of hypergeometric type (%
\ref{nu1}) with%
\begin{eqnarray*}
&&\sigma \left( x\right) =x,\qquad \widetilde{\tau }\left( x\right) =0, \\
&&\widetilde{\sigma }\left( x\right) =\left( \varepsilon ^{2}-1\right)
x^{2}+2\varepsilon \mu x-\nu \left( \nu +1\right) .
\end{eqnarray*}%
The substitution%
\begin{equation}
v_{1}=\varphi \left( x\right) y\left( x\right) ,\qquad \frac{\varphi
^{\prime }}{\varphi }=\frac{\pi \left( x\right) }{\sigma \left( x\right) },
\label{rrc24}
\end{equation}%
where%
\begin{equation}
\pi \left( x\right) =\frac{\sigma ^{\prime }-\widetilde{\tau }}{2}\pm \sqrt{%
\left( \frac{\sigma ^{\prime }-\widetilde{\tau }}{2}\right) ^{2}-\widetilde{%
\sigma }+k\sigma }  \label{rrc25}
\end{equation}%
with $k=\lambda -\pi ^{\prime }$ and $\tau \left( x\right) =\widetilde{\tau }%
\left( x\right) +2\pi \left( x\right) ,$ results in the equation of
hypergeometric type%
\begin{equation}
\sigma \left( x\right) y^{\prime \prime }+\tau \left( x\right) y^{\prime
}+\lambda y=0  \label{rrc26}
\end{equation}%
by the method of \cite{Ni:Uv}; see also Section~5.1. From the four possible
forms of $\pi \left( x\right) :$%
\begin{equation}
\pi \left( x\right) =\frac{1}{2}\pm \left( \sqrt{1-\varepsilon ^{2}}\;x\pm
\left( \nu +\frac{1}{2}\right) \right) ,  \label{rrc27}
\end{equation}%
corresponding to the values of $k$ determined by the condition of the zero
discriminant of the quadratic polynomial under the square root sign in (\ref%
{rrc25}):%
\begin{equation}
k-2\varepsilon \mu =\pm \sqrt{1-\varepsilon ^{2}}\;\left( 2\nu +1\right) ,
\label{rrc28}
\end{equation}%
we select the one when the function $\tau \left( x\right) $ has a negative
derivative and a zero on $\left( 0,+\infty \right) .$ This is true if one
chooses%
\begin{eqnarray*}
k &=&2\varepsilon \mu -a\left( 2\nu +1\right) , \\
\pi \left( x\right) &=&\nu +1-ax, \\
\tau \left( x\right) &=&2\pi \left( x\right) \;=\;2\left( \nu +1-ax\right) ,
\\
\lambda &=&k+\pi ^{\prime }\;=\;2\left( \varepsilon \mu -a\left( \nu
+1\right) \right)
\end{eqnarray*}%
and%
\begin{equation*}
\varphi \left( x\right) =x^{\nu +1}e^{-ax},\qquad \rho \left( x\right)
=x^{2\nu +1}e^{-2ax},
\end{equation*}%
where $a=\sqrt{1-\varepsilon ^{2}}$ and $\nu =$\ $\sqrt{\kappa ^{2}-\mu ^{2}}%
.$ The analysis for (\ref{rrc23}) is similar, one can use the symmetry $\nu
\rightarrow -\nu $ in (\ref{rrc27})--(\ref{rrc28}).\smallskip

From (\ref{ds15}) and (\ref{rrc5})%
\begin{equation}
\int_{0}^{\infty }r^{2}\left( F^{2}\left( r\right) +G^{2}\left( r\right)
\right) \;dr=\beta ^{-3}\int_{0}^{\infty }\left( u_{1}^{2}\left( x\right)
+u_{2}^{2}\left( x\right) \right) \;dx=1.  \label{rrc29}
\end{equation}%
It requires by (\ref{rrc10}) the square integrability of $v_{1}\left(
x\right) $ and $v_{2}\left( x\right) .$ Their boundness at $x=0$ follows
from the asymptotic behavior as $x\rightarrow 0.$ So%
\begin{equation}
\int_{0}^{\infty }v_{1}^{2}\left( x\right) \;dx=\int_{0}^{\infty }\varphi
^{2}\left( x\right) y^{2}\left( x\right) \;dx=\int_{0}^{\infty }x\
y^{2}\left( x\right) \rho \left( x\right) dx<\infty .  \label{rrc29b}
\end{equation}%
For the time being, we replace this condition by%
\begin{equation}
\int_{0}^{\infty }\ y^{2}\left( x\right) \rho \left( x\right) dx<\infty
\label{rrc29a}
\end{equation}%
in order to apply Theorem~1, and will verify the normalization condition (%
\ref{rrc29}) later. Then the corresponding energy levels $\varepsilon
=\varepsilon _{n}$ are determined by%
\begin{equation}
\lambda +n\tau ^{\prime }+\frac{1}{2}n\left( n-1\right) \sigma ^{\prime
\prime }=0\qquad \left( n=0,1,2,\ ...\ \right) ,  \label{rrc30}
\end{equation}%
whence%
\begin{equation}
\varepsilon \mu =a\left( \nu +n+1\right) ,  \label{rrc31}
\end{equation}%
and the eigenfunctions are given by the Rodrigues formula%
\begin{equation}
y_{n}\left( x\right) =\frac{C_{n}}{\rho \left( x\right) }\left( \sigma
^{n}\left( x\right) \rho \left( x\right) \right) ^{\left( n\right)
}=C_{n}\;x^{-2\nu -1}e^{2ax}\frac{d^{n}}{dx^{n}}\left( x^{2\nu
+n+1}e^{-2ax}\right) .  \label{rrc32}
\end{equation}%
The functions $y_{n}\left( x\right) $ are, up to certain constants, the
Laguerre polynomials $L_{n}^{2\nu +1}\left( \xi \right) $ with $\xi =2ax.$%
\smallskip

The previously found eigenvalue $\varepsilon =-\nu /\kappa $ satisfies (\ref%
{rrc31}) with $n=-1.$ Consequently it is natural to replace $n$ by $n-1$ in (%
\ref{rrc31})--(\ref{rrc32}) and define the eigenvalues by%
\begin{equation}
\varepsilon \mu =a\left( \nu +n\right) ,\qquad a=\sqrt{1-\varepsilon ^{2}}%
\qquad \left( n=0,1,2,\ ...\ \right) .  \label{rrc33}
\end{equation}%
Solving for $\varepsilon $ gives the Sommerfeld--Dirac formula (\ref{rc6}).
The corresponding eigenfunctions have the form%
\begin{equation}
v_{1}\left( x\right) =\left\{
\begin{array}{l}
0,\qquad n=0,\medskip \\
A_{n}\xi ^{\nu +1}e^{-\xi /2}L_{n-1}^{2\nu +1}\left( \xi \right) ,\qquad
n=1,2,3,\;...\;.%
\end{array}%
\right.  \label{rrc34}
\end{equation}%
They are square integrable functions on $\left( 0,\infty \right) .$ The
counterparts are%
\begin{equation}
v_{2}\left( x\right) =B_{n}\xi ^{\nu }e^{-\xi /2}L_{n}^{2\nu -1}\left( \xi
\right) ,\qquad n=0,1,2,\;...\;.  \label{rrc35}
\end{equation}%
It is easily seen that our previous solution for $\varepsilon =-\nu /\kappa $
is included in this formula when $n=0.$ By equation (\ref{rrc19}) the other
solutions can be obtain as%
\begin{equation*}
v_{2}\left( x\right) =\frac{1}{1+\kappa \varepsilon /\nu }\left(
v_{1}^{\prime }\left( x\right) +\left( \frac{\nu }{x}-\frac{\varepsilon \mu
}{\nu }\right) v_{1}\left( x\right) \right)
\end{equation*}%
and substituting $v_{1}\left( x\right) $ from (\ref{rrc34}) one gets%
\begin{equation*}
v_{2}\left( x\right) =\xi ^{\nu }e^{-\xi /2}Y\left( \xi \right) ,
\end{equation*}%
where $Y\left( \xi \right) $ is a polynomial of degree $n.$ But function $%
v_{2}\left( x\right) $ satisfies (\ref{rrc23}). By the previous
consideration the substitution%
\begin{equation*}
v_{2}\left( x\right) =x^{\nu }e^{-ax}y\left( x\right)
\end{equation*}%
gives%
\begin{equation*}
xy^{\prime \prime }+\left( 2\nu -2ax\right) y^{\prime }+2any=0,
\end{equation*}%
in view of the quantization rule (\ref{rrc33}). The change of the variable $%
y\left( x\right) =Y\left( \xi \right) $ with $\xi =2ax$ results in%
\begin{equation}
\xi Y^{\prime \prime }+\left( 2\nu -\xi \right) Y^{\prime }+nY=0
\label{rrc36}
\end{equation}%
and the only polynomial solutions are the Laguerre polynomials $L_{n}^{2\nu
-1}\left( \xi \right) ,$ whence (\ref{rrc35}) is correct. Solutions $%
v_{2}\left( x\right) $ are square integrable functions on $\left( 0,\infty
\right) .$\smallskip

To find the relations between the coefficients $A_{n}$ and $B_{n}$ in (\ref%
{rrc34}) and (\ref{rrc35}) we take the limit $x\rightarrow 0$ in (\ref{rrc19}%
) with the help of the following properties of the Laguerre polynomials \cite%
{Ni:Su:Uv}, \cite{Ni:Uv}, \cite{Sze}:%
\begin{equation}
\frac{d}{d\xi }L_{n}^{\alpha }\left( \xi \right) =-L_{n-1}^{\alpha +1}\left(
\xi \right) ,\qquad L_{n}^{\alpha }\left( 0\right) =\frac{\Gamma \left(
\alpha +n+1\right) }{n!\Gamma \left( \alpha +1\right) }.  \label{rrc37}
\end{equation}%
The result is%
\begin{equation*}
2a\left( \nu +1\right) A_{n}L_{n-1}^{2\nu +1}\left( 0\right) =-2a\nu
A_{n}L_{n-1}^{2\nu +1}\left( 0\right) +\left( 1+\frac{\varepsilon \kappa }{%
\nu }\right) B_{n}L_{n}^{2\nu -1}\left( 0\right) ,
\end{equation*}%
whence%
\begin{equation*}
A_{n}=\frac{\nu +\varepsilon \kappa }{an\left( n+2\nu \right) }\;B_{n}\qquad
\left( n=1,2,3,\;...\right) .
\end{equation*}%
Since%
\begin{eqnarray*}
a^{2}n\left( n+2\nu \right) &=&a^{2}\left( \left( n+\nu \right) ^{2}-\nu
^{2}\right) =\mu ^{2}\varepsilon ^{2}-a^{2}\nu ^{2} \\
&=&\mu ^{2}\varepsilon ^{2}-\left( 1-\varepsilon ^{2}\right) \nu ^{2}=\kappa
^{2}\varepsilon ^{2}-\nu ^{2},
\end{eqnarray*}%
we have proved the useful identity%
\begin{equation}
a^{2}n\left( n+2\nu \right) =\varepsilon ^{2}\kappa ^{2}-\nu ^{2},
\label{rrc38}
\end{equation}%
and the final relation is%
\begin{equation}
A_{n}=\dfrac{a}{\kappa \varepsilon -\nu }\;B_{n}.  \label{rrc39}
\end{equation}

By (\ref{rrc10}) and (\ref{rrc18}) we find%
\begin{equation*}
\left(
\begin{array}{c}
u_{1}\medskip \\
u_{2}%
\end{array}%
\right) =C^{-1}\left(
\begin{array}{c}
v_{1}\medskip \\
v_{2}%
\end{array}%
\right) ,\qquad C^{-1}=\frac{1}{2\nu \left( \kappa -\nu \right) }\left(
\begin{array}{cc}
\mu & \smallskip \kappa -\nu \\
\kappa -\smallskip \nu & \mu%
\end{array}%
\right) .
\end{equation*}%
Therefore%
\begin{eqnarray}
xf\left( x\right) &=&\frac{B_{n}}{2\nu \left( \kappa -\nu \right) }\xi ^{\nu
}e^{-\xi /2}\left( f_{1}\xi L_{n-1}^{2\nu +1}\left( \xi \right)
+f_{2}L_{n}^{2\nu -1}\left( \xi \right) \right) ,  \label{rrc40} \\
xg\left( x\right) &=&\frac{B_{n}}{2\nu \left( \kappa -\nu \right) }\xi ^{\nu
}e^{-\xi /2}\left( g_{1}\xi L_{n-1}^{2\nu +1}\left( \xi \right)
+g_{2}L_{n}^{2\nu -1}\left( \xi \right) \right) ,  \label{rrc41}
\end{eqnarray}%
where%
\begin{equation}
f_{1}=\frac{a\mu }{\varepsilon \kappa -\nu },\quad f_{2}=\kappa -\nu ,\quad
g_{1}=\frac{a\left( \kappa -\nu \right) }{\varepsilon \kappa -\nu },\quad
g_{2}=\mu .  \label{rrc42}
\end{equation}%
These formulas remain valid for $n=0;$ in this case the terms containing $%
L_{n-1}^{2\nu +1}\left( \xi \right) $ have to be taken to be zero. Thus we
derive the representation for the radial functions (\ref{rc3}) up to the
constant $B_{n}.$ The normalization condition (\ref{rrc29}) gives the value
of this constant as
\begin{equation}
B_{n}=a\beta ^{3/2}\sqrt{\frac{\left( \kappa -\nu \right) \left( \varepsilon
\kappa -\nu \right) n!}{\mu \Gamma \left( n+2\nu \right) }}.  \label{rrc43}
\end{equation}%
This has been already verified in Section~4.5. Observe that Eq.~(\ref{rrc43}%
) applies when $n=0.$\smallskip

The familiar recurrence relations for the Laguerre polynomials (\ref{a2lr}%
)--(\ref{a2lrr}) allow to present the radial functions (\ref{rc3}) in a
traditional form \cite{Akh:Ber}, \cite{Ber:Lif:Pit}, \cite{Davis} as
\begin{eqnarray}
\left(
\begin{array}{c}
F\left( r\right) \medskip \mathbf{\medskip \bigskip } \\
G\left( r\right)%
\end{array}%
\right) &=&a^{2}\beta ^{3/2}\sqrt{\frac{n!}{\mu \left( \kappa -\nu \right)
\left( \varepsilon \kappa -\nu \right) \Gamma \left( n+2\nu \right) }}\ \xi
^{\nu -1}e^{-\xi /2}  \notag \\
&&\times \left(
\begin{array}{c}
\alpha _{1}\qquad \alpha _{2}\mathbf{\medskip \medskip } \\
\beta _{1}\qquad \beta _{2}%
\end{array}%
\right) \left(
\begin{array}{c}
L_{n-1}^{2\nu }\left( \xi \right) \bigskip \mathbf{\medskip } \\
L_{n}^{2\nu }\left( \xi \right)%
\end{array}%
\right)  \label{rrc44}
\end{eqnarray}%
with%
\begin{equation}
\alpha _{1}=\sqrt{1+\varepsilon }\left( \left( \kappa -\nu \right) \sqrt{%
1+\varepsilon }+\mu \sqrt{1-\varepsilon }\right) ,\quad \alpha _{2}=-\sqrt{%
1+\varepsilon }\left( \left( \kappa -\nu \right) \sqrt{1+\varepsilon }-\mu
\sqrt{1-\varepsilon }\right) ,  \label{maa}
\end{equation}%
\begin{equation}
\beta _{1}=\sqrt{1-\varepsilon }\left( \left( \kappa -\nu \right) \sqrt{%
1+\varepsilon }+\mu \sqrt{1-\varepsilon }\right) ,\quad \beta _{2}=\sqrt{%
1-\varepsilon }\left( \left( \kappa -\nu \right) \sqrt{1+\varepsilon }-\mu
\sqrt{1-\varepsilon }\right) .  \label{mbb}
\end{equation}%
By (\ref{a2}) one can rewrite this representation in terms of the confluent
hypergeometric functions.

\subsection{Nonrelativistic Limit of the Wave Functions}

Throughout the paper we have always used the notation $n=n_{r}$ for the
radial quantum number, which determines the number of zeros of the radial
functions in the relativistic Coulomb problem; see (\ref{rc3}). For the sake
of passing to the limit $c\rightarrow \infty $ in this section, let us
introduce the principal quantum number of the nonrelativistic hydrogen atom
as $n=n_{r}+\left| \kappa \right| =n_{r}+j+1/2$ and temporarily consider $%
N=n_{r}+\nu $ as its ``relativistic analog''. As $c\rightarrow \infty $ one
gets%
\begin{equation}
\nu =\sqrt{\kappa ^{2}-\mu ^{2}}=\left| \kappa \right| -\frac{\mu ^{2}}{%
2\left| \kappa \right| }-\frac{\mu ^{4}}{8\left| \kappa \right| ^{3}}+\text{O%
}\left( \mu ^{6}\right) ,  \label{lim1}
\end{equation}%
\begin{equation}
N=n_{r}+\nu =n_{r}+\left| \kappa \right| -\frac{\mu ^{2}}{2\left| \kappa
\right| }-\frac{\mu ^{4}}{8\left| \kappa \right| ^{3}}+\text{O}\left( \mu
^{6}\right)  \label{lim2}
\end{equation}%
as $\mu =Ze^{2}/\hbar c\rightarrow 0.$ As a result, for the discrete energy
levels%
\begin{equation}
\varepsilon =\left( 1+\frac{\mu ^{2}}{N^{2}}\right) ^{-1/2}  \label{lim3}
\end{equation}%
we arrive at the expansion (\ref{rc6a}) in the nonrelativistic limit $%
c\rightarrow \infty .$\smallskip

In a similar fashion,%
\begin{equation}
a=\sqrt{1-\varepsilon ^{2}}=\frac{\mu }{n_{r}+\left\vert \kappa \right\vert }%
\left( 1+\frac{n_{r}\;\mu ^{2}}{2\left\vert \kappa \right\vert \left(
n_{r}+\left\vert \kappa \right\vert \right) ^{2}}+\text{O}\left( \mu
^{4}\right) \right) ,  \label{lim4}
\end{equation}%
\begin{equation}
\xi =\xi \left( c\right) =2a\;\frac{mc}{\hbar }\;r=\frac{2Ze^{2}m}{n\hbar
^{2}}r\left( 1+\text{O}\left( \mu ^{2}\right) \right)  \label{lim5}
\end{equation}%
as $\mu \rightarrow 0,$ thus giving%
\begin{equation}
\lim_{c\rightarrow \infty }\xi \left( c\right) =\eta =\frac{2Z}{n}\left(
\frac{r}{a_{0}}\right)  \label{lim6}
\end{equation}%
by (\ref{nrc2a}). Also%
\begin{equation}
\kappa -\nu =\left( \kappa -\left\vert \kappa \right\vert \right) +\frac{\mu
^{2}}{2\left\vert \kappa \right\vert }+\text{O}\left( \mu ^{4}\right) ,
\label{lim7}
\end{equation}%
\begin{equation}
\varepsilon \kappa -\nu =\left( \kappa -\left\vert \kappa \right\vert
\right) +\frac{\left( n_{r}+\left\vert \kappa \right\vert \right)
^{2}-\kappa \left\vert \kappa \right\vert }{2\left\vert \kappa \right\vert
\left( n_{r}+\left\vert \kappa \right\vert \right) ^{2}}\;\mu ^{2}+\text{O}%
\left( \mu ^{4}\right)  \label{lim8}
\end{equation}%
as $\mu \rightarrow 0.$ This allows to evaluate the nonrelativistic limit of
the transition matrix:%
\begin{equation}
S=\left(
\begin{array}{c}
f_{1}\qquad f_{2}\mathbf{\medskip } \\
g_{1}\qquad g_{2}%
\end{array}%
\right) =\left(
\begin{array}{c}
\dfrac{a\mu }{\varepsilon \kappa -\nu }\qquad \kappa -\nu \mathbf{\medskip }
\\
\dfrac{a\left( \kappa -\nu \right) }{\varepsilon \kappa -\nu }\qquad \mu%
\end{array}%
\right) .  \label{lim9}
\end{equation}%
There are two distinct cases with the end result%
\begin{equation}
\psi _{\pm }=\left(
\begin{array}{c}
\mathcal{Y}^{\pm }F\medskip \\
i\mathcal{Y}^{\mp }G%
\end{array}%
\right) \rightarrow \left(
\begin{array}{c}
\pm \mathcal{Y}^{\pm }R\medskip \\
0%
\end{array}%
\right) ,\qquad \mu \rightarrow 0.  \label{lim9a}
\end{equation}%
Here $R=R_{nl}\left( r\right) $ are the nonrelativistic radial functions (%
\ref{nrc2})--(\ref{nrc2a}) and $\mathcal{Y}^{\pm }=\mathcal{Y}_{jm}^{\left(
j\pm 1/2\right) }\left( \mathbf{n}\right) $ are the spinor spherical
harmonics (\ref{ang5}).\smallskip

Indeed, if $\kappa =\left| \kappa \right| =j+1/2=l,$%
\begin{equation*}
S=S_{+}\left( \mu \right) =\left(
\begin{array}{c}
\dfrac{2\kappa \left( n_{r}+\kappa \right) }{n_{r}\left( n_{r}+2\kappa
\right) }+\text{O}\left( \mu ^{2}\right) \quad \dfrac{\mu ^{2}}{2\left|
\kappa \right| }+\text{O}\left( \mu ^{4}\right) \mathbf{\medskip } \\
\dfrac{\left( n_{r}+\kappa \right) \;\mu }{n_{r}\left( n_{r}+2\kappa \right)
}+\text{O}\left( \mu ^{3}\right) \qquad \qquad \mu%
\end{array}%
\right) \thicksim \left(
\begin{array}{c}
1\qquad \mu ^{2}\mathbf{\medskip } \\
\mu \qquad \mu%
\end{array}%
\right)
\end{equation*}%
as $\mu \rightarrow 0$ or%
\begin{equation}
\lim_{\mu \rightarrow 0}S_{+}\left( \mu \right) =\dfrac{2nl}{n^{2}-l^{2}}%
\left(
\begin{array}{cc}
1 & 0\medskip \\
0 & 0%
\end{array}%
\right) .  \label{lim10}
\end{equation}%
In this case $\nu \rightarrow l$ and, therefore,%
\begin{equation}
\medskip \mathbf{\medskip \bigskip }\left(
\begin{array}{c}
\mathbf{\medskip }F\left( r\right) \\
G\left( r\right)%
\end{array}%
\right) \rightarrow \left( \frac{Ze^{2}m}{\hbar ^{2}}\right) ^{3/2}\left(
\begin{array}{c}
1\mathbf{\medskip } \\
0%
\end{array}%
\right) \frac{2}{n^{2}}\sqrt{\frac{\left( n-l-1\right) !}{\left( n+l\right) !%
}}\ \eta ^{l}e^{-\eta /2}\;\eta L_{n-l-1}^{2l+1}\left( \eta \right)
\label{lim11}
\end{equation}%
in the limit $c\rightarrow \infty $ thus giving%
\begin{equation}
\psi _{+}=\left(
\begin{array}{c}
\mathcal{Y}^{+}F\medskip \\
i\mathcal{Y}^{-}G%
\end{array}%
\right) \rightarrow \left(
\begin{array}{c}
\mathcal{Y}^{+}R\medskip \\
0%
\end{array}%
\right) ,\qquad \mu \rightarrow 0.  \label{lim12}
\end{equation}

In a similar fashion, when $\kappa =-\left| \kappa \right| =-\left(
j+1/2\right) =-l-1$ one gets%
\begin{equation}
\psi _{-}=\left(
\begin{array}{c}
\mathcal{Y}^{-}F\medskip \\
i\mathcal{Y}^{+}G%
\end{array}%
\right) \rightarrow \left(
\begin{array}{c}
-\mathcal{Y}^{-}R\medskip \\
0%
\end{array}%
\right) ,\qquad \mu \rightarrow 0  \label{lim13}
\end{equation}%
due to the corresponding asymptotic form of the transition matrix $%
S=S_{-}\left( \mu \right) :$%
\begin{equation}
S_{-}\left( \mu \right) =\left(
\begin{array}{c}
-\dfrac{\mu ^{2}}{2\left| \kappa \right| \left( n_{r}+\left| \kappa \right|
\right) }+\text{O}\left( \mu ^{4}\right) \quad -2\left| \kappa \right| +%
\text{O}\left( \mu ^{2}\right) \mathbf{\medskip } \\
\dfrac{\mu }{n_{r}+\left| \kappa \right| }+\text{O}\left( \mu ^{3}\right)
\qquad \qquad \mu%
\end{array}%
\right) \thicksim \left(
\begin{array}{c}
\mu ^{2}\qquad \mathbf{\medskip }1 \\
\mu \qquad \mu%
\end{array}%
\right)  \label{lim14}
\end{equation}%
as $\mu \rightarrow 0$ \cite{Ni:Uv}. This completes the proof of (\ref{lim9a}%
).\smallskip

The representation of the radial functions in the form (\ref{rc3}), due to
Nikiforov and Uvarov \cite{Ni:Uv}, is well adapted for passing to the
nonrelativistic limit since one coefficient of the transition matrix $S$ is
much larger than the others as $\mu \rightarrow 0.$ In the traditional form (%
\ref{maa})--(\ref{mbb}), however, there is an overlap of the orders of these
coefficients and one has to use the recurrence relations (\ref{a2lr})--(\ref%
{a2lrr}) in order to obtain the nonrelativistic wave functions as a limiting
case of relativistic ones.

\section{Method of Separation of Variables and Its Extension}

In this section we give an extension of the method of separation of
variables, that is used in theoretical and mathematical physics for solving
partial differential equations, from a single equation to a system of
partial differential equations which we call Dirac-type system.

\subsection{Method of Separation of Variables}

We follow \cite{Ni:Uv} and give an extension for suitable Dirac's systems.
The method of separation of the variables helps to find particular solutions
of equation%
\begin{equation}
\mathcal{L}u=0  \label{s1}
\end{equation}%
if the operator $\mathcal{L}$ can be represented in the form%
\begin{equation}
\mathcal{L}=\mathcal{M}_{1}\mathcal{N}_{1}+\mathcal{M}_{2}\mathcal{N}_{2}.
\label{s2}
\end{equation}%
Here the operators $\mathcal{M}_{1}$ and $\mathcal{M}_{2}$ act only on one
subset of the variables, and the operators $\mathcal{N}_{1}$ and $\mathcal{N}%
_{2}$ act on the others; a product of operators $\mathcal{M}_{i}\mathcal{N}%
_{k}$ means the result of applying them successively $\left( \mathcal{M}_{i}%
\mathcal{N}_{k}\right) u=\mathcal{M}_{i}\left( \mathcal{N}_{k}u\right) $
with $i,k=1,2$ $;$ it is assumed that the operators $\mathcal{M}_{i}$ and $%
\mathcal{N}_{i}$ are linear operators.\smallskip

We look for solutions of equation (\ref{s1}) in the form%
\begin{equation}
u=f\;g,  \label{s3}
\end{equation}%
where the first unknown function $f$ depends only on the first set of
variables and the second function $g$ depends on the others. Since%
\begin{eqnarray*}
\mathcal{M}_{i}\mathcal{N}_{k}u &=&\left( \mathcal{M}_{i}\mathcal{N}%
_{k}\right) \left( f\;g\right) =\mathcal{M}_{i}\left( \mathcal{N}_{k}\left(
f\;g\right) \right) \\
&=&\mathcal{M}_{i}\left( f\;\left( \mathcal{N}_{k}g\right) \right) =\left(
\mathcal{M}_{i}f\right) \;\left( \mathcal{N}_{k}g\right)
\end{eqnarray*}%
the equation $\mathcal{L}u=0$ can be rewritten in the form%
\begin{equation*}
\frac{\mathcal{M}_{1}f}{\mathcal{M}_{2}f}=-\frac{\mathcal{N}_{2}g}{\mathcal{N%
}_{1}g},
\end{equation*}%
where the left hand side is independent of the second group of the variables
and the right hand side is independent of the first ones. Thus, we must have%
\begin{equation*}
\frac{\mathcal{M}_{1}f}{\mathcal{M}_{2}f}=-\frac{\mathcal{N}_{2}g}{\mathcal{N%
}_{1}g}=\lambda ,
\end{equation*}%
where $\lambda $ is a constant, and one obtains equations%
\begin{equation}
\mathcal{M}_{1}f=\lambda \mathcal{M}_{2}f,\qquad \mathcal{N}_{2}g=-\lambda
\mathcal{N}_{1}g  \label{s4}
\end{equation}%
each containing functions of only some of the variables. Since $\mathcal{L}$
is linear, a linear combination of solutions,%
\begin{equation}
u=\sum_{k}c_{k}\;f_{k}g_{k}  \label{s5}
\end{equation}%
with some constants $c_{k},$ corresponding to all admissible values of $%
\lambda =\lambda _{k},$ will be a solution of the original equation (\ref{s1}%
). Under certain condition of the completeness of the constructed set of
particular solutions, every solution of (\ref{s1}) can be represented in the
form (\ref{s5}). The method of separation of variables is very useful in
theoretical and mathematical physics and partial differential equations ---
including solutions of the nonrelativistic Schr\"{o}dinger equation --- but,
as we have seen in Section 6.2, it should be modified in the case of the
Dirac equation.\smallskip

\noindent \textbf{Example.\/} \ The Schr\"{o}dinger equation in the central
field with the potential energy $U\left( r\right) $ is%
\begin{equation}
\Delta \psi +\frac{2m}{\hbar ^{2}}\left( E-U\left( r\right) \right) \psi =0.
\label{s6}
\end{equation}%
The Laplace operator in the spherical coordinates $r,$ $\theta ,$ $\varphi $
has the form \cite{Ni:Su:Uv}, \cite{Ni:Uv}%
\begin{equation}
\Delta =\Delta _{r}+\frac{1}{r^{2}}\Delta _{\omega }  \label{s7}
\end{equation}%
with%
\begin{equation}
\Delta _{r}=\frac{1}{r^{2}}\frac{\partial }{\partial r}\left( r^{2}\frac{%
\partial }{\partial r}\right) ,\quad \Delta _{\omega }=\frac{1}{\sin \theta }%
\frac{\partial }{\partial \theta }\left( \sin \theta \frac{\partial }{%
\partial \theta }\right) +\frac{1}{\sin ^{2}\theta }\frac{\partial ^{2}}{%
\partial \varphi ^{2}}.  \label{s8}
\end{equation}%
Thus%
\begin{equation}
\mathcal{M}_{1}=\Delta _{r}+\frac{2m}{\hbar ^{2}}\left( E-U\left( r\right)
\right) ,\quad \mathcal{M}_{2}=\frac{1}{r^{2}}  \label{s9}
\end{equation}%
\begin{equation}
\mathcal{N}_{1}=\text{id}=I,\qquad \mathcal{N}_{2}=\Delta _{\omega }
\label{s10}
\end{equation}%
and separation of the variables $\psi =R\left( r\right) Y\left( \theta
,\varphi \right) $ gives%
\begin{equation}
\Delta _{\omega }Y\left( \theta ,\varphi \right) +\lambda Y\left( \theta
,\varphi \right) =0,  \label{s11}
\end{equation}%
\begin{equation}
\frac{1}{r^{2}}\frac{d}{dr}\left( r^{2}\frac{dR\left( r\right) }{dr}\right)
+\left( \frac{2m}{\hbar ^{2}}\left( E-U\left( r\right) \right) -\frac{%
\lambda }{r^{2}}\right) R\left( r\right) =0.  \label{s12}
\end{equation}%
Bounded single-valued solutions of equation (\ref{s11}) on the sphere $S^{2}$
exist only when $\lambda =l\left( l+1\right) $ with $l=0,1,2,\;...\;.$ They
are the spherical harmonics $Y=Y_{lm}\left( \theta ,\varphi \right) .$

\subsection{Dirac-Type Systems}

Let us consider the system of two equations%
\begin{eqnarray}
\mathcal{P}u &=&\alpha v,  \label{b1} \\
\mathcal{P}v &=&\beta u,  \label{b2}
\end{eqnarray}%
where $u=u\left( \mathbf{x}\right) $ and $v=v\left( \mathbf{x}\right) $ are
some unknown (complex) vector valued functions on $\mathbf{R}^{n}$ (or $%
\mathbf{C}^{n}~$). Here operator $\mathcal{P}$ has the following structure
\begin{equation}
\mathcal{P}=\mathcal{N}\left( \mathbf{n}\right) \left( \mathcal{D}_{1}\left(
r\right) \mathcal{L}_{1}\left( \mathbf{n}\right) +\mathcal{D}_{2}\left(
r\right) \mathcal{L}_{2}\left( \mathbf{n}\right) \right) ,\quad \mathcal{N}%
^{2}\left( \mathbf{n}\right) =\text{id}=I,  \label{b3}
\end{equation}%
where $\mathcal{D}_{i}=\mathcal{D}_{i}\left( r\right) ,$ $\mathcal{L}_{k}=%
\mathcal{L}_{k}\left( \mathbf{n}\right) $ and $\mathcal{N=N}\left( \mathbf{n}%
\right) $ are linear operators\ acting with respect to two different subsets
of variables, say ``radial'' $r$ and ``angular'' $\mathbf{n}$ variables,
respectively (in the case of the hyperspherical coordinates in $\mathbf{R}%
^{n}$ \cite{Ni:Su:Uv} one gets $\mathbf{x}=r\mathbf{n}$ and $\mathbf{n}%
^{2}=1,$ which justifies our terminology). The following algebraic
properties hold%
\begin{eqnarray}
&&\left[ \mathcal{D}_{i},\mathcal{L}_{k}\right] =\left[ \mathcal{D}_{i},%
\mathcal{N}\right] =0,  \label{b4} \\
&&\left[ \mathcal{N},\mathcal{L}_{1}\right] =\left[ \mathcal{L}_{1},\mathcal{%
L}_{2}\right] =0,  \label{b5} \\
&&\mathcal{NL}_{2}+\mathcal{L}_{2}\mathcal{N}=\gamma \mathcal{N},  \label{b6}
\end{eqnarray}%
where $\left[ \mathcal{A},\ \mathcal{B}\right] =\mathcal{AB}-\mathcal{BA}$
is the commutator and $\gamma $ is some constant.

We look for solutions of (\ref{b1})--(\ref{b2}) in the form%
\begin{eqnarray}
u &=&\mathcal{Y}\left( \mathbf{n}\right) R\left( r\right) ,  \label{b7} \\
v &=&\left( \mathcal{NY}\left( \mathbf{n}\right) \right) S\left( r\right) ,
\label{b8}
\end{eqnarray}%
where $\mathcal{Y}$ is the common eigenfunction of commuting operators $%
\mathcal{L}_{1}$ and $\mathcal{L}_{2}:$%
\begin{equation}
\mathcal{L}_{1}\mathcal{Y}=\kappa _{1}\mathcal{Y},\qquad \mathcal{L}_{2}%
\mathcal{Y}=\kappa _{2}\mathcal{Y}.  \label{b9}
\end{equation}%
If $w=w\left( \mathbf{x}\right) =F\left( \mathbf{n}\right) G\left( r\right)
, $ we define the action of the ``radial'' and ``angular'' operators in (\ref%
{b3}) as follows%
\begin{equation}
\mathcal{L}_{i}w=\left( \mathcal{L}_{i}F\right) G,\quad \mathcal{N}w=\left(
\mathcal{N}F\right) G,\quad \mathcal{D}_{k}w=F\left( \mathcal{D}_{k}G\right)
.  \label{b9a}
\end{equation}%
The Ansatz (\ref{b7})--(\ref{b8}) results in two equation for our ``radial''
functions $R$ and $S:$%
\begin{eqnarray}
\kappa _{1}\mathcal{D}_{1}R+\kappa _{2}\mathcal{D}_{2}R &=&\alpha S,
\label{b10} \\
\kappa _{1}\mathcal{D}_{1}S+\left( \gamma -\kappa _{2}\right) \mathcal{D}%
_{2}S &=&\beta R.  \label{b11}
\end{eqnarray}

Indeed, in view of (\ref{b1})--(\ref{b2}) and (\ref{b7})--(\ref{b9}) one gets%
\begin{eqnarray*}
\mathcal{P}u &=&\mathcal{N}\left( \mathcal{D}_{1}\mathcal{L}_{1}+\mathcal{D}%
_{2}\mathcal{L}_{2}\right) \mathcal{Y}R \\
&=&\mathcal{N}\left( \left( \mathcal{L}_{1}\mathcal{Y}\right) \left(
\mathcal{D}_{1}R\right) +\left( \mathcal{L}_{2}\mathcal{Y}\right) \left(
\mathcal{D}_{2}R\right) \right) \\
&=&\left( \mathcal{NY}\right) \left( \kappa _{1}\mathcal{D}_{1}R+\kappa _{2}%
\mathcal{D}_{2}R\right) \\
&=&\alpha \left( \mathcal{NY}\right) S=\alpha v,
\end{eqnarray*}%
which gives (\ref{b10}). In a similar fashion, with the aid of (\ref{b6})%
\begin{eqnarray*}
\mathcal{P}v &=&\mathcal{N}\left( \mathcal{D}_{1}\mathcal{L}_{1}+\mathcal{D}%
_{2}\mathcal{L}_{2}\right) \left( \mathcal{NY}\right) S \\
&=&\mathcal{N}\left( \left( \mathcal{L}_{1}\mathcal{NY}\right) \left(
\mathcal{D}_{1}S\right) +\left( \mathcal{L}_{2}\mathcal{NY}\right) \left(
\mathcal{D}_{2}S\right) \right) \\
&=&\left( \mathcal{NL}_{1}\mathcal{NY}\right) \left( \mathcal{D}_{1}S\right)
+\left( \mathcal{NL}_{2}\mathcal{NY}\right) \left( \mathcal{D}_{2}S\right) \\
&=&\left( \mathcal{N}^{2}\mathcal{L}_{1}\mathcal{Y}\right) \left( \mathcal{D}%
_{1}S\right) +\left( \left( \gamma -\mathcal{L}_{2}\right) \mathcal{N}^{2}%
\mathcal{Y}\right) \left( \mathcal{D}_{2}S\right) \\
&=&\mathcal{Y}\left( \kappa _{1}\mathcal{D}_{1}S+\left( \gamma -\kappa
_{2}\right) \mathcal{D}_{2}S\right) =\beta \mathcal{Y}R=\beta u,
\end{eqnarray*}%
which results in the second equation (\ref{b11}) and our proof is
complete.\smallskip

\noindent \textbf{Example.\/} \ The original Dirac system (\ref{ds4})--(\ref%
{ds5}) has%
\begin{equation*}
\psi =\left(
\begin{array}{c}
\mathbf{u\medskip } \\
\mathbf{v}%
\end{array}%
\right) =\left(
\begin{array}{c}
\mathbf{\varphi \medskip } \\
\mathbf{\chi }%
\end{array}%
\right)
\end{equation*}%
and%
\begin{equation}
\mathcal{P}=c\mathbf{\sigma p}=\hbar c\left( \mathbf{\sigma n}\right) \left(
\frac{1}{i}\frac{\partial }{\partial r}+\frac{i}{r}\mathbf{\sigma l}\right) .
\label{ed1}
\end{equation}%
Here%
\begin{equation}
\mathcal{N}=\mathbf{\sigma n},\quad \mathcal{D}_{1}=\frac{\hbar c}{i}\frac{%
\partial }{\partial r},\quad \mathcal{L}_{1}=\text{id}=I,\quad \mathcal{D}%
_{2}=\frac{i\hbar c}{r},\quad \mathcal{L}_{2}=\mathbf{\sigma l}  \label{ed2}
\end{equation}%
and%
\begin{equation}
\alpha \left( r\right) =E+mc^{2}-U\left( r\right) ,\quad \beta \left(
r\right) =E-mc^{2}-U\left( r\right) ,\quad \gamma =-2  \label{ed3}
\end{equation}%
by (\ref{ds9}). Moreover, $\kappa _{1}=1,$ $\kappa _{2}=-\left( 1+\kappa
\right) $ and we use $R=F\left( r\right) ,$ $S=-iG\left( r\right) .$ The
Anzats (\ref{ds10})--(\ref{ds11}) gives the familiar radial equations (\ref%
{ds12})--(\ref{ds13}).

\section{Appendix: Useful Formulas}

This section contains some relations involving the generalized
hypergeometric series, the Laguerre and Hahn polynomials, the spherical
harmonics and Clebsch--Gordan coefficients, which are used throughout the
paper.\smallskip

The generalized hypergeometric series is \cite{An:As:Ro}, \cite{Ba}, \cite%
{BooleDelta}, \cite{Ga:Ra}
\begin{eqnarray}
&&\ _{p}F_{q}\left( a_{1},\ a_{2},\ ...\ ,\ a_{p};\;b_{1},\ b_{2},\ ...\ ,\
b_{q};\ z\right)  \label{a1} \\
&&\quad =\ _{p}F_{q}\left(
\begin{array}{c}
a_{1},\ a_{2},\ ...\ ,\ a_{p}\medskip \\
b_{1},\ b_{2},\ ...\ ,\ b_{q}\medskip%
\end{array}%
;\ z\right) =\sum_{n=0}^{\infty }\frac{\left( a_{1}\right) _{n}\left(
a_{2}\right) _{n}...\left( a_{p}\right) _{n}\ z^{n}}{\left( b_{1}\right)
_{n}\left( b_{2}\right) _{n}...\left( b_{q}\right) _{n}n!},  \notag
\end{eqnarray}%
where $\left( a\right) _{n}=a\left( a+1\right) ...\left( a+n-1\right)
=\Gamma \left( a+n\right) /\Gamma \left( a\right) .$ By the ratio test, the $%
{}_{p}F_{q}$ series converges absolutely for all complex values of $z$ if $%
p\leq q,$ and for $|z|<1$ if $p=q+1.$ By an extension of the ratio test (%
\cite{Brom}, p.~241), it converges absolutely for $|z|=1$ if $p=q+1$ and $%
z\neq 0$ or $p=q+1$ and $\text{Re}\,\left[ b_{1}+\ldots +b_{q}-\left(
a_{1}+\ldots +a_{p}\right) \right] >0.$ If $p>q+1$ and $z\neq 0$ or $p=q+1$
and $|z|>1,$ then this series diverges, unless it terminates.\smallskip

The Laguerre polynomials are defined as \cite{An:As:Ro}, \cite{Ni:Su:Uv},
\cite{Ni:Uv}, \cite{Sze}%
\begin{equation}
L_{n}^{\alpha }\left( x\right) =\frac{\Gamma \left( \alpha +n+1\right) }{%
n!\;\Gamma \left( \alpha +1\right) }\ _{1}F_{1}\left(
\begin{array}{c}
-n\medskip \\
\alpha +1\medskip%
\end{array}%
;\ x\right) .  \label{a2}
\end{equation}%
It is a consequence of Theorem 3. The differentiation formulas \cite%
{Ni:Su:Uv}, \cite{Ni:Uv}%
\begin{equation}
\frac{d}{dx}L_{n}^{\alpha }\left( x\right) =-L_{n-1}^{\alpha +1}\left(
x\right) ,  \label{a2ld}
\end{equation}%
\begin{equation}
x\frac{d}{dx}L_{n}^{\alpha }\left( x\right) =nL_{n}^{\alpha }\left( x\right)
-\left( \alpha +n\right) L_{n-1}^{\alpha }\left( x\right)  \label{a2ldd}
\end{equation}%
imply a recurrence relation%
\begin{equation}
xL_{n-1}^{\alpha +1}\left( x\right) =\left( \alpha +n\right) L_{n-1}^{\alpha
}\left( x\right) -nL_{n}^{\alpha }\left( x\right) .  \label{a2lr}
\end{equation}%
The simplest case of the connecting relation (\ref{i10a}) is%
\begin{equation}
L_{n}^{\alpha }\left( x\right) =L_{n}^{\alpha +1}\left( x\right)
-L_{n-1}^{\alpha +1}\left( x\right) .  \label{a2lrr}
\end{equation}

The Hahn polynomials are \cite{Ni:Su:Uv}, \cite{Ni:Uv}%
\begin{equation}
h_{n}^{\left( \alpha ,\ \beta \right) }\left( x,N\right) =\left( -1\right)
^{n}\frac{\Gamma \left( N\right) \left( \beta +1\right) _{n}}{n!\;\Gamma
\left( N-n\right) }\ _{3}F_{2}\left(
\begin{array}{c}
-n\medskip ,\ \alpha +\beta +n+1,\ -x \\
\beta +1\medskip ,\quad 1-N%
\end{array}%
;\ 1\right) .  \label{a3}
\end{equation}%
We usually omit the argument of the hypergeometric series $_{3}F_{2}$ if it
is equal to one. An asymptotic relation with the Jacobi polynomials is%
\begin{equation}
\frac{1}{\left. \widetilde{N}\right. ^{n}}\;h_{n}^{\left( \alpha ,\ \beta
\right) }\left( \frac{\widetilde{N}}{2}\left( 1+s\right) -\frac{\beta +1}{2}%
,\;N\right) =P_{n}^{\left( \alpha ,\ \beta \right) }\left( s\right) +\text{O}%
\left( \frac{1}{\left. \widetilde{N}\right. ^{2}}\right) ,  \label{a3ha}
\end{equation}%
where $\widetilde{N}=N+\left( \alpha +\beta \right) /2$ and $N\rightarrow
\infty ;$ see \cite{Ni:Su:Uv} for more details.\smallskip

Thomae's transformation \cite{Ba}, \cite{Ga:Ra} is%
\begin{equation}
_{3}F_{2}\left(
\begin{array}{c}
-n\medskip ,\ a,\ b \\
c\medskip ,~d%
\end{array}%
;\ 1\right) =\frac{\left( d-b\right) _{n}}{\left( d\right) _{n}}\
_{3}F_{2}\left(
\begin{array}{c}
-n\medskip ,\ c-a,\ b \\
c\medskip ,~b-d-n+1%
\end{array}%
;\ 1\right)  \label{a4}
\end{equation}%
with $n=0,1,2,\ ...\ .$\smallskip

The summation formula of Gauss \cite{An:As:Ro}, \cite{Ba}, \cite{Ga:Ra}%
\begin{equation}
_{2}F_{1}\left(
\begin{array}{c}
a\medskip ,\ b \\
c\medskip%
\end{array}%
;\ 1\right) =\frac{\Gamma \left( c\right) \Gamma \left( c-a-b\right) }{%
\Gamma \left( c-a\right) \Gamma \left( c-b\right) },\qquad \text{Re}\left(
c-a-b\right) >0.  \label{a4a}
\end{equation}

The gamma function is defined as \cite{An:As:Ro}, \cite{Erd}, \cite{Ni:Uv}%
\begin{equation}
\Gamma \left( z\right) =\int_{0}^{\infty }e^{-t}t^{z-1}\;dt,\qquad \text{Re\
}z>0.  \label{gamma1}
\end{equation}%
It can be continued analytically over the whole complex plane except the
points $z=0,-1,-2,\;...$ at which it has simple poles. Functional equations
are%
\begin{equation}
\Gamma \left( z+1\right) =z\Gamma \left( z\right) ,  \label{gamma2}
\end{equation}%
\begin{equation}
\Gamma \left( z\right) \Gamma \left( 1-z\right) =\frac{\pi }{\sin \pi z},
\label{gamma3}
\end{equation}%
\begin{equation}
2^{2z-1}\Gamma \left( z\right) \Gamma \left( z+1/2\right) =\sqrt{\pi }\Gamma
\left( 2z\right) .  \label{gamma4}
\end{equation}

The generating function for the Legendre polynomials and the addition
theorem for spherical harmonics give rise to the following expansion formula
\cite{Ni:Uv}, \cite{Var:Mos:Kher}
\begin{equation}
\frac{1}{\left| \mathbf{r}_{1}-\mathbf{r}_{2}\right| }=\sum_{l=0}^{\infty
}\sum_{m=-l}^{l}\frac{4\pi }{2l+1}\frac{r_{<}^{l}}{r_{>}^{l+1}}\
Y_{lm}\left( \theta _{1},\varphi _{1}\right) Y_{lm}^{\ast }\left( \theta
_{2},\varphi _{2}\right) ,  \label{a5}
\end{equation}%
where $r_{<}=\min \left( r_{1},r_{2}\right) $ and $r_{>}=\max \left(
r_{1},r_{2}\right) .$\smallskip

The Clebsch--Gordan series for the spherical harmonics is \cite{Ni:Su:Uv},
\cite{Rose}, \cite{Var:Mos:Kher}%
\begin{eqnarray}
Y_{l_{1}m_{1}}\left( \theta ,\varphi \right) \ Y_{l_{2}m_{2}}\left( \theta
,\varphi \right) &=&\sum_{l=\left| l_{1}-l_{2}\right| }^{l_{1}+l_{2}}\sqrt{%
\frac{\left( 2l_{1}+1\right) \left( 2l_{2}+1\right) }{4\pi \left(
2l+1\right) }}  \label{a6} \\
&&\times C_{l_{1}m_{1}l_{2}m_{2}}^{l,\ m_{1}+m_{2}}\ C_{l_{1}0l_{2}0}^{l,\
0}\ Y_{l,m_{1}+m_{2}}\left( \theta ,\varphi \right) ,  \notag
\end{eqnarray}%
where $C_{l_{1}m_{1}l_{2}m_{2}}^{lm}$ are the Clebsch--Gordan coefficients.
The special case $l_{2}=1$ reads \cite{Fermi}%
\begin{eqnarray}
&&-\sin \theta e^{i\varphi }\ Y_{l,\ m-1}=\sqrt{\frac{\left( l+m\right)
\left( l+m+1\right) }{\left( 2l+1\right) \left( 2l+3\right) }}\ Y_{l+1,\ m}-%
\sqrt{\frac{\left( l-m\right) \left( l-m-1\right) }{\left( 2l+1\right)
\left( 2l-1\right) }}\ Y_{l-1,\ m},  \label{a7} \\
&&\ \sin \theta e^{-i\varphi }\ Y_{l,\ m+1}=\sqrt{\frac{\left( l-m\right)
\left( l-m+1\right) }{\left( 2l+1\right) \left( 2l+3\right) }}\ Y_{l+1,\ m}-%
\sqrt{\frac{\left( l+m\right) \left( l+m+1\right) }{\left( 2l+1\right)
\left( 2l-1\right) }}\ Y_{l-1,\ m},  \label{a8} \\
&&\ \qquad \quad \cos \theta \ Y_{lm}=\sqrt{\frac{\left( l+1\right)
^{2}-m^{2}}{\left( 2l+1\right) \left( 2l+3\right) }}\ Y_{l+1,\ m}+\sqrt{%
\frac{l^{2}-m^{2}}{\left( 2l-1\right) \left( 2l+1\right) }}\ Y_{l-1,\ m},
\label{a9}
\end{eqnarray}%
where%
\begin{equation}
\sqrt{\frac{8\pi }{3}}\ Y_{1,\ \pm 1}=\mp \sin \theta e^{\pm i\varphi
},\qquad \sqrt{\frac{4\pi }{3}}\ Y_{10}=\cos \theta .  \label{a10}
\end{equation}%
These relations allow to prove (\ref{ang11}) by a direct
calculation.\smallskip

\noindent \textbf{Acknowledgment.\/} The authors thank Dieter Armbruster and
Carlos Castillo-Chavez for support. One of us (SKS) is grateful to George
Andrews, Dick Askey and Mizan Rahman for valuable comments.%
We thank the referee for corrections.

\end{document}